\documentclass[paperpaper,twocolumn,english,superscriptaddress,aps,prb,floatfix,amsfonts,amssymb]{revtex4}
\usepackage{float}
\usepackage{epsfig}
\usepackage{graphicx}% Include figure files
\usepackage{dcolumn}% Align table columns on decimal point
\usepackage{bm}% bold math
\usepackage{appendix}
\usepackage{subfigure}
\usepackage{color}
\usepackage{natbib}

\begin{document}

%%%%%%%%%%%%%%%%%%%%%%%%%%%%%%%%%%%%%%%%%%%%%%%%%%%%%%%%%%%%%%%%%%%%%%%%%%
%                               Title                                    
%%%%%%%%%%%%%%%%%%%%%%%%%%%%%%%%%%%%%%%%%%%%%%%%%%%%%%%%%%%%%%%%%%%%%%%%%%

\title{Effects of finite-range interactions on the one-electron spectral properties of TTF-TCNQ}
\author{Jos\'e M. P. Carmelo}
\affiliation{Boston University, Department of Physics, 590 Commonwealth Ave, Boston, Massachusetts 02215, USA}
\affiliation{Massachusetts Institute of Technology, Department of Physics, Cambridge, Massachusetts 02139, USA}
\affiliation{Center of Physics of University of Minho and University of Porto, P-4169-007 Oporto, Portugal}
\affiliation{Department of Physics, University of Minho, Campus Gualtar, P-4710-057 Braga, Portugal}
\author{Tilen \v{C}ade\v{z}}
\affiliation{Center of Physics of University of Minho and University of Porto, P-4169-007 Oporto, Portugal}
\affiliation{CAS Key Laboratory of Theoretical Physics, Institute of Theoretical Physics,
Chinese Academy of Sciences, Beijing 100190, China}
\affiliation{Beijing Computational Science Research Center, Beijing 100193, China}
\author{David K. Campbell}
\affiliation{Boston University, Department of Physics, 590 Commonwealth Ave, Boston, Massachusetts 02215, USA}
\author{Michael Sing}
\affiliation{Experimentelle Physik 4, Universit\"at W\"urzburg, Am Hubland, D-97074 W\"urzburg, Germany}
\author{Ralph Claessen}
\affiliation{Physikalisches Institut and W\"urzburg-Dresden Cluster of Excellence Complexity and Topology in Quantum Matter, 
Universitat W\"urzburg, D-97074 W\"urzburg, Germany}

\date{28 July 2019}
%\date{\today}

%%%%%%%%%%%%%%%%%%%%%%%%%%%%%%%%%%%%%%%%%%%%%%%%%%%%%%%%%%%%%%%%%%%%%%%%%%
%                              abstract                                  
%%%%%%%%%%%%%%%%%%%%%%%%%%%%%%%%%%%%%%%%%%%%%%%%%%%%%%%%%%%%%%%%%%%%%%%%%%

\begin{abstract}
The electronic dispersions of the  quasi-one-dimensional organic conductor TTF-TCNQ are studied by angle-resolved 
photoelectron spectroscopy (ARPES) with higher angular resolution and accordingly smaller step width than in previous studies. 
Our experimental results suggest that a refinement of the single-band 1D Hubbard model that includes finite-range interactions 
is needed to explain these photoemission data. To account for the effects of these finite-range interactions we employ a mobile 
quantum impurity scheme that  describes the scattering of fractionalized particles at energies above the standard Tomonaga-Luttinger 
liquid limit. Our theoretical predictions agree quantitatively with the location in the $(k,\omega)$ plane of the experimentally 
observed ARPES structures at these higher energies. The nonperturbative microscopic mechanisms that control the spectral 
properties are found to simplify in terms of the exotic scattering of the charge fractionalized particles.
We find that the scattering occurs in the unitary limit of (minus) infinite scattering length, which limit occurs within 
neutron-neutron interactions in shells of neutron stars and in the scattering of ultracold atoms but not in perturbative electronic 
condensed-matter systems. Our results provide important physical information on the exotic processes involved in the finite-range 
electron interactions that control the high-energy spectral properties of TTF-TCNQ. Our results also apply to
a wider class of 1D and quasi-1D materials and systems that are of theoretical and potential technological interest.  
\end{abstract}


\maketitle
%%%%%%%%%%%%%%%%%%%%%%%%%%%%%%%%%%%%%%%%%%%%%%%%%%%%%%%%%%%%%%%%%%%%%%%%%%
%                              body of paper                             
%%%%%%%%%%%%%%%%%%%%%%%%%%%%%%%%%%%%%%%%%%%%%%%%%%%%%%%%%%%%%%%%%%%%%%%%%%

\section{Introduction}
\label{SECI}

The organic quasi-one-dimensional (1D) conductor tetrathiafulvalene-tetracyanoquinodimethane 
(TTF-TCNQ) was the first material for which angle-resolved photoemission spectroscopy (ARPES) 
was able to identify charge-spin separation on the energy scale of the band width \cite{Claessen_02,Sing_03}. 
Its stacks of the TCNQ and TTF planar molecules are effectively doped and 
become conducting by charge transfer from TTF to TCNQ. The single-band 1D Hubbard model
provides a preliminary description of the experimental data \cite{Sing_03}. However, it does not
explain some of the TTF-TCNQ properties consistently \cite{Sing_07}.

In this paper we study the electronic structure of TTF-TCNQ using ARPES with {\it 
higher angle resolution} and accordingly smaller step width than in previous 
studies \cite{Claessen_02,Sing_03}. We show that inconsistencies and 
unrealistic parameter choices in previous attempts to describe the experimental 
dispersions can be resolved by including finite-range interactions 
in a single-band 1D Hubbard model description of each of the TTF and TCNQ stacks.

We employ a mobile quantum impurity model (MQIM) \cite{Imambekov_09,Imambekov_12,Carmelo_19} 
to describe the microscopic mechanisms underlying the experimental ARPES data at energies above the 
standard Tomonaga-Luttinger liquid (TTL) limit \cite{Blumenstein_11,Voit_95,Tomonaga_50,Luttinger_63}. 
Our theoretical predictions agree quantitatively with the location in the $(k,\omega)$ plane of the experimentally 
observed ARPES structures at such energies. The nonperturbative microscopic mechanisms that control the TTF-TCNQ 
spectral properties are found to simplify in terms of the exotic scattering of the charge fractionalized particles.

We find that the scattering occurs in the unitary limit of (minus) infinite scattering length, which limit 
plays an important role in the physics of several well-known systems, including the dilute neutron 
matter in shells of neutron stars \cite{Dean_03} and the atomic scattering in systems of trapped cold atoms 
\cite{Zwerger_12,Horikoshi_17}, but not in perturbative electronic 
condensed-matter systems. Our results thus provide important physical information on the exotic processes involved in the finite-range 
electron interactions that control the high-energy spectral properties of TTF-TCNQ. Our results also apply 
to a wider class of 1D and quasi-1D materials and systems that are of theoretical and potential technological 
interest \cite{Blumenstein_11}.

A complication relative to the ARPES data of the 1D metallic states in simpler systems 
(such as the line defects of MoSe$_2$ \cite{Ma_17,Cadez_19} 
and in those of Bi/InSb(001) \cite{Carmelo_19}) is that TTF-TCNQ contains two stacks of 
molecules. Metallicity arises through charge transfer of about $0.59$ electrons per TTF to each TCNQ molecule. It follows that the electronic density of the 
TCNQ stack of molecules is $n_e^Q = 0.59$ and that of the TTF stack of molecules is $n_e^F = 2 - 0.59 = 1.41$. 
For the theoretical study of the one-electron spectral features associated with the TTF stack of molecules, we rely on a particle-hole 
symmetry. This allows us to determine the spectral function for one-electron {\it removal} at density $2 - 0.59 = 1.41$ from the corresponding 
spectral function for one-electron {\it addition} at density $2 - 1.41 = 0.59$.

The MQIM theoretical approach used in our study involves a uniquely defined transformation
from 1D Hubbard model and its pseudofermion dynamical theory (PDT) \cite{Carmelo_05,Carmelo_18}. 
Such a transformation adds to that model finite-range interactions. The corresponding model  
accounts for the effects of the {\it higher-order} (HO) terms in the 
effective-range expansion \cite{Bethe_49,Blatt_49} of the fractionalized particles
charge-charge phase shifts; hence we call the model MQIM-HO. The  charge-spin separation 
in the MQIM \cite{Imambekov_09,Imambekov_12,Carmelo_19}, 
implies that the part of the one-electron spectral-function spectrum associated with the spin degrees
of freedom remains invariant under the transformation.
Beyond the studies of Ref. \onlinecite{Carmelo_19}, useful related representations for the Hamiltonian of the lattice 
electronic model with finite-range interactions in a relevant one-electron subspace 
are used to access different aspects of the microscopic mechanisms that control the spectral properties.

For low energies, 1D correlated electronic metallic systems show universal properties captured by the 
TLL description\cite{Blumenstein_11,Voit_95,Tomonaga_50,Luttinger_63}.  
An important low-energy property of such systems is the universal power-law scaling of the spectral 
intensity $I (\omega,T)$, such that $I (\omega,0)\propto\vert\omega\vert^{\alpha}$.
Here the exponent $\alpha$ controls the suppression of the density of states (SDS) and $\omega$ is a small excitation 
energy near the ground-state level. The value of the SDS exponent, $\alpha = (1-K_{\rho})^2/(4K_{\rho})$, is determined by 
that of the TLL charge parameter $K_{\rho}$ \cite{Blumenstein_11,Voit_95,Schulz_90}.

In the case of TTF-TCNQ, the exponent $\alpha$ is very difficult to be accessed  experimentally\cite{Sing_03}.
For the metallic states of other 1D and quasi-1D electronic systems, the experimental values of  $\alpha$ lie in the range $0.5-0.8$
\cite{Claessen_02,Blumenstein_11,Voit_95,Ma_17,Schulz_90,Kim_06,Schoenhammer_93,Ohtsubo_15,Ohtsubo_17}. 
For the TTF-TCNQ system, we predict that  $\alpha = \alpha_C =0.50$ and $\alpha = \alpha_F =0.74$ for the TCNQ and TTF and stacks of molecules, respectively.
If the interaction between the two stacks is weak, the leading contribution would then be $I (\omega,0)\propto\vert\omega\vert^{\alpha}$ 
where $\alpha = \alpha_C = 0.50$. If otherwise, the SDS exponent $\alpha$ will have a single value in the range  $\in [0.50,0.74]$ and its
expression will involve both $\alpha_C$ and $\alpha_F$. Those values could not be reached within the 
1D Hubbard model whose TLL charge parameter $K_{\rho}$ is larger than $1/2$ thus requiring $\alpha$ to be smaller than $1/8$. 

Our results explain the {\it seeming} lower degree of charge-spin splitting of the TTF derived part of the 
ARPES data, although the electronic states of the (i) TTF stacks are more strongly correlated than those of 
the (ii) TCNQ stacks. This is confirmed by our prediction for their TLL charge parameters 
(i) $K_{\rho} = K_{\rho}^F = 0.21$ and (ii) $K_{\rho} = K_{\rho}^C = 0.27$,
respectively, and known experimental properties \cite{Sing_03,Sing_07,Takahashi_84}.

In order to achieve a preliminary description of the TCNQ experimental data within the 1D Hubbard model,
a renormalization of the hopping integral $t= 0.40$ eV at the surface by a factor of $2$ with respect to the bulk value 
from density-functional theory or estimates from bulk-sensitive measurements had to be assumed \cite{Sing_03}.
However, there is some evidence that the observed transfer of spectral weight at $k_F$ over the entire conduction 
band width with increasing temperatures cannot be reconciled within the use of $t= 0.40$ eV \cite{Sing_07}. 
Our results confirm that these data can be consistently interpreted incorporating finite-range interactions, in addition to the 
onsite Coulomb energy $U$.

In this paper we employ units of $\hbar =1$ and $k_B =1$. 
The use of units of lattice spacing $a_0=1$ is limited to Secs. IV and V.  
In Sec. \ref{SECII} we provide the experimental and technical details of the high resolution ARPES measurements 
performed for TTF-TCNQ. The model Hamiltonian, the $\xi_c\rightarrow {\tilde{\xi}}_c$ transformation,
and the universal properties of the potential $V_c (x)$ associated with the fractionalized particles
charge-charge interactions induced by the electronic potential $V_e (r)$ for 
interaction distances $r>0$ are the topics addressed in Sec. \ref{SECIII}. In Sec. \ref{SECIV} several
useful representations for the Hamiltonian of the lattice electronic model with finite-range interactions 
used in our studies and the corresponding one-electron {\it singularities} subspace are introduced
and discussed. The line shape near the $(k,\omega)$-plane singularities of the model
one-electron removal and addition spectral functions is the subject of Sec. \ref{SECV}.
In Sec. \ref{SECVI} the ARPES data are presented. In addition, 
we determine the parameter values such that the experimentally observed high-energy ARPES structures
in the $(k,\omega)$ plane  agree with those deduced from the 
theoretical spectral-function singularities. Finally, we summarize our results and present concluding remarks in Sec. \ref{SECVII}. 
Three Appendices provide useful information needed for the studies of this paper.

\section{Experimental details}
\label{SECII}

The ARPES data were recorded in our laboratory in Experimentelle Physik 4, 
University of W\"urzburg, with a SPECS Phoibos 100 
electron spectrometer equipped with a two-dimensional charge-coupled device 
(CCD) detector. Photoelectrons were excited by using nonmonochromatized He~I 
radiation from a duoplasmatron discharge lamp (SPECS UVS 300) with the 
operation conditions optimized for maximal flux of $21.22$\,eV photons 
(He~\textsc{I}$_\alpha$) and at the same time negligible contributions from the 
satellite lines at higher energies (He~\textsc{I}$_\beta$, 
He~\textsc{I}$_\gamma$). The energy resolution was set to 60\,meV, while 
the angle resolution along the analyzer slit, i.e., parallel to the 1D chains 
of TTF-TCNQ, amounted to $<0.2^\circ$. Note that the energy widths of the 
spectral features in the ARPES spectra are not resolution-limited. The Fermi 
energy was calibrated to the Fermi cut-off of a freshly 
sputtered Ag foil. 

The measurements were performed at 60\,K, i.e., in the 
metallic phase above the charge-density wave transition at 54\,K. The 
single crystals were grown by diffusion in pure acetonitrile and fresh, clean 
surfaces parallel to the $(\textbf{a},\textbf{b})$ plane. They were exposed by 
\textit{in situ} cleavage at the measuring temperature and a pressure $<3 
\times 10^{-10}$\,mbar. Cleavage was accomplished by knocking-off a post glued 
on the top of the $2 \times 5 \times 0.2$\,mm$^3$ TTF-TCNQ platelets with the 
chain axis $\textbf{b}$ oriented along the long crystal dimension. Special 
attention was paid to collect data on short enough time scales so as not to 
spoil the spectra owing to photoinduced sample degradation \cite{Sing_03}. 

\section{The model, the $\xi_c\rightarrow {\tilde{\xi}}_c$ transformation, and the
induced MQIM-HO potential $V_c (x)$}
\label{SECIII}

\subsection{The model Hamiltonian and the $\xi_c\rightarrow {\tilde{\xi}}_c$ transformation}
\label{SECIIIA}

The 1D model Hamiltonian associated with the MQIM-HO for electronic density $n_e\in ]0,1[$ is given by \cite{Carmelo_19}
\begin{eqnarray}
{\hat{H}} & = & t\,\hat{T} + \hat{V}\hspace{0.20cm}{\rm where}
\nonumber \\
\hat{T} & = & - \sum_{\sigma=\uparrow,\downarrow}\sum_{j=1}^{L}\left(c_{j,\sigma}^{\dag}\,
c_{j+1,\sigma} + c_{j+1,\sigma}^{\dag}\,c_{j,\sigma}\right)
\nonumber \\
\hat{V} & = & U\sum_{j=1}^{L}\hat{\rho}_{j,\uparrow}\hat{\rho}_{j,\downarrow} 
\nonumber \\
& + & \sum_{r=1}^{L/2-1}V_e (r)
\sum_{\sigma=\uparrow,\downarrow}\sum_{\sigma'=\uparrow,\downarrow}\sum_{j=1}^{L}\hat{n}_{j,\sigma}\hat{n}_{j+r,\sigma'} \, .
\label{equ1}
\end{eqnarray}
Here $t$ is the transfer integral, $\hat{\rho}_{j,\sigma} = \left(\hat{n}_{j,\sigma} - {1\over 2}\right)$, $\hat{n}_{j,\sigma} = c_{j,\sigma}^{\dag}\,c_{j,\sigma}$,
$V_e (r) = U\,F_e (r)/r$ for $r>0$ where $U$ is the interaction, and $F_e (r)$ is a continuous 
screening function. It is such that $F_e (r)\leq 1/4$. At large $r$ it vanishes as some inverse power of $r$
whose exponent is larger than one, $\lim_{r\rightarrow\infty}F_e (r)=0$. 
Different values of $t$ and $U$ and potentials $V_e (r)$ with different $r$ dependence
are found in Secs. \ref{SECVI} and \ref{SECVII} to describe the stacks of the TCNQ and 
TTF molecules, respectively.

At $F_e (r)=0$ the interaction $U$ is simply an onsite repulsion and
the model in Eq. (\ref{equ1}) becomes the 1D Hubbard model. The charge parameter 
that for the latter model is denoted here by $\xi_c$ and for the model, Eq. (\ref{equ1}),
by ${\tilde{\xi}}_c$, plays an important role in our study. It is directly related 
to the TLL charge parameter \cite{Blumenstein_11,Voit_95} as 
$K_{\rho}^0 = \xi_c^2/2$ and $K_{\rho} = {\tilde{\xi}}_c^2/2$, respectively.
For the 1D Hubbard model its dependence on $u=U/4t$ and $n_e \in ]0,1[$ 
is defined by Eq. (\ref{equA12}) of Appendix \ref{APPA} and is such that $\xi_c \in ]1,\sqrt{2}[$. Upon 
increasing $u$, it decreases from $\xi_c=\sqrt{2}$ for $u\rightarrow 0$ and reaches its smallest 
value $\xi_c=1$ in the $u\rightarrow\infty$ limit. 

Within the MQIM-HO \cite{Carmelo_19}, there is a {\it $\xi_c\rightarrow {\tilde{\xi}}_c$ transformation} 
for $n_e\in ]0,1[$ such that ${\tilde{\xi}}_c\in ]1/2,1[\,;]1,\xi_c]$ is the above renormalized charge 
parameter of the model, Eq. (\ref{equ1}). This transformation maps the 1D Hubbard model onto that model, 
upon gently turning on $F_e (r)$. Consistently, $\lim_{{\tilde{\xi}}_c\rightarrow\xi_c}F_e (r)\rightarrow 0$ for 
$r\in [0,\infty]$. In this paper we use the following ${\tilde{\xi}}_c$-related deviation parameter,
\begin{eqnarray}
{\tilde{\delta}}_c & \equiv & (\xi_c - {\tilde{\xi}}_c) \,\in ]0,{\tilde{\delta}}_c^{(1)}[\,;]{\tilde{\delta}}_c^{(1)},{\tilde{\delta}}_c^{({1\over 2})}[ 
\hspace{0.20cm}{\rm where}
\nonumber \\
{\tilde{\delta}}_c^{(1)} & = & (\xi_c - 1) \in \,]0,(\sqrt{2}-1)[\hspace{0.20cm}{\rm and}
\nonumber \\
{\tilde{\delta}}_c^{({1\over 2})} & = & (\xi_c - 1/2) \in \,]1/2,(\sqrt{2}-1/2)[ \, .
\label{equ2}
\end{eqnarray}
The advantage of using this parameter is that $\lim_{{\tilde{\delta}}_c\rightarrow 0}F_e (r)\rightarrow 0$, as
it reads ${\tilde{\delta}}_c=0$ at the boundary value ${\tilde{\xi}}_c = \xi_c$ that refers to the bare
1D Hubbard model. Its value increases upon decreasing the renormalized charge parameter ${\tilde{\xi}}_c$
under the $\xi_c\rightarrow {\tilde{\xi}}_c$ transformation. 
Hence in this paper the latter model is called {\it ${\tilde{\delta}}_c =0$ bare model}.

The MQIM-HO expressions near the singularities  
of the one-electron spectral function of the model, Eq. (\ref{equ1}),
\begin{eqnarray}
B_{-1} (k,\,\omega) & = & \sum_{\sigma}\sum_{\nu^-}\vert\langle\nu^-\vert\, c_{k,\sigma}\vert \,GS\rangle\vert^2 
\nonumber \\
& \times & \delta (\omega + (E_{\nu^-}^{N_e -1}-E_{GS}^{N_e})) \hspace{0.5cm} \omega \leq 0 
\nonumber \\
B_{+1} (k,\,\omega) & = & \sum_{\sigma}\sum_{\nu^+}\vert\langle\nu^+\vert\, c^{\dagger}_{k,\sigma} \vert\,GS\rangle\vert^2
\nonumber \\
& \times & \delta (\omega - (E_{\nu^+}^{N_e +1}-E_{GS}^{N_e}))  \hspace{0.5cm} \omega \geq 0 \, ,
\label{equ3}
\end{eqnarray}
which are studied below in Sec. \ref{SECV}, are used in this paper to describe the $(k,\omega)$-plane location 
of the related TTF and TCNQ stacks of molecules ARPES spectral features. 
Here $\gamma=-1$ (and $\gamma=+1$) for one-electron removal (and addition)
in $B_{\gamma} (k,\,\omega)$, $c_{k,\sigma}$ and $c^{\dagger}_{k,\sigma}$ are electron
annihilation and creation operators, respectively, of momentum $k$ and spin
projection $\sigma$, $\vert GS\rangle$ denotes the
initial $N_e$-electron ground state of energy $E_{GS}^{N_e}$, the $\nu^-$ and $\nu^+$
summations run over the $N_e -1$ and $N_e +1$-electron excited 
energy eigenstates, respectively, and $E_{\nu^-}^{N_e -1}$ and 
$E_{\nu^+}^{N_e +1}$ are the corresponding energies. As noted in Sec. \ref{SECI},
the one-electron removal spectral function for 
the TTF electronic density $n_e^F = 1.41$ is accessed through the one-electron removal 
spectral function at $n_e = 2 - n_e^F = 0.59$.

For noncommensurate electronic densities, no $T=0$ broken-symmetry transition  
{\it at} ${\tilde{\xi}}_c = \sqrt{2}\,n_e$ is expected \cite{Hohenadler_12}
for the model, Eq. (\ref{equ1}). For some long-range potentials, lattice fermions have a metallic ground state for all 
${\tilde{\xi}}_c$ values \cite{Fano_99}. Independent of the nature of the ground state of the model, 
Eq. (\ref{equ1}), which is determined by the unknown precise form of $V_e (r)$, we use its lowest 
metallic energy eigenstate as a reference ``ground state''. Our goal is to employ a model Hamiltonian of
general form, Eq. (\ref{equ1}), but with specific transfer integral $t$ and interaction $U$ values and potential $V_e (r)$ 
to describe the metallic one-electron spectral properties at $T=60$ K of the TTF and TCNQ stacks of molecules, respectively. 

Our theoretical scheme uses a {\it rotated-electron representation} for the model Hamiltonian, Eq. (\ref{equ1}).
It is a suitable representation for the description of the separation of the degrees of freedom
at all MQIM energy scales. Such rotated electrons are related to the electrons
by a unitary transformation. As described in Appendices \ref{APPB} and \ref{APPC},
in the one-electron subspace, the fractionalization of the 
rotated-electron occupancy configurations leads to a representation in terms of
charge and spin fractionalized particles. Within the MQIM-HO, they are called
$c$ (charge) and $s$ (spin) particles, respectively \cite{Carmelo_19}. The corresponding $c$ and $s$ bands
include $c$ and $s$ holes, respectively, which also play an active role in the physics. 

The suitability of the rotated-electron related $c$ and $s$ 
particle representations used in this paper is justified by the corresponding $c$ and $s$ band occupancy configurations 
generating states that in an important subspace of the one-electron subspace defined below in 
Sec. \ref{SECIVA} are either exact energy eigenstates of the Hamiltonian, Eq. (\ref{equ1}),
or states with quantum overlap primarily with one of its energy eigenstates.

\subsection{Properties of the potential $V_c (x)$ induced by $V_e (r)$ that
controls the $\xi_c\rightarrow {\tilde{\xi}}_c$ transformation}
\label{SECIIIB}

The 1D charge-spin separation that occurs at all energy scales in the MQIM
class of systems \cite{Imambekov_09,Imambekov_12,Carmelo_19}, leads to
the electronic potential $V_e (r)$ in Eq. (\ref{equ1}), giving rise to an attractive 
potential $V_c (x)$. It is associated with fractionalized particles charge-charge interactions \cite{Carmelo_19}.
In the case of one-electron removal excitations, it is associated 
with the interaction between $c$ particles and a $c$ (hole) mobile 
scattering center at distance $x$. In this paper we find
that for one-electron addition excitations it is rather associated 
with the interaction between $c$ holes and a $c$ (particle) 
mobile scattering center at distance $x$. Both for one-electron removal 
and addition excitations such an interaction is attractive and thus associated with 
a negative scattering length. The  $c$ 
mobile scattering center involved in the interactions is created by such excitations. 
The $c$ particle and $c$ hole scatterers preexisted in the ground state.

The use of the expressions for the $c$ (charge-charge) phase shifts given in Eq. (\ref{equA4}) of
Appendix \ref{APPA} for both one-electron removal and addition in the corresponding effective-range 
expansion \cite{Carmelo_19,Bethe_49,Blatt_49} provided in Eq. (\ref{equA3}) of that Appendix, gives the negative 
renormalized and bare scattering lengths ${\tilde{a}}$ and $a$, respectively, 
Eq. (\ref{equA5}) of Appendix \ref{APPA}. This confirms that their negativity applies both to one-electron 
removal and addition excitations and that $V_c (x)$ is attractive for both of them.

The values as given in Eq. (\ref{equA5}) of Appendix \ref{APPA},
${\tilde{a}} = - \infty$ and $a = - \infty$ both for one-electron removal 
and addition are known as the unitary limit \cite{Carmelo_19,Zwerger_12,Horikoshi_17}.
The MQIM-HO is valid for that limit whose existence implies that ${\tilde{\delta}}_c\neq {\tilde{\delta}}_c^{(1)}$,
${\tilde{\delta}}_c\neq {\tilde{\delta}}_c^{({1\over 2})}$, $\xi_c \neq 1$, and thus that the relative momentum 
$k_r$ in the effective-range expansion, Eq. (\ref{equA3}) of Appendix \ref{APPA}, obeys the inequality
$\vert k_r\vert\ll \tan (\pi\,n_e)/(4u)$ \cite{Carmelo_19}. This excludes electronic densities very near $n_e=0$ and $n_e=1$
for all $u$ values and excludes large $u$ values for the remaining electronic densities.

The properties of the potential $V_c (x)$ are determined by those of the electronic
potential $V_e (r)$ in Eq. (\ref{equ1}). For the class of lattice electronic systems with finite-range
interactions to which the MQIM-HO applies\cite{Carmelo_19}, the general properties of $V_c (x)$ play an important role. 
We discuss them briefly here.  $V_c (x)$ is negative for $x>x_0$, where $x_0$ is a nonuniversal distance
that either vanishes or is much smaller than the lattice spacing $a_0$.
The scattering energy of the residual interactions of the $c$ particles or $c$ holes with the $c$ mobile scattering center is 
smaller than the depth $\vert V_c (x_1)\vert = -V_c (x_1)$ of the potential $V_c (x)$ well. Here $x_1$ is a small nonuniversal 
potential-dependent value of $x$ such that $x_0<x_1<a_0$ at which $\partial V_c (x)/\partial x =0$ and $-V_c (x)$ 
reaches its maximum value\cite{Carmelo_19}.

For small distances, the potential $V_c (x)$ has a non universal form, 
determined by the specific small-$r$ form of $V_e (r)$ itself. A universal
property \cite{Carmelo_19} is its behavior at  {\it large $x$}, for which it vanishes as $V_c^{\rm asy} (x) = - C_c/(x/2r_l)^l$,
Eq. (\ref{equA7}) of Appendix \ref{APPA}. Here $1/C_c =(2r_l)^2\mu$, with
$\mu$ a nonuniversal reduced mass, and $l$ an integer determined by the large-$r$ behavior of $V_e (r)$.
The effective range of the interactions associated with $V_c (x)$ considered below converges only if 
$l>5$ \cite{Carmelo_19,Flambaum_99,Burke_11}. The $l$ dependence of the length scale $2r_l$
is provided in Eq. (\ref{equA8}) of  Appendix \ref{APPA} for the interval ${\tilde{\delta}}_c > {\tilde{\delta}}_c^{(1)}$
of interest for TTF-TCNQ. It reads $5.95047\,a_0$ at the value $l=6$ at which $2r_l$ is twice the van der Waals length,
reaches a maximum $6.48960\,a_0$ at $l=10$, and decreases to $4.93480\,a_0$ as $l\rightarrow\infty$. 

Despite its nonuniversal form (except for large $x$), $V_c (x)$ obeys
two universal sum rules. In the interval $x\in [x_0,\infty]$ the positive ``momentum'' $\sqrt{2\mu (-V_c (x))}$ obeys a first sum rule,
$\Phi= \int_{x_0}^{x_2}dx\sqrt{2\mu (-V_c (x))}$, Eq. (\ref{equA9}) of Appendix \ref{APPA}. Here $\Phi$ is a zero-energy phase,
$x_2$ is a length scale defined in that equation, and the relative fluctuation $\Delta a/{\tilde{a}}$ in the expression 
$\tan (\Phi) = - (\Delta a/{\tilde{a}})\cot (\pi/[l-2])$ also given in that equation
involves the difference $\Delta a = a - {\tilde{a}}$. Although both  $a=-\infty$ and ${\tilde{a}}=-\infty$, their ratio ${\tilde{a}}/a$ 
is finite, as given in Eq. (\ref{equA6}) of Appendix  \ref{APPA}. 
The corresponding finite relative fluctuation $\Delta a/{\tilde{a}}$ controls the effects of the finite-range 
interactions \cite{Carmelo_19}. These are stronger for the range 
${\tilde{\delta}}_c\in [{\tilde{\delta}}_c^{\oslash},{\tilde{\delta}}_c^{({1\over 2})}[$ where 
${\tilde{\delta}}_c^{\oslash} = \xi_c(1 - 1/\xi_c^2) \in ]0,1/\sqrt{2}[$.

As justified below in Sec. \ref{SECIV}, the second sum rule reads,
$\int_{0}^{\infty}(-V_c (x)) = {\pi\over 4}\,[(\xi_c^4 - {\tilde{\xi}}_c^4)/{\tilde{\xi}}_c^4]\,{\breve{v}}_{Fc}$.
The relation of the velocity ${\breve{v}}_{Fc} = {\breve{v}}_c (2k_F)$ appearing here
to the ${\tilde{\delta}}_c =0$ bare $c$ band Fermi velocity $v_{Fc} = v_c (2k_F)$ is given 
by Eqs. (\ref{equC31}) and (\ref{equC35}) of Appendix \ref{APPC}.

The effective range of the interactions ($R_{\rm eff}$) of the $c$ particles and $c$ holes with
the $c$ mobile scattering center plays an important role in the MQIM-HO physics
\cite{Carmelo_19,Flambaum_99}. For ${\tilde{\delta}}_c > {\tilde{\delta}}_c^{(1)}$,
$R_{\rm eff} = a_0 (1 - c_1\,({\tilde{a}}/a) + c_2\,({\tilde{a}}/a)^2)$,
Eq. (\ref{equA10}) of Appendix \ref{APPA}. Here $c_1$ and $c_2$ only
depend on the integer $l>5$, as given in Eq. (\ref{equA11}) of that Appendix.
They decrease from $c_1=c_2=2$ at $l=6$ to $c_1=1$ and $c_2=1/3$ 
for $l\rightarrow\infty$. The effective range $R_{\rm eff}$ appears in the expressions of the spectral-function exponents
given below in Sec. \ref{SECVA}. This occurs through the charge-charge phase shift $2\pi{\tilde{\Phi}}_{c,c} (\pm 2k_F,q)$, as shown by
Eq. (\ref{equA2}) of Appendix \ref{APPA}. Its value $R_{\rm eff}=\infty$ for ${\tilde{\delta}}_c={\tilde{\delta}}_c^{({1\over 2})}$ is excluded, as 
${\tilde{\delta}}_c={\tilde{\delta}}_c^{({1\over 2})}$ is outside the range of validity of the unitary limit \cite{Carmelo_19}.

The intervals ${\tilde{\delta}}_c\in]0,{\tilde{\delta}}_c^{(1)}[$ and ${\tilde{\delta}}_c\in]{\tilde{\delta}}_c^{(1)},{\tilde{\delta}}_c^{({1\over 2})}[$ for 
which the SDS exponent is such that $\alpha <1/8$ and $\alpha >1/8$, respectively, refer to two {\it qualitatively 
different} physical problems. The ${\tilde{\xi}}_c$ value in the $\xi_c\rightarrow {\tilde{\xi}}_c$ transformation
is uniquely defined for {\it each} of the two such intervals solely by the integer quantum number $l>5$ in the potential $V_c  (x)$ 
large-$x$ expression, $\tan (\Phi)$, and the initial value of $\xi_c = \xi_c (n_e,u)$, Eqs. (\ref{equA7}), (\ref{equA9}), and 
(\ref{equA12}) of Appendix \ref{APPA}, respectively. Specifically \cite{Carmelo_19},
\begin{eqnarray}
{\tilde{\xi}}_c & = & \eta_c (\xi_c,\Phi,l)\left(1 + \sqrt{1 - {1\over \eta_c^2 (\xi_c,\Phi,l)}}\right)
\hspace{0.075cm}{\rm for}\hspace{0.20cm}{\tilde{\delta}}_c\in [0,{\tilde{\delta}}_c^{(1)}[
\nonumber \\
& = & \eta_c (\xi_c,\Phi,l) \left(1 - \sqrt{1 - {1\over \eta_c^2 (\xi_c,\Phi,l)}}\right) 
\nonumber \\
&& {\rm for}\hspace{0.20cm}{\tilde{\delta}}_c\in]{\tilde{\delta}}_c^{(1)},{\tilde{\delta}}_c^{({1\over 2})}[ \, ,
\label{equ4}
\end{eqnarray}
where,
\begin{equation}
\eta_c (\xi_c,\Phi,l) = 1 + {1\over 2\pi}\arctan\left({\tan \left({(\xi_c -1)^2\pi\over \xi_c}\right)\over
1 + \tan \left({\pi\over l-2}\right) \tan (\Phi) }\right)  \, .
\label{equ5}
\end{equation}

\section{Useful representations of the model Hamiltonian}
\label{SECIV}

The universal properties of the potential $V_c (x)$ induced by $V_{e} (r)$ 
(through the related rotated-electron potential $V_{re} (r)$ considered in Appendix \ref{APPC})
reported above have been introduced and used in the studies
of Ref. \onlinecite{Carmelo_19}. However, the expressions of the Hamiltonian, Eq. (\ref{equ1}), 
in which the Fourier transform of such potentials emerges, have neither been 
given nor studied in that reference. The derivation of such expressions from the Hamiltonian written in 
terms of rotated-electron operators provides important physical information. It is needed for the further 
clarification of the microscopic mechanisms that describe our high-resolution ARPES data of TTF-TCNQ.

Two other problems also require additional information on the rotated-electron representation and related fractionalized
particles representations beyond that provided in Refs. \onlinecite{Ma_17,Cadez_19,Carmelo_19}:
1) The extension of the MQIM-HO to one-electron addition excitations and 2) accounting
for the renormalization of the line shape near the corresponding one-electron spectral-function
branch lines and spectral features called boundary lines considered below in Sec. \ref{SECV}.

We address all these issues in the ensuing sections. These results involve several related $c$ particle representations
that are associated with the Fourier transform of corresponding potentials.
These representations are developed from the rotated-electron representation. 
Due to the charge-spin separation at all energy scales of the MQIM, the $s$ (spin) particle terms of the Hamiltonian in all 
such representations remain invariant under the $\xi_c\rightarrow{\tilde{\xi}}_c$ transformation.

The corresponding developments of the MQIM-HO require accounting for properties of the rotated-electron 
representation and corresponding $c$ and $s$ particle representations in the ${\tilde{\delta}}_c =0$ bare limit
that {\it have not} been studied elsewhere. Such properties are reported and briefly discussed in 
Appendix \ref{APPB}. In the case of the ${\tilde{\delta}}_c =0$ bare model, one can extract from the Bethe 
ansatz solution all quantities of its Hamiltonian expression in the $c$ and $s$ particle representation
\cite{Carmelo_18,Carmelo_17} that diagonalizes it in the one-electron subspace, Eq. (\ref{equB14}) of Appendix \ref{APPB}.

In that Appendix we show how to derive that Hamiltonian expression from
the corresponding rotated-electron expression with an infinite number of
Hamiltonian terms. The rotated-electron operators are related to the electron operators by the unitary 
transformation, Eq. (\ref{equB1}) of Appendix \ref{APPB}. The corresponding unitary operator 
$\hat{U} = e^{\hat{S}}$ is uniquely defined in Ref. \onlinecite{Carmelo_17}
in terms of the $4^L\times 4^L$ matrix elements between all the model energy eigenstates.

The relation between the diagonal expression of the Hamiltonian, Eq. (\ref{equ1}), 
at ${\tilde{\delta}}_c =0$ in the one-electron subspace, Eq. (\ref{equB14}) of Appendix \ref{APPB},
to the same Hamiltonian expressed in an alternative $c$ particle representation where
it is nondiagonal, Eq. (\ref{equB18}) of that Appendix, provides essential information.
In the latter Hamiltonian expression, ${\cal{V}}_c^1 (k)$ is the Fourier transform of an effective potential 
$V_c^1 (x)$. It is associated with the interaction of the $c$ particles/holes scatterers with the
$c$ mobile scattering center. It controls the dependence of the one-electron matrix elements on the 
phase shifts of such scatterers. 

The information on the relation between the rotated-electron representation and the
$c$ and $s$ particle representations provided in Appendix \ref{APPB} for ${\tilde{\delta}}_c =0$
is a first needed step for their use for the ${\tilde{\delta}}_c >0$ model, Eq. (\ref{equ1}).
As ${\tilde{\delta}}_c$ is smoothly turned on, the potential 
$V_c^1 (x)$ evolves into a related potential ${\tilde{V}}_c^1 (x)$. The part of the latter potential that
accounts for most of the renormalization of $V_c^1 (x)$ by the finite-range interactions 
is the potential $V_c (x)$ whose universal properties were reported in Sec. \ref{SECIIIB}.

In this section, we discuss the partial generalization of the use of the ${\tilde{\delta}}_c =0$ operator 
representations to the model in Eq. (\ref{equ1}) for 
${\tilde{\delta}}_c\in [0,{\tilde{\delta}}_c^{(1)}[\,;]{\tilde{\delta}}_c^{(1)},{\tilde{\delta}}_c^{({1\over 2})}[$. 
We examine the corresponding technicalities in Appendix \ref{APPC}.
A qualitative difference between the ${\tilde{\delta}}_c =0$ and ${\tilde{\delta}}_c >0$ quantum problems
is that their Hamiltonians in the one-electron subspace when expressed in terms of $c$ and $s$ particle operators
can and cannot be diagonalized, respectively. 

Indeed, except in the ${\tilde{\delta}}_c =0$ bare limit, for the model,
Eq. (\ref{equ1}), the rotated-electron occupancy configurations and corresponding 
$c$ and $s$ particle occupancy configurations that generate the states of the representations used in this 
paper are not in general exact energy eigenstates.
Consistent with the properties of the general MQIM \cite{Imambekov_09,Imambekov_12,Carmelo_19},
they become exact energy eigenstates when generated by processes 
that lead to the line shape near some types of singular spectral features: specifically, near
the low-energy TLL spectral features and in the vicinity of high-energy spin spectral features called $s$ and $s'$ branch lines.
In the case of the latter lines, this follows from conservation laws and for ${\tilde{\delta}}_c >0$
applies to momentum intervals for which the exponent in the spectral-function expression near them 
is negative. Such laws are due to these lines coinciding with edges of support 
of the one-electron spectral function. These lines separate $(k,\omega)$-plane regions without and with finite spectral weight. 

On the other hand, in the present representation, 
some excited states (specifically those generated by
processes that contribute to the line shape near two types of spectral features called 
$c-s$ boundary lines and $c$, $c'$, $c''$, and $c'''$
branch lines, respectively) have quantum overlap mainly with a single excited energy eigenstate.
The corresponding branch lines occur in that region of the $(k,\omega)$ plane in which 
there is a continuum distribution of spectral weight; this is 
represented by light grey regions in the sketch of Fig. \ref{figure1}.
For them this applies again to the $k$ intervals in that plane for which
the corresponding exponent in the spectral-function expression is negative. 

The line shape of the one-electron spectral function
in the vicinity of the $(k,\omega)$-plane regions where it displays these different types of
singularities is controlled by transitions to a particular class of excited states.
They span a smaller subspace, contained in the one-electron subspace.
Fortunately, in that {\it singularities subspace} the ${\tilde{\delta}}_c >0$ Hamiltonian, Eq. (\ref{equ1}), can be
diagonalized when expressed in terms of $c$ and $s$ particle operators. In the next section,  we begin by introducing
that subspace.

\subsection{The one-electron singularities subspace}
\label{SECIVA}

The singularities subspace considered here refers to one-electron removal and addition at electronic 
densities $n_e \in ]0,1[$. Indeed, as noted previously, the spectral features of the ARPES for the TTF 
stack of molecules with electronic density $n_e^F = 1.41$ can under a suitable and well-defined transformation 
be described by the spectral function for one-electron addition at density $n_e = 2 - n_e^F = 0.59$. 

The one-electron subspace is generated from application of one-electron 
annihilation or creation operators onto the $N_e$-electron ground state. On the other hand, the 
singularities subspace is spanned by the one-electron excited
states that contribute to the line shape near the above mentioned singularities of
the spectral function $(k,\omega)$-plane. Those
are identified with the ARPES structures associated with the TTF and TCNQ stacks 
of molecules.

\begin{figure}
\begin{center}
\centerline{\includegraphics[width=8.0cm]{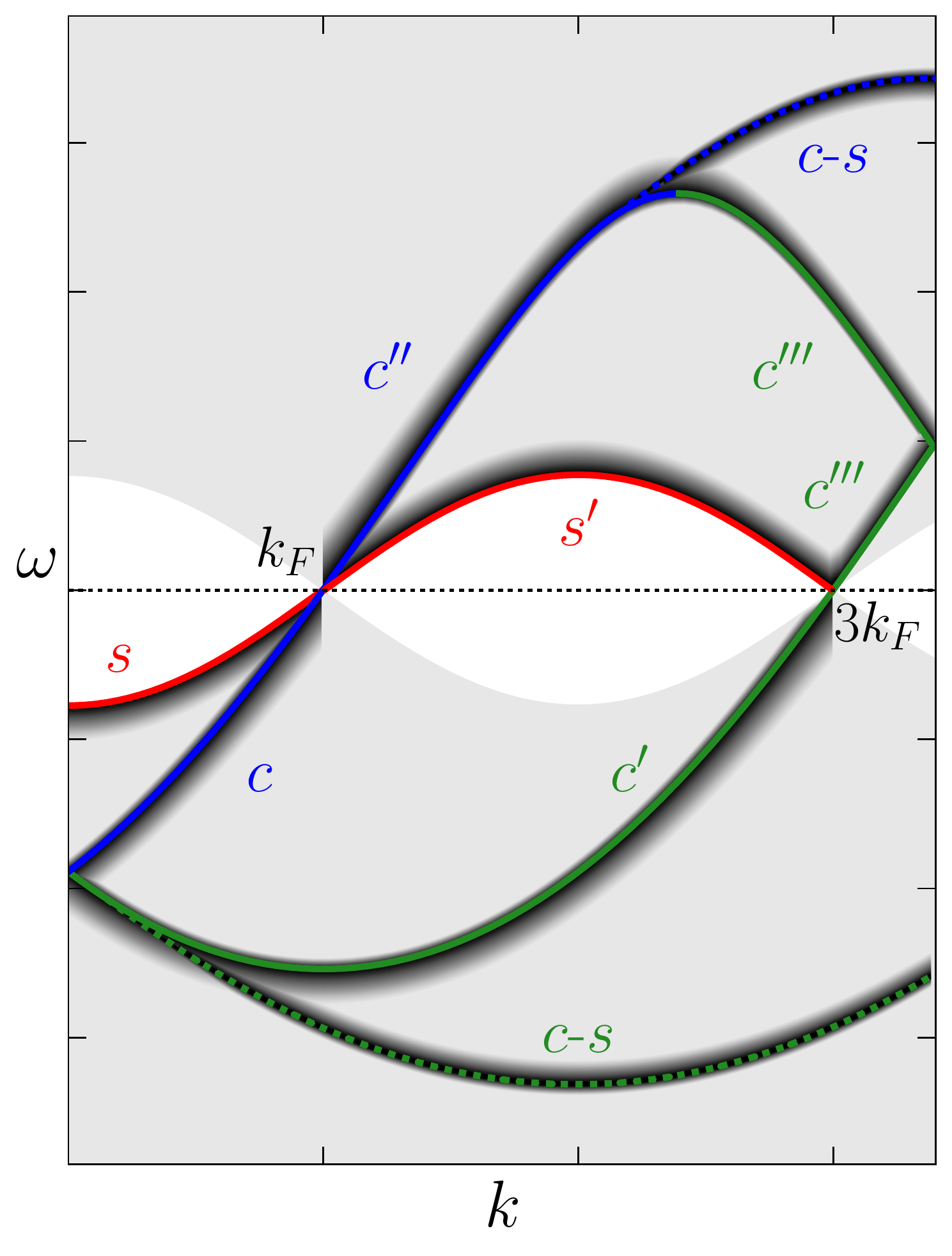}}
\caption{Sketch of the (i) $\omega <0$ $s$ (spin) and $c$ and $c'$ (charge) branch lines, (ii) 
$\omega >0$ $s'$ (spin), $c''$ and $c'''$ (charge) branch 
lines, and (iii) $\omega <0$ and $\omega >0$ $c-s$ boundary lines in the one-electron removal and addition spectral functions
for momentum values $k>0$ of the models discussed in this paper. 
Their spectra are defined below in Eqs. (i) (\ref{equ17}), (ii) (\ref{equ18}), and (iii) (\ref{equ20}), respectively.
(The branch and boundary lines correspond to the solid and dotted lines, respectively.)
The soft grey region refers to the small spectral-weight distribution 
continuum. The darker grey regions below the one-electron removal branch lines,
above the one-electron addition branch lines, and both below and above the boundary
lines typically display more weight. In the actual spectral function this applies in the case of the 
branch lines to $k$ intervals for which the exponents that control the line shape near them are negative.}
\label{figure1}
\end{center}
\end{figure}

The $c$ and $s$ particles occupy a $c$ band and
an $s$ band whose momentum values $q_j$ and $q_j'$, respectively, are such that
$q_{j+1}-q_j = 2\pi/L$ and $q_{j+1}'-q_j' = 2\pi/L$ \cite{Carmelo_19}. In the thermodynamic limit, one often
uses a continuum representation in terms of continuum momentum
variables $q$ and $q'$, respectively. Their ground-state occupancies are
$q \in [-2k_F,2k_F]$ and $q' \in [-k_F,k_F]$, respectively, where $2k_F=\pi n_e$.
In Sec. \ref{SECV} we discuss the expression in terms of corresponding $q$ and $q'$ occupancies 
of the physical momentum $k$ of the one-electron excitations 
whose spectra are sketched in Fig. \ref{figure1}.

For one-electron {\it removal}, most of the spectral function weight is generated in the thermodynamic limit by transitions from the ground 
state to excited states involving the following processes: 1) creation of one $c$ hole at some $c$ band momentum in the interval
$q \in ]-2k_F,2k_F[$; 2) one $s$ hole in the $s$ band interval $q' \in ]-k_F,k_F[$; plus 3) low-energy particle-hole processes in such bands. 
For one-electron {\it addition}, most of the spectral function weight is generated in that limit by the
following processes under transitions to excited states: 
1) creation of one $c$ particle at some $c$ band momentum in the intervals
$q \in ]-\pi,-2k_F[$ or $q \in ]2k_F,\pi[$; 2) one $s$ hole in the $s$ band interval $q' \in ]-k_F,k_F[$; plus 3) 
low-energy particle-hole processes in such bands. 
In the case particle-hole processes in the $s$ band, both for one-electron removal and addition
correct results are achieved under addition of a small virtual magnetic field
that is made to vanish in the end of the calculations.
This gives $q' \in ]-k_{F\downarrow},k_{F\downarrow}[$
and $\vert q'\vert  \in ]k_{F\downarrow},k_{F\uparrow}[$ for the occupied and 
unoccupied $s$ band Fermi seas, respectively. 
\begin{table}
\begin{center}
\begin{tabular}{|c|} 
\hline
{\bf Momentum intervals of scattering centers}\\
\hline
{\bf $s$ hole} (low-energy ``spinon'') \\
\hline
$q' = - k_F + p'$ where $p' \in [0,\delta p_{Fs}]$ \\
\hline
$q' = + k_F + p'$ where  $p' \in [-\delta p_{Fs},0]$\\
\hline
{\bf $s$ hole} (single $s$ mobile scattering center) \\
\hline
$q' \in [-k_F + \delta p_{Fs},k_F - \delta p_{Fs}]$ \\
\hline
{\bf $c$ hole} (low-energy ``holon'') \\
\hline
$q = - 2k_F + p$ where $p \in [0,\delta p_{Fc}]$ \\
\hline
$q = + 2k_F + p$ where  $p \in [-\delta p_{Fc},0]$\\
\hline
{\bf $c$ hole} (single $c$ mobile scattering center) \\
\hline
$q \in [-2k_F + \delta p_{Fc},2k_F - \delta p_{Fc}]$ \\
\hline
{\bf $c$ hole and $s$ hole} (both mobile scattering centers) \\
\hline
$q \in [-q_c^h,q_c^h]$ where $\vert{\tilde{v}}_c (\pm q_c^h)\vert = v_{Fs}^-$ and $q_c^h < 2k_F$ \\
\hline
$q'$ such that $v_s (q') = {\tilde{v}}_c (q)$ \\
\hline
{\bf $c$ particle} (low-energy ``holon'' removal) \\
\hline
$q = - 2k_F + p$ where $p \in [-\delta p_{Fc},0]$ \\
\hline
$q = + 2k_F + p$ where $p \in [0,\delta p_{Fc}]$\\
\hline
{\bf $c$ particle} (single $c$ mobile scattering center) \\
\hline
$q \in [-\pi,-2k_F -\delta p_{Fc}]\,;[2k_F +\delta p_{Fc},\pi]$ \\
\hline
{\bf $c$ particle and $s$ hole} (both mobile scattering centers) \\
\hline
$q \in [-\pi,-q_c]\,;[q_c,\pi]$ where $\vert{\tilde{v}}_c (\pm q_c^h)\vert = v_{Fs}^-$ and $q_c > 2k_F$\\ 
\hline
$q'$ such that $v_s (q') = {\tilde{v}}_c (q)$ \\
\hline
\end{tabular}
\caption{The momentum intervals for the $c$ and $s$ bands of scattering centers created under 
one-electron removal and addition involving processes in the singularities subspace.
Here $v_{Fs}^- \equiv v_s (k_F - \delta p_{Fs})$.}
\label{table1}
\end{center}
\end{table} 

Importantly, {\it only} the above reported processes contribute to the line shape near the 
singularities of spectral-function in the $(k,\omega)$ plane that describe the ARPES 
structures. Processes where both the $c$ and $s$ holes
for one-electron removal and the  $c$ particle and the $s$ hole for one-electron addition
are created away from the $c$ band and $s$ band Fermi points $\pm 2k_F$ and $\pm k_F$, respectively,
contribute to the spectral-function continuum away from such singularities.
The exception refers to the processes (4-Rem) and (4-Add) defined below. They contribute to
a particular type of such singularities, called {\it boundary lines}.

The $s$ hole and (i) the $c$ hole and (ii) the $c$ particle created under one-electron 
(i) removal and (ii) addition excitations, respectively, refer to scattering centers. The corresponding
scatterers are the $c$ and $s$ particles and $c$ holes that preexisted in the ground state. 
The singularities subspace is spanned by excited states reached
by transitions from the ground state within the processes 
defined in the following. By high energy we
%CORR - meant REPLACED BY mean in the following
mean in the following energy scales beyond the reach of the low-energy TLL.
All the following processes are {\it dressed} by low-energy and
small momentum particle-hole processes near the $c$ and $s$ bands
Fermi points $\pm 2k_F$ and $\pm k_{F\downarrow}\rightarrow \pm k_F$, respectively:

(i) (1-Rem) and (ii) (1-Add) - Low-energy TLL processes where (i) one $c$ hole (``holon'') and (ii) one $c$ particle
(``holon'' removal), respectively, is created  in the vicinity of one of the $c$ band
Fermi points at $q = \pm 2k_F + p$, and one $s$ hole (``spinon'') is created near one of the $s$ band Fermi points 
at $q' = \pm k_F + p'$. The small momenta $p$ and $p'$ intervals are here and in the following given in Table \ref{table1}.
There $\delta  q_{Fc}$ such that $\delta  q_{Fc}/2k_F\ll 1$ and $\delta  q_{Fs}$ such that $\delta  q_{Fs}/k_F\ll 1$ 
are very small for a large finite system and may vanish in the thermodynamic limit. Such processes 
contribute to the low-energy spectral weight distribution near $k=\pm k_F$ and $k=\pm 3k_F$. See
the sketch of the spectral features in Fig. \ref{figure1} for $k>0$.

(i) (2-Rem) and (ii) (2-Add) - High-energy processes where one $s$ hole is created at momentum values spanning a $s$ band subinterval 
of the interval $q' \in [-k_F + \delta p_{Fs},k_F - \delta p_{Fs}]$ for which the spectral function displays singularities controlled 
by negative $k$ dependent exponents and (i) one $c$ hole (``holon'') and (ii) one $c$ particle
(``holon'' removal), respectively, is created near one of the $c$ band Fermi points at $q = \pm 2k_F + p$. 
Those processes contribute to the spectral weight distribution in the
vicinity of the subintervals of the (i) $s$ and (ii) $s'$ branch lines. Their spectra run 
in the thermodynamic limit in the intervals (i) $k\in ]-k_F,k_F[$ and (ii) $k\in ]-3k_F,-k_F[\,;]k_F,3k_F[$, respectively,
shown in the sketch of Fig. \ref{figure1} for $k>0$. We call the $s$ band hole created away from the corresponding 
$s$ band Fermi points a {\it $s$ (hole) mobile scattering center}.

(i) (3-Rem) and (ii) (3-Add) - High-energy processes where one $s$ hole (``spinon'') is created near one of the $s$ band Fermi points 
at $q' = \pm k_F + p'$ and one (i) $c$ hole and (ii) one $c$ particle is created at momentum values spanning a $c$ band subinterval 
of the intervals (i) $q \in [-2k_F + \delta p_{Fc},2k_F - \delta p_{Fc}]$ and (ii)
$q \in [-\pi,-2k_F -\delta p_{Fc}]\,;[2k_F +\delta p_{Fc},\pi]$, respectively,
for which the spectral function displays singularities controlled by negative $k$ dependent exponents.
For (i) one-electron removal, such processes contribute to the spectral weight distribution near the corresponding 
subintervals of the $c$ and $c'$ branch lines. Their spectra run in the thermodynamic limit in the intervals 
$k\in ]-k_F,k_F[$ and $]-3k_F,3k_F[$, respectively, shown in
the sketch of Fig. \ref{figure1} for $k>0$. For (ii) one-electron addition,
they contribute to the weight distribution in the vicinity of the 
subintervals of the $c''$, $c'''$ (branch I), and $c'''$ (branch II) branch lines.
Their spectra run in the thermodynamic limit in the intervals 
$k\in ]-(\pi -k_F),-k_F[\,;]k_F,(\pi -k_F)[$, $k\in ]-\pi,-3k_F[\,;]3k_F,\pi[$,
and $k\in ]-\pi,-(\pi -k_F)[\,;](\pi -k_F),\pi[$, respectively, also shown in
the sketch of Fig. \ref{figure1} for $k>0$. We call the (i) $c$ hole or (ii) $c$ particle created away from the $c$ band Fermi points a
{\it $c$ (hole or particle) mobile scattering center}. 

(i) (4-Rem) and (ii) (4-Add) - High-energy processes where (i) one $c$ hole and (ii) 
one $c$ particle is created at a $c$ band momentum (i) $q \in [-q_c^h,q_c^h]$
and (ii) $q \in [-\pi,-q_c]\,;[q_c,\pi]$, respectively, and one $s$ hole is created
at a $s$ band momentum $q'$ such that  $v_{s}(q') = {\tilde{v}}_{c}(q)$. 
Here $0 <q_c^h < 2k_F$ and $2k_F <q_c <\pi$ are such that $\vert{\tilde{v}}_c (\pm q_c^h)\vert = v_{Fs}^-$ and 
%CORR - q_c^h REPLACED BY q_c
$\vert{\tilde{v}}_c (\pm q_c)\vert = v_{Fs}^-$, respectively. In these
expressions $v_{Fs}^- \equiv v_s (k_F - p_{Fs})$,
$v_s (q')$ is the ${\tilde{\delta}}_c = 0$ bare $s$ band group velocity
\cite{Carmelo_19}, and ${\tilde{v}}_{c}(q)$ is the renormalized
$c$ band group velocity given in Eq. (\ref{equC11}) of Appendix \ref{APPC}.
Such processes contribute to the spectral weight distribution near well-defined 
spectral features called $c-s$ boundary lines. In the sketch of Fig. \ref{figure1}
they are represented by dotted lines.

As discussed below in Sec. \ref{SECV}, for the ${\tilde{\delta}}_c > 0$ model, Eq. (\ref{equ1}), in the singularities subspace 
only the phase shifts imposed by creation of the $c$ and $s$ scattering centers 
onto $c$ particles, $c$ holes, and $s$ particles scatterers with momentum values near their bands Fermi points 
play an active role. Indeed, only such scatterers contribute to the line shape in the vicinity of the branch lines that display singularities. 
Their $c$ and $s$ band momentum values intervals are provided in Table \ref{table2}.
\begin{table}
\begin{center}
\begin{tabular}{|c|} 
\hline
{\bf Momentum intervals of active scatterers} \\
\hline
{\bf $s$ particle} (electron removal and addition) \\
\hline
$q' = - k_F + p'$ where $p' \in [0,\delta p_{Fs}]$ \\
\hline
$q' = + k_F + p'$ where  $p' \in [-\delta p_{Fs},0]$\\
\hline
{\bf $c$ particle} (electron removal) \\
\hline
$q = - 2k_F + p$ where $p \in [0,\delta p_{Fc}]$ \\
\hline
$q = + 2k_F + p$ where  $p \in [-\delta p_{Fc},0]$\\
\hline
{\bf $c$ hole} (electron addition) \\
\hline
$q = - 2k_F + p$ where $p \in [-\delta p_{Fc},0]$ \\
\hline
$q = + 2k_F + p$ where $p \in [0,\delta p_{Fc}]$\\
\hline
\end{tabular}
\caption{The momentum intervals in the $c$ and $s$ bands for processes involving active scatterers
within the singularities subspace.}
\label{table2}
\end{center}
\end{table} 

\subsection{The $c$ and $s$ particle representation associated with the potential $V_c (x)$}
\label{SECIVB}

As shown in Appendix \ref{APPC}, for the ${\tilde{\delta}}_c >0$ model, Eq. (\ref{equ1}),
there is a $c$ and $s$ particle representation whose nondiagonal Hamiltonian in the one-electron subspace
involves the Fourier transform ${\tilde{\cal{V}}}_c^1 (k)$ of 
the potential ${\tilde{V}}_c^1 (x)$. As given in Eq. (\ref{equC12}) of that Appendix, in the
${\tilde{\delta}}_c\rightarrow 0$ bare limit, ${\tilde{V}}_c^1 (x)$ and ${\tilde{\cal{V}}}_c^1 (k)$ 
become the effective potential $V_c^1 (x)$ and
its Fourier transform ${\cal{V}}_c^1 (k)$, Eq. (\ref{equB25})
of Appendix \ref{APPB}, of the ${\tilde{\delta}}_c =0$ bare model.

The free term of the ${\tilde{\delta}}_c =0$ bare Hamiltonian in Eq. (\ref{equB23}) of Appendix \ref{APPB}, which corresponds
to the potential $V_c^1 (x)$ representation, involves the velocity $v_{Fc}^1= (\xi_c^2/2)\,v_{Fc}$.
It is a charge elementary current that controls the charge stiffness of the ${\tilde{\delta}}_c =0$ bare model, 
$D_{\rho}^0 = v_{Fc}^1/\pi$, in the real part of the conductivity, 
$\sigma_{\rho}^0 (\omega) = 2\pi D_{\rho}^0 \delta (\omega) + \sigma_{\rho}^{\rm reg} (\omega)$ \cite{Carmelo_18}.
Similarly, the renormalized potential ${\tilde{V}}_c^1 (x)$ is associated with
a charge elementary current ${\tilde{v}}_{Fc}^1= ({\tilde{\xi}}_c^2/2)\,{\tilde{v}}_{Fc}$ that controls
the corresponding renormalized charge stiffness $D_{\rho}$. At $k=0$ its Fourier transform ${\tilde{\cal{V}}}_c^1 (k)$ 
can be expressed in terms of it as ${\tilde{\cal{V}}}_c^1 (0) = \pi ({1\over{\tilde{\xi}}_c^4} - {1\over 4})\,2\pi D_{\rho}$.

With the finite-range interactions, the SDS 
exponent $\alpha$, compressibility $\chi_{\rho}$, and charge stiffness $D_{\rho}$ read,
\begin{equation}
\alpha = {(2 - {\tilde{\xi}}_c^2)^2\over 8{\tilde{\xi}}_c^2} \, ; \hspace{0.20cm} 
\chi_{\rho} = {2\over\pi\,n_e^2\,{\tilde{v}}_{Fc}^0} \, ;
\hspace{0.20cm} D_{\rho} = {{\tilde{v}}_{Fc}^1\over\pi} \, ,
\label{equ6}
\end{equation}
respectively. Here ${\tilde{v}}_{Fc}^0 = (2/{\tilde{\xi}}_c^2)\,{\tilde{v}}_{Fc}$, and
${\tilde{v}}_{Fc} = \sqrt{{\tilde{v}}_{Fc}^0\times {\tilde{v}}_{Fc}^1}$ is the renormalized 
Fermi velocity associated with the $c$ energy dispersion given below in Sec. \ref{SECIVC}.

The effects of the finite-range interactions upon increasing ${\tilde{\delta}}_c$ 
enhance the SDS exponent $\alpha$ from $\alpha_0 \in (2 - \xi_c^2)^2/(8\xi_c^2)\in ]0,1/8[$
to $\alpha\in [\alpha_0,1/8[\,;]1/8,49/32[$. In contrast, as follows from Eqs. (\ref{equC31}) and (\ref{equC35}) of 
Appendix \ref{APPC}, both the compressibility $\chi_{\rho}$ and the charge stiffness $D_{\rho}$ tend 
to be suppressed by  such interactions. 

As noted above, the potential $V_c (x)$ can be viewed as the part of ${\tilde{V}}_c^1 (x)$ that accounts 
for most of the renormalization of 
$V_c^1 (x)$ by the ${\tilde{\delta}}_c >0$ finite-range interactions. 
Indeed, ${\tilde{V}}_c^1 (x)$ includes both contributions that
stem from $V_c^1 (x)$ and $V_{e} (r)$. On the other hand, the related potential $V_c (x)$
is mostly induced by the electron potential $V_{e} (r)$ and thus by 
the corresponding rotated-electron potential $V_{re} (r)$ considered in Appendix \ref{APPC}.
In contrast to ${\tilde{V}}_c^1 (x)$, both $V_c (x)$ and $V_{re} (r)$
and their Fourier transforms ${\cal{V}}_c (k)$ and ${\cal{V}}_{re} (k)$, respectively,
vanish in the ${\tilde{\delta}}_c\rightarrow 0$ limit, as given in Eq. (\ref{equC12}) of Appendix \ref{APPC}.

However, $V_c (x)$ and $V_{re} (r)$ describe different microscopic mechanisms. 
The former controls the renormalization by the finite-range interactions of the $c$ particle/hole phase shifts
in Eqs. (\ref{equA1}) and (\ref{equA2}) of Appendix \ref{APPA}. It thus controls the 
renormalization of the corresponding quantum overlaps of the matrix elements
in the one-electron spectral function, Eq. (\ref{equ3}). In turn, the potential $V_{re} (r)$ controls the 
renormalization of the excitation energy spectra,
$(E_{\nu^{\mp}}^{N_e \mp 1}-E_{GS}^{N_e})$, in the same spectral function expression
provided in that equation. Their $k$ and $\omega$ dependence near the singularities is given below in Sec. \ref{SECV}.

Three related $c$ particle representations that describe complementary aspects of
the quantum problem microscopic mechanisms are associated with the potentials,
$V_{re} (r)$, ${\tilde{V}}_c^1 (x)$, and $V_c (x)$, respectively. The dependence on $k$ of the 
Fourier transforms of such potentials [${\cal{V}}_{re} (k)$, ${\tilde{\cal{V}}}_c^1 (k)$, and ${\cal{V}}_c (k)$, respectively,
Eq. (\ref{equC15}) of Appendix \ref{APPC}], is only
universal in the limit of small $k$. This follows from the corresponding 
potentials $V_{re} (r)$, ${\tilde{V}}_c^1 (x)$, and $V_c (x)$ having a nonuniversal form,  except at large distances. It is determined by 
that of the electronic potential $V_e (r)$ itself for general $r \in [0,\infty]$ values.

Having as starting point the Hamiltonian, Eq. (\ref{equ1}), expressed in terms of rotated-electron operators,
Eqs. (\ref{equC1}) and (\ref{equC2}) of  Appendix \ref{APPC}, we derive
in that Appendix three corresponding nondiagonal expressions of that
Hamiltonian in the one-electron subspace in terms of the $c$ and $s$ particle representations.
They are given in Eq. (\ref{equC16}) of that Appendix.
[The common notation used in that equation for the three representations 
associated with the Fourier transforms ${\cal{V}}_{re} (k)$, ${\tilde{\cal{V}}}_c^1 (k)$,
and ${\cal{V}}_c (k)$ is defined in Eq. (\ref{equC17}) of Appendix \ref{APPC}.]

The three alternative expressions for the Hamiltonians cannot be diagonalized in the one-electron subspace.
This is due to their terms denoted by $\check{H}_{re}$ in Eq. (\ref{equC16}) of Appendix \ref{APPC}. 
Fortunately, their complicated form is not needed for our studies. 
The emergence of such terms follows from, in contrast to the ${\tilde{\delta}}_c =0$ bare model, 
the states generated by occupancy configurations of the $c$ and $s$ particles in general not being energy 
eigenstates. 

The use of the choices ${\check{\cal{V}}} (k)={\cal{V}}_c (k)$, ${\check{f}}_{q,c}^{\dag} = {\breve{f}}_{q,c}^{\dag}$, 
and ${\check{\varepsilon}}_c (q) = {\breve{\varepsilon}}_c (q)$ in the general expression given in Eq. (\ref{equC28}) 
of Appendix \ref{APPC} leads to the following Hamiltonian expression in the 
specific representation of $V_c (x)$,
\begin{eqnarray}
:\hat{H}: & = & :\hat{H}_{c}: + :\hat{H}_{s}:
\nonumber \\
:\hat{H}_{c}: & = & \sum_{q=-\pi}^{\pi}{\breve{\varepsilon}}_c (q):{\breve{f}}_{q,c}^{\dag}{\breve{f}}_{q,c}: + \,\hat{V}_{c,\gamma}
\nonumber \\
\hat{V}_{c,-1} & = & {1\over L}\sum_{\iota =\pm}\sum_{k,p,q} {\cal{V}}_c (k)
\nonumber \\
& \times & {\breve{f}}_{\iota 2k_F + p,c}^{\dag}{\breve{f}}_{\iota 2k_F+ p + k,c}\,{\breve{f}}_{q-k,c}{\breve{f}}_{q,c}^{\dag} 
\nonumber \\
\hat{V}_{c,+1} & = & {1\over L}\sum_{\iota =\pm}\sum_{k,p,q} {\cal{V}}_c (k)
\nonumber \\
& \times & {\breve{f}}_{\iota 2k_F+ p + k,c}{\breve{f}}_{\iota 2k_F + p,c}^{\dag}\,{\breve{f}}_{q,c}^{\dag}{\breve{f}}_{q-k,c} 
\nonumber \\
:\hat{H}_s: & = & \sum_{q'=-k_F}^{k_F}\varepsilon_s (q'):f_{q',s}^{\dag}f_{q',s}: \, .
\label{equ7}
\end{eqnarray}
Here $:\hat{O}:$ stands for the (standard) normal ordering of an operator $\hat{O}$ and
the $k$, $p$, $q$ summations in the interaction terms $\hat{V}_{c,\gamma}$ 
suitable to one-electron removal $(\gamma = -1)$ and addition $(\gamma = +1)$ run in 
different intervals given in Table \ref{table3}. This is consistent with the singularities subspace processes 
defined in Sec. \ref{SECIVA}. 

The charge term $:\hat{H}_c:$ in Eq. (\ref{equ7}) describes the interaction of the $c$ particle or 
$c$ hole scatterers with the $c$ (hole or particle) mobile scattering center.
Specifically, in the expressions of its interacting term $\hat{V}_{c,\gamma}$ for one-electron (i) removal
$(\gamma = -1)$ and (ii) addition $(\gamma = +1)$, the operators (i) ${\breve{f}}_{\iota 2k_F + p,c}^{\dag}{\breve{f}}_{\iota 2k_F+ p + k,c}$ 
and (ii) ${\breve{f}}_{\iota 2k_F+ p + k,c}{\breve{f}}_{\iota 2k_F + p,c}^{\dag}$ refer to the (i) $c$ particle
and (ii) $c$ hole scatterer, respectively. On the other hand, the operators (i)
${\breve{f}}_{q-k,c}{\breve{f}}_{q,c}^{\dag}$ and (ii) ${\breve{f}}_{q,c}^{\dag}{\breve{f}}_{q-k,c}$
correspond to the $c$ mobile (i) hole and (ii) particle scattering center, respectively.
\begin{table}
\begin{center}
\begin{tabular}{|c|} 
\hline
$k \in [-\delta p_{Fc}-p,-p]$ for $\iota\,\gamma = -1$ \\
\hline
$k \in [-p,\delta p_{Fc}-p]$ for $\iota\,\gamma = +1$ \\
\hline
$p \in [-\delta p_{Fc},0]$ for $\iota\,\gamma = -1$ \\
\hline
$p \in [0,\delta p_{Fc}]$ for $\iota\,\gamma = +1$\\
\hline
$q \in [-2k_F + \delta p_{Fc},2k_F - \delta p_{Fc}]$ for $\gamma = -1$ \\
\hline
$q \in [-\pi,-2k_F - \delta p_{Fc}]\,;[2k_F + \delta p_{Fc},\pi]$ for $\gamma = +1$ \\
\hline
\end{tabular}
\caption{Intervals of $\sum_{k,p,q}$ in $\hat{V}_{c,\gamma}$ for $\gamma =\mp 1$ and
$\iota = \pm$, Eq. (\ref{equ7}).}
\label{table3}
\end{center}
\end{table} 

The spin term $:\hat{H}_s:$ in Eq. (\ref{equ7}) remains invariant under the $\xi_c\rightarrow {\tilde{\xi}}_c$
transformation. It equals that of the ${\tilde{\delta}}_c = 0$ model whose 
dispersion $\varepsilon_s (q')$ is defined in Ref. \onlinecite{Carmelo_19}. The renormalized $c$ particle
energy dispersion ${\breve{\varepsilon}}_c (q)$ in the $:\hat{H}_c:$ expression is given in Eq. (\ref{equC24}) of Appendix \ref{APPC} 
in terms of the ${\tilde{\delta}}_c = 0$ bare dispersion $\varepsilon_c (q)$ \cite{Carmelo_19}. 

Since $\delta p_{Fc}$ in Table \ref{table3} is very small and may vanish in the thermodynamic limit,
the same applies to the the $k$ intervals given in that table. Hence in the expression, Eq. (\ref{equ7}), 
of the Hamiltonian, Eq. (\ref{equ1}), in the singularities subspace, the $k=0$ value of ${\cal{V}}_c (k)$ plays a major role. 
Combining the information provided in Eqs. (\ref{equC23}) and (\ref{equC25}) 
of Appendix \ref{APPC}, one finds
\begin{eqnarray}
&& {\cal{V}}_c (k) = - {\pi\over 2}\,{\breve{\alpha}}_c (k)\,{\breve{v}}_{Fc} 
\hspace{0.20cm}{\rm where}\hspace{0.20cm}
{\breve{\alpha}}_c (k) = {\breve{\alpha}}_c + {\cal{O}} (k^2)
\nonumber \\
&& {\breve{\alpha}}_c = {\xi_c^4 - {\tilde{\xi}}_c^4\over {\tilde{\xi}}_c^4} 
\hspace{0.20cm}{\rm and}\hspace{0.20cm}
{\breve{\beta}}_c = {{\tilde{v}}_{Fc} - {\breve{v}}_{Fc}\over {\breve{v}}_{Fc}} = {\xi_c^2 - {\tilde{\xi}}_c^2\over {\tilde{\xi}}_c^2} \, ,
\label{equ8}
\end{eqnarray}
for the whole interval ${\tilde{\delta}}_c \in [0,{\tilde{\delta}}_c^{(1)}[\,;]{\tilde{\delta}}_c^{(1)},{\tilde{\delta}}_c^{({1\over 2})}[$.
The potential $V_c (x)$ sum rule reported in Sec. \ref{SECIIIB}, 
$\int_{0}^{\infty}(-V_c (x)) = {\pi\over 4}\,[(\xi_c^4 - {\tilde{\xi}}_c^4)/{\tilde{\xi}}_c^4]\,{\breve{v}}_{Fc}$, 
follows directly from the $k=0$ value ${\cal{V}}_c (0)$ given in Eq. (\ref{equ8}). 

The corresponding parameter values for the representation associated with the related potential ${\tilde{V}}_c^1 (x)$ are
${\tilde{\alpha}}_c^1 = (4 - {\tilde{\xi}}_c^4)/{\tilde{\xi}}_c^4$ and ${\tilde{\beta}}_c^1 = (2 - {\tilde{\xi}}_c^2)/{\tilde{\xi}}_c^2$,
Eq. (\ref{equC25}) of Appendix \ref{APPC}. In the ${\tilde{\delta}}_c\rightarrow 0$ bare limit they
become the parameters $\alpha_c^1$ and $\beta_c^1$, Eqs. (\ref{equB25}) and
(\ref{equB26}) of Appendix \ref{APPB}, respectively. Their expressions are the same under the replacement
of ${\tilde{\xi}}_c$ by $\xi_c$. In contrast, ${\breve{\alpha}}_c$ and ${\breve{\beta}}_c$ in Eq. (\ref{equ8})
vanish in that limit.

The representation, Eq. (\ref{equ7}), of the ${\tilde{\delta}}_c >0$ Hamiltonian, Eq. (\ref{equ1}), in the singularities subspace 
is that which explicitly displays the microscopic processes that
control the renormalization of the $c$ particle/hole phase shifts. We will define and discuss these briefly in the following. 
They appear in the expressions given below in Sec. \ref{SECVA} of the $k$ dependent exponents
that control the quantum overlaps in the matrix elements of the spectral function, Eq. (\ref{equ3}).

For one-electron removal, $-2\pi{\tilde{\Phi}}_{c,s}(\pm 2k_F,q')$ and $-2\pi{\tilde{\Phi}}_{c,c}(\pm 2k_F,q)$ 
are the phase shifts imposed on a $c$ particle scatterer of $c$ band momentum $\pm 2k_F + p$
at and near the $c$ band Fermi points by creation of one $s$ mobile scattering center at 
momentum $q'$ and one $c$  mobile scattering center at $q$, respectively. Here the interval of $p$ is given in
Table \ref{table2} and those of the momenta $q'$ and $q$ of the created $s$ and $c$ (hole) mobile scattering centers,
respectively, are provided in Table \ref{table1}. For electron addition, the same applies to $-2\pi{\tilde{\Phi}}_{c,s}(\pm 2k_F,q')$,
whereas $2\pi{\tilde{\Phi}}_{c,c}(\pm 2k_F,q)$ is the phase shift imposed on a $c$ hole scatterer of $c$ band momentum 
$\pm 2k_F + p$ at and near the $c$ band Fermi points. The intervals of the momentum $q$ of the created $c$ (particle) 
mobile scattering center and of $p$ are again given in Tables \ref{table1} and \ref{table2}, respectively. 

The phase shift $2\pi{\tilde{\Phi}}_{c,s}(\pm 2k_F,q')$ has the same expression,
Eq. (\ref{equA1}) of Appendix \ref{APPA}, for both one-electron removal and addition.
In contrast, the $2\pi{\tilde{\Phi}}_{c,c}(\pm 2k_F,q)$ expression, Eq. (\ref{equA2}) of that Appendix,
has different $q$ intervals for one-electron removal and addition. A second difference is the factor
$\gamma $ in the expression of phase shift term $2\pi{\tilde{\Phi}}_{c,c}^{R_{\rm eff}} (k_r)$
in the latter equation, which reads $\gamma = -1$ and $\gamma = +1$ for electron removal and addition, respectively.

How does the Hamiltonian expression, Eq. (\ref{equ7}), controls the renormalization of the phase shifts? 
The renormalization of $2\pi{\tilde{\Phi}}_{c,s} (\pm 2k_F,q')$ and of the phase-shift term $2\pi{\tilde{\Phi}}_{c,c}^{{\tilde{a}}} (\pm 2k_F,q)$ 
in Eqs. (\ref{equA1}) and (\ref{equA2}) of Appendix \ref{APPA}, respectively, involves only the renormalized and 
bare charge parameters ${\tilde{\xi}}_c$ and $\xi_c$, respectively. Indeed, ${\cal{V}}_c (k)$, Eq. (\ref{equ8}), in 
that Hamiltonian expression only depends on these two parameters. 

Moreover, for the $c$ particle operators in the Hamiltonian expression, Eq. (\ref{equ7}), the relative momentum reads 
$k_r = (q - \iota 2k_F) - (k + p)$. For a large finite system, one has according to the $k$ and $p$ intervals provided in 
Table \ref{table3} that $(k + p)$ is very small or vanishes. Indeed, $p_{Fc}$ is very small and may vanish in the thermodynamic limit.
This is why $k_r = (q - \iota 2k_F)$ where $\iota =\pm$. Some of the processes associated with the higher-order contributions 
${\cal{O}} (k^2)$ in Eq. (\ref{equ8}) are accounted for the dependence on $k_r = (q \mp 2k_F)$ of the phase shift term 
$2\pi{\tilde{\Phi}}_{c,c}^{R_{\rm eff}} (k_r)$ in Eq. (\ref{equA2}) of Appendix \ref{APPA}. In its expression, $P_c (k_r)=0$ 
in the present unitary limit \cite{Carmelo_19}. The effective range $R_{\rm eff}$ in the expression of $2\pi{\tilde{\Phi}}_{c,c}^{R_{\rm eff}} (k_r)$ 
depends on the ratio ${\tilde{a}}/a$. And again as ${\tilde{\alpha}}_c = {\tilde{\alpha}}_c (0)$ in the ${\cal{V}}_c (k)$ 
expression, Eq. (\ref{equ8}), that ratio depends only on the parameters ${\tilde{\xi}}_c$ and $\xi_c$, as given in Eq. (\ref{equA6}) 
of Appendix \ref{APPA}.

\subsection{The rotated-electron potential $V_{re} (r)$ and related $c$ and $s$ particle representation}
\label{SECIVC}

The expression of the Hamiltonian, Eq. (\ref{equ1}), in the one-electron subspace that involves the 
Fourier transform ${\cal{V}}_{re} (k)$ is given in Eq. (\ref{equC9}) of Appendix \ref{APPC} 
in terms of $c$ and $s$ particle operators. It has again been
derived from the expression in terms of rotated-electron operators, Eqs. (\ref{equC1}) and (\ref{equC2})
of that Appendix. That specific expression explicitly displays the microscopic processes that
control the renormalization of the spectra $(E_{\nu^{\mp}}^{N_e \mp 1}-E_{GS}^{N_e})$
in the spectral function, Eq. (\ref{equ3}).

The corresponding simplified Hamiltonian expression in the smaller singularities subspace that can be diagonalized 
in terms of $c$ and $s$ particle operators is given by 
the choices ${\check{\cal{V}}} (k)={\cal{V}}_{re} (k)$, ${\check{f}}_{q,c}^{\dag} = f_{q,c}^{\dag}$, and
${\check{\varepsilon}}_c (q) = \varepsilon_c (q)$ in the general expression, Eq. (\ref{equC28}) of 
Appendix \ref{APPC}. 

As discussed in Appendix \ref{APPC3}, the deformation of the $c$ band
energy dispersion by the finite-range interactions must preserve its energy bandwidth, 
${\tilde{W}}_{c} = W_c = 4t$. The compressibility in Eq. (\ref{equ6}) is largest in the ${\tilde{\delta}}_c=0$ limit and 
tends to be suppressed by the finite-range interactions. The degree of that deformation is limited by its 
largest ${\tilde{\delta}}_c = 0$ bare value through the inequality ${\tilde{v}}_{Fc} \leq v_{Fc}^0$. 
Here $v_{Fc}^0$ is the velocity, Eq. (\ref{equB27}) of Appendix \ref{APPB}, in the ${\tilde{\delta}}_c=0$ 
compressibility expression. The corresponding effects on the $k=0$ value of ${\cal{V}}_{re} (k)$ are 
then found in Appendix \ref{APPC3} to lead to
\begin{eqnarray}
{\cal{V}}_{re} (0) & = & {\pi\over 2}\,\alpha_c\,v_{Fc}
\hspace{0.20cm}{\rm where}\hspace{0.20cm}
\alpha_c = {\xi_c^4- \chi_c^4 ({\tilde{\xi}}_c)\over \chi_c^4 ({\tilde{\xi}}_c)}
\hspace{0.20cm}{\rm and}
\nonumber \\
\chi_c ({\tilde{\xi}}_c) & = & {\tilde{\xi}}_c \hspace{0.20cm}{\rm for}\hspace{0.20cm}
{\tilde{\delta}}_c \leq {\tilde{\delta}}_c^{\,\,\breve{}} - \delta
\nonumber \\
& = & {\tilde{\xi}}_c^{\,\,\breve{}} \hspace{0.20cm}{\rm for}\hspace{0.20cm}
{\tilde{\delta}}_c \leq {\tilde{\delta}}_c^{\,\,\breve{}} + \delta \hspace{0.20cm}{\rm with}\hspace{0.20cm}
\delta \ll {\tilde{\delta}}_c^{\,\,\breve{}} \, ,
\label{equ9}
\end{eqnarray}
so that, $\int_0^{\infty}dr V_{re} (r) = {\pi\over 4}\,\alpha_c\,v_{Fc}$. Here, 
\begin{eqnarray}
{\tilde{\xi}}_c^{\,\,\breve{}} & = & {\xi_c^2\over\sqrt{2}} \in ]1/\sqrt{2},1[
\hspace{0.20cm}{\rm and}
\nonumber \\
{\tilde{\delta}}_c^{\,\,\breve{}} & = & \xi_c \left(1 - {\xi_c\over\sqrt{2}}\right) \in ]0,(\sqrt{2}-1)/\sqrt{2}[ \, .
\label{equ10}
\end{eqnarray}
The only property of $\chi_c ({\tilde{\xi}}_c)$ in the small interval 
${\tilde{\delta}}_c \in [{\tilde{\delta}}_c^{\,\,\breve{}} - \delta,{\tilde{\delta}}_c^{\,\,\breve{}} + \delta]$,
where $\delta \ll {\tilde{\delta}}_c^{\,\,\breve{}}$, needed for our studies is that its derivative with 
respect to ${\tilde{\xi}}_c$ has no discontinuity. 

Diagonalizing the Hamiltonian, Eq. (\ref{equ1}), in the singularities subspace is achieved 
by diagonalizing its terms in Eq. (\ref{equC9}) of Appendix \ref{APPC} except ${\hat{H}}_{re}$
under the transformation, Eq. (\ref{equC10}) of that Appendix. One then
limits the $c$ band momentum summations to the intervals 
given in Table \ref{table3} that refer to such a subspace. This gives
\begin{equation}
:\hat{H}: = \sum_{q=-\pi}^{\pi}{\tilde{\varepsilon}}_c (q):{\tilde{f}}_{q,c}^{\dag}{\tilde{f}}_{q,c}: +
\sum_{q'=-k_F}^{k_F}{\tilde{\varepsilon}}_s (q'):{\tilde{f}}_{q,s}^{\dag}{\tilde{f}}_{q,s}: \, ,
\label{equ11}
\end{equation}
where the spin term $:\hat{H}_s:$ remains that in Eq. (\ref{equ7}).
The $c$ and $s$ band energy dispersions are given by
\begin{eqnarray}
{\tilde{\varepsilon}}_c (q) & = & \left(1 + \beta_c\right)\varepsilon_c (q) \hspace{0.20cm}{\rm for}\hspace{0.20cm} 
q \in [-2k_F,2k_F]
\nonumber \\
& = &  \left(1 + \beta_c \left\{1 - {4t\over W_c^h} \left({\varepsilon_c (q)\over W_c^h}\right)\right\}\right)\varepsilon_c (q) 
\nonumber \\
& & \hspace{2.25cm}{\rm for}\hspace{0.20cm} \vert q\vert \in [2k_F,\pi]  
\nonumber \\
{\tilde{\varepsilon}}_s (q') & = & \varepsilon_s (q') \hspace{0.20cm}{\rm for}\hspace{0.20cm} 
q' \in [-k_F,k_F] \, .
\label{equ12}
\end{eqnarray}
Here $\varepsilon_c (q)$ and $\varepsilon_s (q)$ are ${\tilde{\delta}}_c = 0$ bare  
energy dispersions \cite{Carmelo_19}, $W_c^h = 4t - W_c^p = \varepsilon_c (\pm\pi)$
and $W_c^p = - \varepsilon_c (0)$ are corresponding $c$ hole and $c$ particle occupancy
energy bandwidths, respectively, and the renormalized ${\tilde{\varepsilon}}_c (q)$ expression is obtained 
in Appendix \ref{APPC}. The $\beta_c$ expression, 
\begin{equation}
\beta_c = {\xi_c^2 - \chi_c^2 ({\tilde{\xi}}_c)\over \chi_c^2 ({\tilde{\xi}}_c)} \, ,
\label{equ13}
\end{equation}
is derived from that of $\alpha_c$ in Eq. (\ref{equ9}) by use of the relation, $\beta_c = \sqrt{1 + \alpha_c} -1$. 
This ${\tilde{\varepsilon}}_c (q)$ expression is valid under the $\xi_c\rightarrow {\tilde{\xi}}_c$ 
transformation for the interval $\xi_c \in [\xi_c^0,\sqrt{2}[$ where $\xi_c^0 = \sqrt{1 + W_c^p/4t}$.
Consistent with the unitary-limit MQIM-HO regime, this excludes electronic densities very near 
$n_e=1$ for all $u$ values and excludes large $u$ values for the remaining electronic densities.
For $\xi_c < \xi_c^0$ the ${\tilde{\varepsilon}}_c (q)$ expression is slightly different, as the renormalized energy bandwidth 
${\tilde{W}}_c^p = - {\tilde{\varepsilon}}_c (0)$ maximum enhancement parameter is smaller
than $(1 + \beta_c)=2/\xi_c^2$. The latter is that of ${\tilde{v}}_{Fc}$, Eq. (\ref{equC35}) of Appendix \ref{APPC}.

That the $s$ (spin) energy dispersion, ${\tilde{\varepsilon}}_s (q')=\varepsilon_s (q')$, 
in Eq. (\ref{equ12}) and corresponding $s$ (spin) band Fermi velocity, ${\tilde{v}}_{Fs} = v_{Fs} = v_s (k_F)$,
remain invariant under the effects of the finite-range interactions whereas
the $c$ (charge) band Fermi velocity, ${\tilde{v}}_{Fc} = {\tilde{v}}_c (2k_F)$, is slightly increased as the range of 
interactions increases, is known from numerical studies. 
(See the related charge and spin spectra in Fig. 7 of Ref. \onlinecite{Hohenadler_12} and 
the corresponding discussion.)

Finally, combining the parameters expressions in Eqs. (\ref{equ8}) and (\ref{equ9}) one finds
that at $k=0$ the Fourier transform of the potential $V_c (x)$ is related to that of $V_{re} (r)$ as
\begin{eqnarray}
{\cal{V}}_c (0) & = & - C_{ce}\,{\cal{V}}_{re} (0)
\hspace{0.20cm}{\rm where}
\nonumber \\
C_{ce} & = & 1
\hspace{0.20cm}{\rm for}\hspace{0.20cm} {\tilde{\delta}}_c \leq {\tilde{\delta}}_c^{\,\,\breve{}} - \delta
\nonumber \\
& = & {2\,(\xi_c^4-{\tilde{\xi}}_c^4)\over {\tilde{\xi}}_c^2\,\xi_c^2(4-\xi_c^4)}
\hspace{0.20cm}{\rm for}\hspace{0.20cm} {\tilde{\delta}}_c \geq {\tilde{\delta}}_c^{\,\,\breve{}} + \delta \, .
\label{equ14}
\end{eqnarray}
Hence $\int_0^{\infty}dx (-V_{c} (x)) = C_{ce}\int_0^{\infty}dr V_{re} (r)$ where $C_{ce} \geq 1$.

\section{The one-electron spectral function}
\label{SECV}

As noted in previous sections, in the case of the spectral structures associated with the stacks of TTF molecules,
our analysis relies on the following exact symmetry relation between 
the one-electron removal and addition spectral functions, Eq. (\ref{equ3}),
for electronic densities $n_e\in ]1,2[$ and $\bar{n}_e = 2- n_e \in ]0,1[$,
respectively,
\begin{eqnarray}
B_{-1} (k,\omega)\vert_{n_e = 1.41} & = & B_{+1} (-k,-\omega)\vert_{n_e = 2-1.41=0.59} 
\nonumber \\
& = & B_{+1} (k,-\omega)\vert_{n_e =0.59} \, .
\label{equ15}
\end{eqnarray}
Here we used the symmetry, $B_{\gamma} (k,\omega) = B_{\gamma} (-k,\omega)$. 

\subsection{The one-electron spectral function near the branch lines}
\label{SECVA}

Within the MQIM-HO, one finds that for small energy deviations $[{\tilde{\omega}}_{c_c} (k)-\omega]\gamma >0$ 
near the (i) $c_c =c,c'$ and (ii) $c_c =c'',c'''$ branch lines and $[{\tilde{\omega}}_{s_s} (k)-\omega]\gamma >0$ 
in the vicinity of the (i) $s_s = s$ and (ii) $s_s = s'$ branch lines for (i) $\gamma =-1$ and
(ii) $\gamma =+1$, respectively, the spectral function, Eq. (\ref{equ3}), behaves as \cite{Carmelo_19}
\begin{eqnarray}
B_{\gamma} (k,\omega) & \approx & \sum_{\iota=\pm 1}C_{c_c,\gamma,\iota}
\nonumber \\
& \times & {\rm Im}\left\{\left({(\iota)\over{[\tilde{\omega}}_{c_c} (k)-\omega]\gamma - {i\over 2\tau_{c_c} (k)}}\right)^{-\zeta_{c_c} (k)}\right\} 
\nonumber \\
B_{\gamma} (k,\omega) & = & C_{s_s,\gamma} ([{\tilde{\omega}}_{s_s} (k)-\omega]\gamma)^{\zeta_{s_s} (k)} \, .
\label{equ16}
\end{eqnarray}
Here $C_{c_c,\gamma\,\iota}$ and $C_{s_s,\gamma}$ are $n_e$, $u=U/4t$, and ${\tilde{\xi}}_c$ dependent constants
for energy and momentum values corresponding to such small energy deviations 
and ${\tilde{\omega}}_{c_c} (k)\gamma >0$, ${\tilde{\omega}}_{s_s} (k)\gamma >0$, and
$\omega\gamma >0$ are high energies beyond the reach of the TLL. For ${\tilde{\delta}}_c >0$
the expressions, Eq. (\ref{equ16}), near the branch lines apply only to
$k$ intervals for which the exponents are negative. For the $c_c = c,c',c'',c'''$ branch lines 
the $c$ mobile scattering center lifetime $\tau_{c_c} (k)$ is very large for such $k$ intervals \cite{Carmelo_19} and
the singularities in Eq. (\ref{equ16}) refer to peak structures with very small widths.

That the $s_s =s,s'$ branch lines coincide with edges of the support for the spectral function 
ensures that near them the line shape is power-law like. This applies to $k$ intervals for which $\zeta_{s_s} (k) <0$.
The $c_c = c,c',c'',c'''$ branch likes run within the weight continuum. 
The width of the $k$ intervals for which the lifetime $\tau_{c_c} (k)$ in Eq. (\ref{equ16}) is large
and $\zeta_{c_c} (k)$ negative tends to decrease upon increasing ${\tilde{\delta}}_c$ in the 
range ${\tilde{\delta}}_c\in [{\tilde{\delta}}_c^{\oslash},{\tilde{\delta}}_c^{({1\over 2})}[$.
Here ${\tilde{\delta}}_c^{\oslash} = \xi_c(1 - 1/\xi_c^2)$. This is due to the effects
of the relaxation processes induced by the finite-range interactions then becoming more pronounced. Actually, upon 
increasing ${\tilde{\delta}}_c$ towards ${\tilde{\delta}}_c^{({1\over 2})}$, such processes 
progressively wash out all peak structures \cite{Carmelo_19}.

The spectra of the $s$ and $c,c'$ branch lines in the expressions for the $\gamma =-1$ spectral function 
in Eq. (\ref{equ16}) involve the $s$ and $c$ dispersions in Eq. (\ref{equ12}). They read
\begin{eqnarray}
{\tilde{\omega}}_{s} (k) & = & {\tilde{\varepsilon}}_s (k) = \varepsilon_{s} (k) \leq 0\hspace{0.20cm}{\rm for}
\hspace{0.20cm} k = -q' \in [-k_F,k_F] 
\nonumber \\
{\tilde{\omega}}_c (k) & = & {\tilde{\varepsilon}}_c (\vert k\vert + k_F)\leq 0
\hspace{0.20cm}{\rm for}
\nonumber \\
k & = & -{\rm sgn}\{k\} k_F - q \in [-k_F,k_F] \hspace{0.20cm}{\rm with}
\nonumber \\
q & \in & [-2k_F,-k_F] \hspace{0.20cm}{\rm for}\hspace{0.20cm}k \in [0,k_F]
\nonumber \\
q & \in & [k_F,2k_F] \hspace{0.20cm}{\rm for}\hspace{0.20cm}k \in [-k_F,0]
\nonumber \\
{\tilde{\omega}}_{c'} (k) & = & {\tilde{\varepsilon}}_c (\vert k\vert - k_F) \leq 0
\hspace{0.20cm}{\rm for}
\nonumber \\
k & = & {\rm sgn}\{k\} k_F - q \in [-3k_F,3k_F] \hspace{0.20cm}{\rm with}
\nonumber \\
q & \in & [-k_F,2k_F] \hspace{0.20cm}{\rm for}\hspace{0.20cm}k \in [-3k_F,0]
\nonumber \\
q & \in & [-2k_F,k_F] \hspace{0.20cm}{\rm for}\hspace{0.20cm}k \in [0,3k_F]\, .
\label{equ17}
\end{eqnarray}
Using the relation, Eq. (\ref{equ15}), minus the spectra of the $s'$ and $c'',c'''$ branch lines in the expressions 
for the $\gamma =+1$ spectral function at density $n_e =0.59$, we find the following spectra for the
$\gamma =-1$ spectral function at density $n_e^F = 2 - n_e = 1.41$,
\begin{eqnarray}
- {\tilde{\omega}}_{s'} (k) & = & {\tilde{\varepsilon}}_s (\vert k\vert - 2k_F) \leq 0\hspace{0.20cm}{\rm for}
\nonumber \\
k & = & {\rm sgn}\{k\} 2k_F - q' \in [-3k_F,-k_F]\,;[k_F,3k_F]
\nonumber \\
& & {\rm with}\hspace{0.20cm}q' \in [-k_F,k_F] 
\nonumber \\
- {\tilde{\omega}}_{c''} (k) & = & - {\tilde{\varepsilon}}_c (\vert k\vert + k_F)\leq 0
\hspace{0.20cm}{\rm for}
\nonumber \\
k & = & -{\rm sgn}\{k\} k_F + q \in [-(\pi - k_F),-k_F]
\nonumber \\
& \in & [k_F, (\pi - k_F)]\hspace{0.20cm}{\rm with}
\nonumber \\
q & \in & [-\pi,-2k_F] \hspace{0.20cm}{\rm for}\hspace{0.20cm}k \in [-(\pi - k_F),-k_F]
\nonumber \\
q & \in & [2k_F,\pi] \hspace{0.20cm}{\rm for}\hspace{0.20cm}k \in  [k_F, (\pi - k_F)]
\nonumber \\
- {\tilde{\omega}}_{c'''} (k) & = & - {\tilde{\varepsilon}}_c (\vert k\vert - k_F) \leq 0
\hspace{0.20cm}{\rm branch}\hspace{0.10cm}{\rm I}\hspace{0.20cm}{\rm for}
\nonumber \\
k & = & {\rm sgn}\{k\} k_F + q \in [-\pi,-3k_F]
\nonumber \\
& \in & [3k_F,\pi]\hspace{0.20cm}{\rm with}
\nonumber \\
q & \in & [-(\pi - k_F),-2k_F] \hspace{0.20cm}{\rm for}\hspace{0.20cm}k \in  [-\pi,-3k_F]
\nonumber \\
q & \in & [2k_F, (\pi - k_F)] \hspace{0.20cm}{\rm for}\hspace{0.20cm}k \in [3k_F,\pi] 
\nonumber \\
- {\tilde{\omega}}_{c'''} (k) & = & - {\tilde{\varepsilon}}_c (\vert k\vert - (2\pi - k_F)) \leq 0
\hspace{0.20cm}{\rm branch}\hspace{0.10cm}{\rm II}\hspace{0.20cm}{\rm for}
\nonumber \\
k & = & - {\rm sgn}\{k\}(2\pi -k_F) + q \in [-\pi,-(\pi -k_F)]
\nonumber \\
& \in & [(\pi -k_F),\pi]\hspace{0.20cm}{\rm with}
\nonumber \\
q & \in & [(\pi - k_F),\pi] \hspace{0.20cm}{\rm for}\hspace{0.20cm}k \in [-\pi,-(\pi -k_F)]
\nonumber \\
q & \in &[-\pi,-(\pi -k_F)]\hspace{0.20cm}{\rm for}\hspace{0.20cm}k \in [(\pi -k_F),\pi] .
\label{equ18}
\end{eqnarray}
The branch-line spectra, Eqs. (\ref{equ17}) and (\ref{equ18}), are represented  
in the sketch of Fig. \ref{figure1} by solid lines. Their expressions in these equations are defined for the
$q$ and $q'$ intervals given in the equations.

The exponents $\zeta_{s_s} (k)$ and $\zeta_{c_c} (k)$ in Eq. (\ref{equ16}) that control the line
shape near the (i) $s_s = s$ and $c_c = c,c'$ branch lines for $\gamma = -1$ and (ii)
$s_s = s'$ and $c_c = c'',c'''$ branch lines for $\gamma = +1$ are given by
\begin{eqnarray}
\zeta_{s_s} (k) & = & -1 + \sum_{\iota=\pm}\left({(1+\gamma){\tilde{\xi}}_c\over 2}
+ {\iota\over 2{\tilde{\xi}}_c} - \gamma\,{\tilde{\Phi}}_{c,s}(\iota 2k_F,q')\right)^2
\nonumber \\
\zeta_{c_c} (k) & = & -{1\over 2} + \sum_{\iota=\pm}\left({{\tilde{\xi}}_c\over 4} 
+ \gamma\,{\tilde{\Phi}}_{c,c}(\iota 2k_F,q)\right)^2 \, .
\label{equ19}
\end{eqnarray}
Here $q'$ and $q$ belong to the intervals defined from those in Eqs. (\ref{equ17}) and (\ref{equ18}) 
under the replacement of $\pm k_F$ and $\pm 2k_F$ by $\pm (k_F - p_{Fs})$ and $\pm (2k_F + \gamma\,p_{Fc})$,
respectively. This is consistent with the intervals in Table \ref{table1} for processes (2-Rem), (3-Rem), (2-Add), and (3-Add). 

On the other hand, within the low-energy TLL processes (1-Rem) and (1-Add), the scattering centers are created
in the same very small intervals given in Table \ref{table2} as the scatterers. Hence the $c$ and $s$ scattering centers lose
their identity. Indeed, they cannot be distinguished from the TLL ``holons``and
``spinons'' in the low-energy and small-momentum particle-hole
processes. As a result, the spectral-function TLL exponents have different expressions 
than those in Eq. (\ref{equ19}) \cite{Carmelo_19}.

The contribution of the $s$ particle scatterers phase shifts to the exponents expressions given in Eq. (\ref{equ19})
has been accounted for. They do not appear explicitly in these expressions because 
they simplify to $\gamma\,2\pi{\tilde{\Phi}}_{s,c}(\iota k_F,q) = - \gamma\iota\pi/\sqrt{2}$ and
$-2\pi{\tilde{\Phi}}_{s,s} (\iota k_F,q') = - \iota\pi/\sqrt{2}$. Here 
$\iota = \pm $ for both $\gamma = \pm 1$ and the $c$ and $s$ scattering centers
$q$ and $q'$ intervals, respectively, are provided in Table \ref{table1}. Their simplicity is due both to the 
global spin $SU(2)$ symmetry and their invariance under the $\xi_c\rightarrow {\tilde{\xi}}_c$ transformation.

\subsection{The one-electron spectral function near the $c-s$ boundary lines}
\label{SECVB}

Another type of spectral-function singularity is located on the $c-s$ boundary lines shown in Fig \ref{figure1}.
The spectral-function expressions near them are for ${\tilde{\delta}}_c=0$ 
provided by the PDT \cite{Carmelo_17}. They involve an exponent $-1/2$ that remains invariant 
under the $\xi_c\rightarrow{\tilde{\xi}}_c$ transformation. The corresponding ${\tilde{\delta}}_c >0$ 
expressions are then obtained under the replacement of the ${\tilde{\delta}}_c=0$ bare $c$ particle energy dispersion
and group velocity by ${\tilde{\varepsilon}}_c (q)$, Eq. (\ref{equ12}), and ${\tilde{v}}_{c}(q)$,
Eq. (\ref{equC11}) of Appendix \ref{APPC}, respectively.

The spectral weight distribution in the vicinity of such lines results from
the processes (4-Rem) and (4-Add) defined in Sec. \ref{SECIVA}. 
As given in Table \ref{table1}, under such processes the $c$ hole 
is created at $q \in [-q_c^h,q_c^h]$ where $\vert{\tilde{v}}_c (\pm q_c^h)\vert = v_{Fs}^-$, $0 < q_c^h < 2k_F$,
and $v_{Fs}^- \equiv v_s (k_F - \delta p_{Fs})$ for $\gamma = -1$. For $\gamma = +1$,
the $c$ particle is created at $q \in [-\pi,-q_c]\,;[q_c,\pi]$ where 
$\vert{\tilde{v}}_c (\pm q_c)\vert = v_{Fs}^-$ 
%CORR - q_c^h REPLACED BY q_c
and $2k_F < q_c < \pi$.
Here $q_c^h = q_c$ for $u\rightarrow 0$ and $q_c^h = 0$ and $q_c = \pi$ for $u\rightarrow\infty$.
The $s$ hole is created both for $\gamma = -1$ and $\gamma = +1$ at a $s$ band momentum $q'$ 
such that ${\tilde{v}}_{c}(q) = v_{s}(q')$. Hence $q$ and $q'$ are not independent of each other. 
The corresponding $c-s$ boundary line $(k,\omega)$-plane spectrum is of the general form,
\begin{eqnarray}
{\tilde{\omega}}_{c-s} (k) & = & 
\left(\gamma\,{\tilde{\varepsilon}}_c (q) - \varepsilon_s (q')\right)\,\delta_{{\tilde{v}}_{c}(q),{\tilde{v}}_{s}(q')}
\nonumber \\
k & = & \mp (\gamma - 1)k_F + \gamma\,q - q'\hspace{0.20cm}{\rm for}\hspace{0.20cm}\gamma = \pm 1 \, .
\label{equ20}
\end{eqnarray}
%CORR - "Here ${\tilde{v}}_{s}(q')=v_{s}(q')$" added
Here ${\tilde{v}}_{s}(q')=v_{s}(q')$.
The $c-s$ boundary spectra are represented in the sketch of Fig. \ref{figure1} by dotted lines. 

In the vicinity of such a line the one-electron spectral function, Eq. (\ref{equ3}), has the following behavior,
\begin{eqnarray}
B_{\gamma} (k,\omega) & = & C_{c-s}^{\iota}\Bigl([{\tilde{\omega}}_{c-s} (k)-\omega]\gamma\iota\Bigr)^{-1/2}
\nonumber \\
{\rm for} && {\rm small}\hspace{0.20cm}[{\tilde{\omega}}_{c-s} (k)-\omega]\gamma\iota \, .
\label{equ21}
\end{eqnarray}
Here $C_{c-s}^{\iota}$ where $\iota =+$ refers to $[{\tilde{\omega}}_{c-s} (k)-\omega]\gamma >0$
and $\iota =-$ to $[{\tilde{\omega}}_{c-s} (k)-\omega]\gamma <0$
are $n_e$, $u=U/4t$, and ${\tilde{\xi}}_c$ dependent constants for energy and momentum values 
corresponding to the small energy deviations $[{\tilde{\omega}}_{c-s} (k)-\omega]\gamma$ 
and ${\tilde{\omega}}_{c-s} (k)\gamma >0$ and $\gamma\omega >0$ are high energies beyond 
those of the TLL.

\section{ARPES in TTF-TCNQ}
\label{SECVI}

\subsection{ARPES data}
\label{SECVIA}

In the top panel of Fig. \ref{figure2} we display the raw ARPES band map along 
the 1D direction of TTF-TCNQ corresponding to the $\Gamma$Z high-symmetry line 
of the Brillouin zone. The $\Gamma$ point in the zone center at zero momentum 
as well as the Z point at the zone boundary are indicated by solid vertical 
lines. The energies are referenced to the chemical potential at zero energy, 
marked by a horizontal solid line.  In contrast to our previous 
photoemission data \cite{Claessen_02,Sing_03}, already in the raw data dispersive features are clearly 
discernible due to the better momentum resolution and finer  $k$-grid, 
e.g., the V-shape like intensity distribution centered around zero momentum and 
a structure, shifting away from close to the chemical potential at about 
0.23\AA$^{-1}$ towards the zone boundary. Even more details become apparent in 
the negative second derivative along the energy axis, clipped at zero 
intensity, which is plotted in the lower panel of Fig. \ref{figure2}. These are indicated 
by solid lines and denoted by the characters $s$, $c$, $c'$, and $c''$ when
associated with the theoretical branch lines and by dotted lines and denoted by
$c-s$ in the case of boundary lines. Such branch and boundary lines were introduced in Sec. \ref{SECIVA}. 
They have been observed previously but with less 
clarity and are discussed in the following paragraphs in more detail based on 
our theoretical analysis.

\subsection{Agreement between ARPES and the theory predictions}
\label{SECVIB}

The use of the one-electron spectral function expressions provided by the MQIM-HO,
Eqs. (\ref{equ16}) and (\ref{equ21}), allows the prediction of (i) the location in 
the $(k,\omega)$ plane of the experimentally observed ARPES structures 
at energy scales beyond the reach of the TLL and (ii) the values of the low-energy TLL SDS exponent 
$\alpha$ in Eq. (\ref{equ6}). Indeed, the latter only depends on the charge parameter
${\tilde{\xi}}_c$ in the high-energy exponent expressions, Eq. (\ref{equ19}).
In using our $T=0$ theoretical results to describe high-energy ARPES data taken at $60$ K, we expect (and observe)
that the corresponding predicted peak structures are slightly smeared by thermal 
fluctuations and coupling to phonons. 
\begin{figure}
\begin{center}
\subfigure{\includegraphics[width=8.00cm]{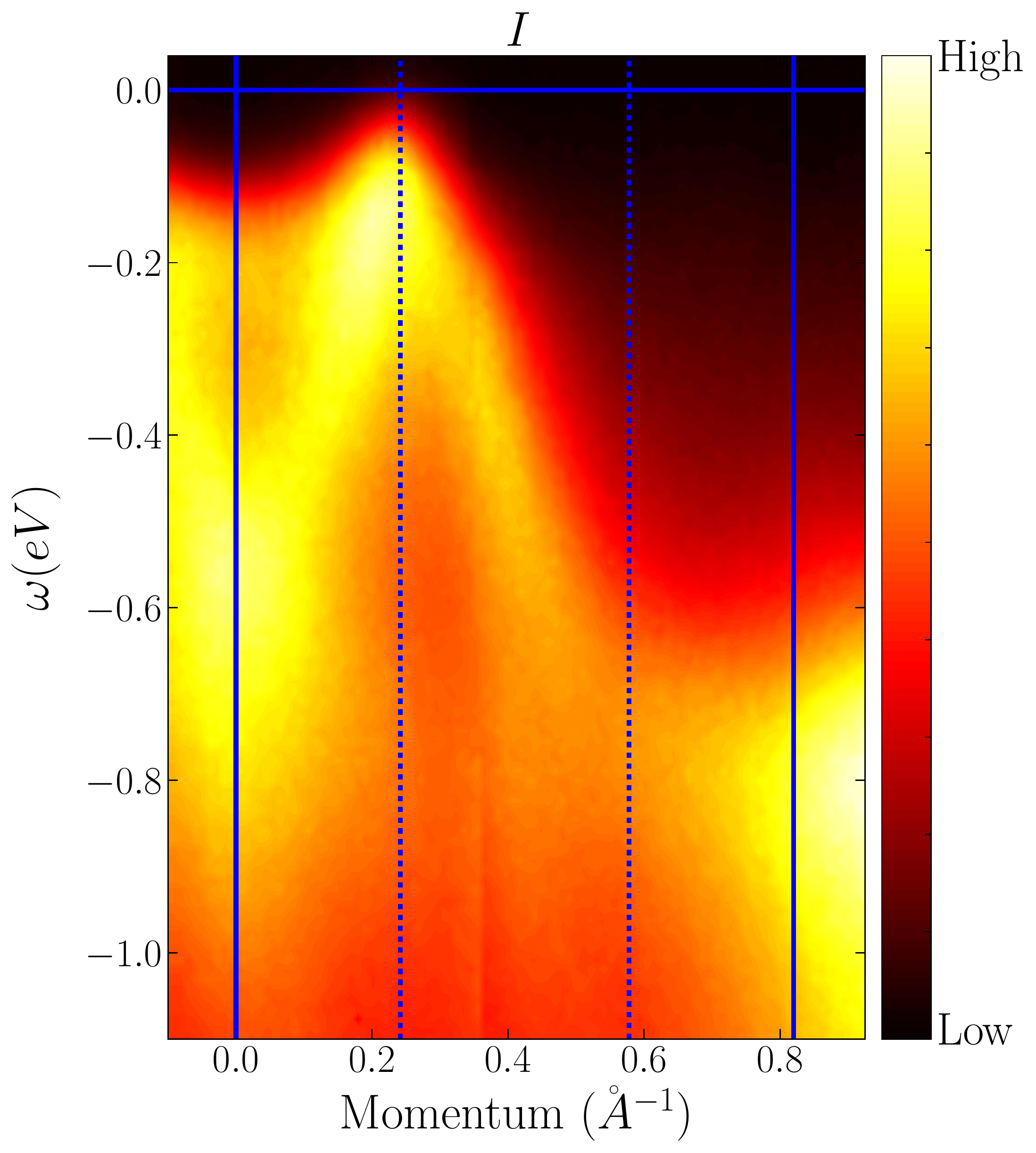}}
\subfigure{\includegraphics[width=8.00cm]{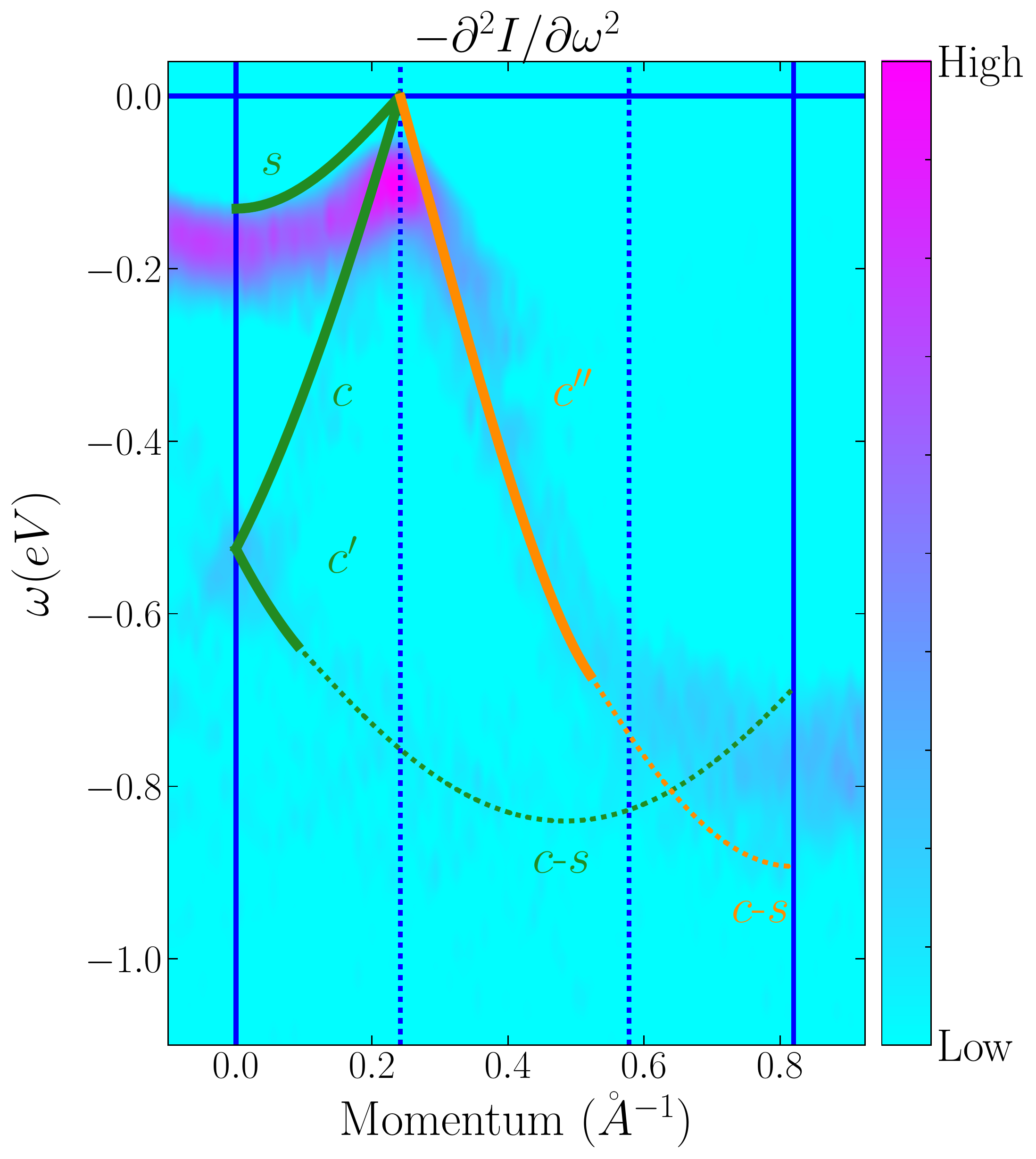}}
\caption{Raw TTF-TCNQ ARPES data and second-derivative ARPES map
showing the experimental dispersions obtained along the easy-transport axis 
and matching theoretical boundary lines and branch lines for the $k>0$ intervals for which the 
exponents of the latter are negative. Their choice is justified in Sec. \ref{SECVIB}. 
The theoretical lines refer to (i) $u=U/4t=0.800$ and $t=0.29$ eV 
and (ii) $u=1.225$ and $t=0.24$ eV for the (i) TTF (orange) and (ii) TCNQ 
(green) stacks of molecules, respectively.
The branch and boundary lines are represented by solid and dotted lines, respectively.
The spectral function line shape near and below the branch lines has the singular form
given in Eq. (\ref{equ16}) for the $k$ intervals for which the corresponding exponents
are negative. Its line shape near both sides of the boundary lines is provided in Eq. (\ref{equ21}). 
(The TTF related $s'$ and $c'''$ branch lines do not appear in the figure because their exponents are positive.)
The TCNQ and TTF stacks of molecules related boundary lines emerge from the $c'$ and $c''$ branch lines
at $k = k_{cs}^Q = 0.115\,\textrm{\AA}^{-1}$ and $k = k_{cs}^F = 0.520\,\textrm{\AA}^{-1}$, respectively,
and run up to $k = \pi/a_0 \approx 0.823\,\textrm{\AA}^{-1}$.}
\label{figure2}
\end{center}
\end{figure}

To access the two sets of parameter values suitable for the description of the spectral features related to
the TTF and TCNQ stacks of molecules, two entangled criteria associated with
the spectra and the matrix elements in the spectral function,
Eq. (\ref{equ3}), respectively, are used. 

The first criterion refers to the overall agreement between the $(k,\omega)$-plane spectra shape of the theoretical 
branch lines and $c-s$ boundary lines and that of the ARPES maps shown in 
Fig. \ref{figure2}. The latter criterion involves the exponents in Eq. (\ref{equ19}) that control the quantum overlap 
in the spectral-function matrix elements near the branch lines and the $k$ intervals, Eq, (\ref{equ20}), for which 
boundary lines exist. It refers to the agreement between the location in the 
$(k,\omega)$ plane of (i) the $k$ intervals of both boundary-line singularities and those of the branch-line singularities 
for which their exponents are negative and (ii) that of the corresponding high-energy 
ARPES structures in Fig. \ref{figure2}. 

\begin{figure}
\begin{center}
\centerline{\includegraphics[width=8.75cm]{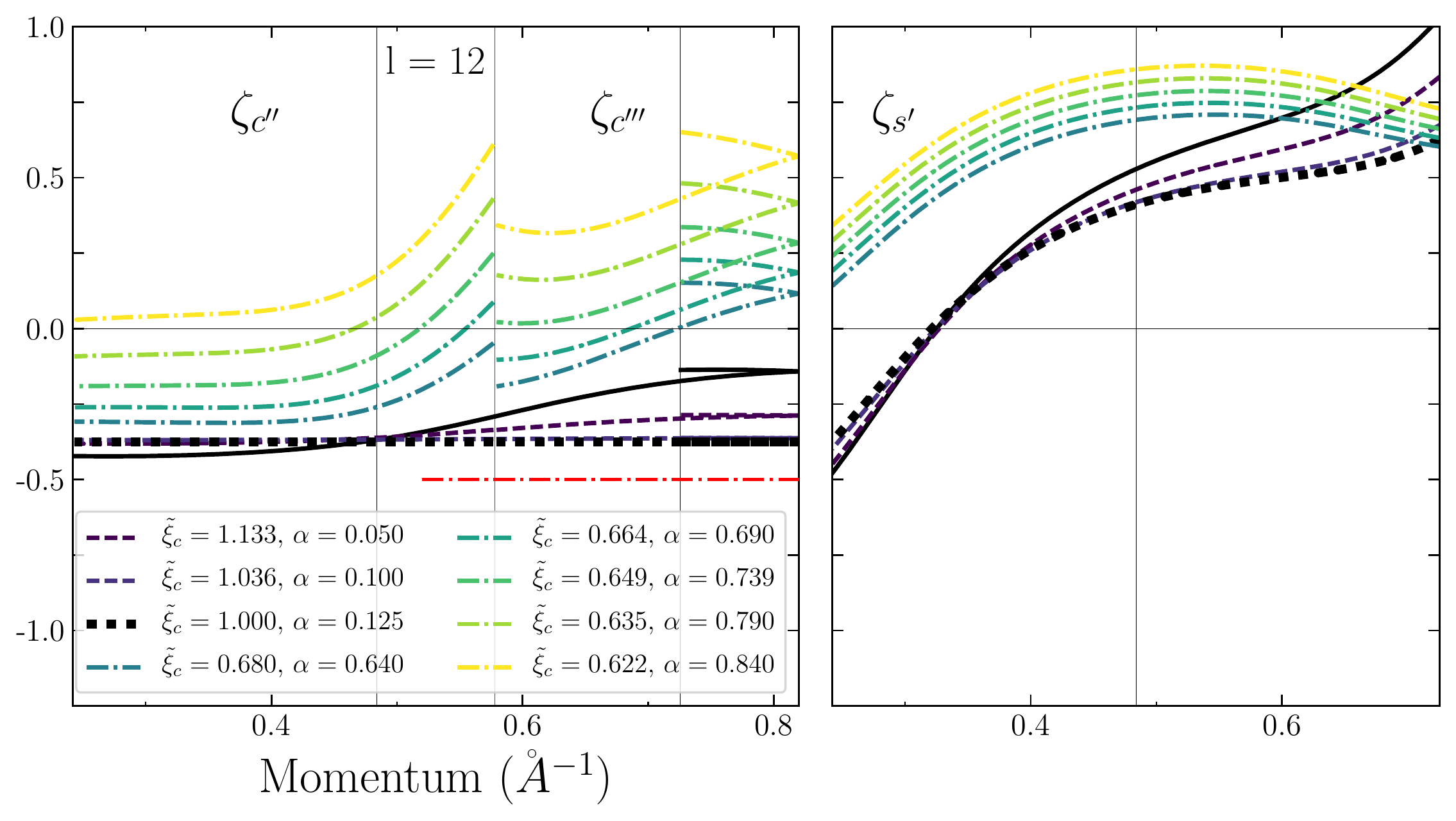}}
\caption{The exponents, Eq. (\ref{equ19}), in the
spectral function, Eq. (\ref{equ16}), that control the line shape
near the theoretical $c''$ branch line, two branches of the $c'''$ branch line, and $s'$ branch line in Fig. \ref{figure1}
for $n_e^F = 2- n_e=1.41$, $u=U/4t=0.80$, and $l=12$. 
The $k$ intervals for which these exponents are negative
describe the $(k,\omega)$-plane location of the corresponding experimental 
high-energy structures in the ARPES maps of Fig. \ref{figure2} related to the 
TTF stacks of molecules. This holds
for the specific exponent lines corresponding to the parameter
values ${\tilde{\xi}}_c= {\tilde{\xi}}_c^F = 0.649$ and $\alpha = \alpha^F = 0.739$ for which
$\zeta_{c''} (k)$ crosses zero at $k = k_{cs}^F = 0.520\,\textrm{\AA}^{-1}$
and all exponents meet the agreement criteria.
The $-1/2$ exponent dashed-dotted line refers to the $c-s$ boundary line.}
\label{figure3}
\end{center}
\end{figure}

For the TTF stacks of molecules, the analysis of the problem involves
the $c-s$ boundary line running in the interval $k \in [k_{cs}^F,\pi/a_0]$ where $k_{cs}^F < \pi/a_0 - k_F$
and the $k$ intervals for which the branch-line exponents $\zeta_{c''} (k)$, $\zeta_{c'''} (k)$, and $\zeta_{s'} (k)$ 
are negative and positive, respectively. They are plotted in Fig. \ref{figure3} for the parameter values
found below to be suitable for TTF. In the case of the TCNQ stacks of molecules, 
the $c-s$ boundary line running for $k \in [k_{cs}^Q,\pi/a_0]$ where $k_{cs}^Q < k_F$
and the $k$ dependence of the exponents $\zeta_c (k)$, $\zeta_{c'} (k)$, and $\zeta_{s} (k)$ are
those involved in such an analysis. They are plotted in Fig. \ref{figure4} for the TCNQ parameter values whose 
choice is justified below.

In Figs. \ref{figure3} and \ref{figure4} different curves  are associated with different values of the 
renormalized charge parameter ${\tilde{\xi}}_c$ and thus of the TLL charge parameter 
$K_{\rho} = {\tilde{\xi}}_c^2/2$, SDS exponent $\alpha = (2-{\tilde{\xi}}_c^2)^2/(8{\tilde{\xi}}_c^2)$ in Eq. (\ref{equ6}),
and effective range $R_{\rm eff}$, Eq. (\ref{equA10}) of Appendix \ref{APPA}. 
The black solid and black dashed lines refer to the ${\tilde{\delta}}_c =0$ bare limit and 
the interval ${\tilde{\delta}}_c \in ]0,{\tilde{\delta}}_c^{(1)}[$, respectively. 

For simplicity, in the following we limit our analysis regarding the fulfillment of
the agreement criteria to momentum values $k>0$. However, similar results hold for $k<0$.
One finds from analysis of the experimentally observed high-energy ARPES 
structures in Fig. \ref{figure2}
associated with the TTF stacks of molecules that 
the $c''$ branch line exponent should be negative for 
the interval $k \in [k_F + \delta p_{Fc},k_{cs}^F]$ and positive for 
the small interval $k \in [k_{cs}^F,\pi - k_F]$.
In contrast, the exponents of the two branches of the $c'''$ branch line 
and that of the $s'$ branch line should be positive for all their $k$ intervals. 
The $c-s$ branch line that emerges from the
$c''$ branch line at $k = k_{cs}^F$ and runs until $k = \pi/a_0$
should separate a $(k,\omega)$-plane region above it with finite spectral weight from
a region below it with very little weight. The spectral function expression given in
Eq. (\ref{equ16}) for the $c_c=c''$ branch line refers to the region just below it.
However, that line runs within the spectral-weight continuum and thus can have finite
weight above it. That spectral-function expression does not apply thought to that region. 

On the other hand, in the case of the TCNQ
high-energy ARPES structures in Fig. \ref{figure2}, the $c$ and $s$ branch 
lines exponents should be negative for 
their whole intervals $k \in [0,k_F - \delta p_{Fc}]$ and $k \in [0,k_F-\delta p_{Fs}]$, respectively.
The exponent of the $c'$ branch line should be negative for the small interval $k \in [0,k_{cs}^Q]$ and positive for 
$k \in [k_{cs}^Q,3k_F-\delta p_{Fc}]$. Here $k_{cs}^Q < k_F$ is the momentum at which a $c-s$ boundary emerges 
from the $c'$ branch line. The $s$ branch line {\it must} coincide with the edge of support of the one-electron spectral function.
It separates $(k,\omega)$-plane regions without and with finite spectral weight above
and below that line, respectively. And again the spectral function expression given in
Eq. (\ref{equ16}) for the $c_c=c,c'$ branch lines refers to the region just below them.
These lines run within the spectral-weight continuum yet 
there is no imposition that there is or there is not a significant amount of weight above them.
In the region above them the spectral function expression is not given by Eq. (\ref{equ16}).

For the (i) TTF stacks of molecules with density $n_e^F = 2 - n_e = 1.41$ 
and the (ii) TCNQ stacks of molecules with density $n_e^Q = n_e = 0.59$ the best agreement concerning {\it both} the $(k,\omega)$-plane 
spectra shape and location of the singularities is reached (i) for $u=U/4t = 0.8$ and $t=0.29$ eV ($\xi_c = 1.228$) and  (ii) 
for $u= 1.225$ and $t=0.24$ eV ($\xi_c = 1.171$), respectively. 
This gives interactions $U\approx 1$ eV for both systems, (i) $U = 0.928\approx 0.9$ eV and (ii) $U = 1.176\approx 1.2$ eV for TTF and TCNQ,
respectively.

At such fixed $n_e$ and $u=U/4t$ values, those of $l>5$ and ${\tilde{\xi}}_c$ are determined by
the criterion involving the $(k,\omega)$-plane location of the singularities. In the case of the TFF related spectral features,
the best agreement is reached for $l=12$ at the value ${\tilde{\xi}}_c^F = 0.649$ for which
$\zeta_{c''} (k)$ crosses zero at $k = k_{cs}^F = 0.520\,\textrm{\AA}^{-1}$ in Fig. \ref{figure3}. This
corresponds to $K_{\rho}^F = 0.211$, $R_{\rm eff}^F = 6.173\,a_0 = 23.575\,\textrm{\AA}$, and $\alpha_F = 0.739$. 
At $l = 11$ the agreement is poorer but the other
parameters have nearly the same values. For $l<11$ some of the agreement criteria are not met.

For the TCNQ related spectral features, the best agreement is reached for $l=6$ at the value ${\tilde{\xi}}_c^C = 0.734$ for
which $\zeta_{c'} (k)$ crosses zero at $k = k_{cs}^Q = 0.115\,\textrm{\AA}^{-1}$ in Fig. \ref{figure4}. This
corresponds to $K_{\rho}^C = 0.269$, $R_{\rm eff}^C = 24.828\,a_0 = 94.818\,\textrm{\AA}$, 
and $\alpha_C = 0.495$. For $l>7$ some of the agreement criteria are not met.
At $l = 7$ the agreement is poorer, but all other parameters except $R_{\rm eff}^C$ 
have nearly the same values. Due to its dependence on $l$, Eqs. (\ref{equA10}) and (\ref{equA11}) of 
Appendix \ref{APPA}, it drops to $R_{\rm eff}^C = 12.818\,a_0 = 48.952\,\textrm{\AA}$. 

\section{Discussion and concluding remarks}
\label{SECVII}

In this paper we have reported new high-resolution ARPES data for TTF-TCNQ
and used an extended version of the MQIM-HO \cite{Carmelo_19} to describe
the microscopic mechanisms behind the spectral properties of that conductor at $T=60$ K. 
This involved the use of a model Hamiltonian of general form, Eq. (\ref{equ1}), with transfer 
integral $t$, interaction $U$, and potential $V_e (r)$ specific to the TTF and TCNQ stacks 
of molecules, respectively, to describe their one-electron spectral properties. 

The best agreement between the theory and the high-resolution ARPES data was obtained
for the spectral features related to (i) TTF and (ii) TCNQ stacks of molecules 
for (i) $u=U/4t = 0.80$ with $t= 0.29$ eV and $U= 0.928$ eV and for (ii) $u= 1.225$ with $t=0.24$ eV
and $U= 1.176$ eV, respectively. Despite the smaller $u=U/4t$, $U$, and effective range $R_{\rm eff}$
values, we found the TTF stacks of molecules to be more correlated 
then those of TCNQ, in as far as their smaller charge parameter value ${\tilde{\xi}}_c^F = 0.649$ and
corresponding TLL charge parameter $K_{\rho}^F = 0.21$ (and thus larger SDS exponent $\alpha_F = 0.74$) is concerned.
Indeed, for the TTF stacks of molecules such parameters were found to be given by
${\tilde{\xi}}_c^C = 0.734$, $K_{\rho}^C = 0.27$, and $\alpha_C = 0.50$, respectively. 

The effective ranges of the fractionalized particles charge-charge interaction 
found for TTF and TCNQ read $R_{\rm eff}^F = 6.2\,a_0=23.6\,\textrm{\AA}$ 
and $R_{\rm eff}^C = 24.8\,a_0=94.8\,\textrm{\AA}$,
respectively. However, at $l=7$ for which the agreement with the experimental data is poorer for TCNQ, 
the values of its other parameters remain nearly the same except that 
the effective range drops to $R_{\rm eff}^C = 12.8\,a_0=49.0\,\textrm{\AA}$. In any case, the results reveal that a first-neighbor 
interaction $V$ is not sufficient to describe the physics of TTF-TCNQ. Indeed, $R_{\rm eff} > 6\,a_0$
for both its stacks of molecules with $R_{\rm eff}$ also applying to the related rotated-electron interactions.

A necessary condition for the occurrence of a low-temperature $4k_F$ 
charge-density wave (CDW) is that ${\tilde{\xi}}_c < 1$ and $K_{\rho}<1/2$ 
\cite{Schulz_90,Hohenadler_12}. That for the TTF stacks of molecules
the charge parameters ${\tilde{\xi}}_c$ and $K_{\rho}$ are quite smaller 
than for those of TCNQ is consistent with its $4k_F$ CDW phase observed at temperatures
$T< 49$ K \cite{Kagoshima_88}. 

However, ${\tilde{\xi}}_c < 1$ {\it is not a sufficient condition}
for a $4k_F$ CDW \cite{Hohenadler_12}. Indeed, lattice fermions have for some 
forms of {\it long-range} potentials a metallic ground state for the whole ${\tilde{\xi}}_c<1$ range \cite{Fano_99}. In addition to the 
requirement that ${\tilde{\xi}}_c < 1$, the occurrence of a $4k_F$ CDW depends 
on the specific form of the electronic potential $V_e (r)$. The lack of a low-$T$ $4k_F$ CDW phase for the TCNQ stacks 
of molecules then reveals that, besides the different parameters values reported above,
the type of $r$ dependence of the electronic potentials $V_e (r)$ of the two stacks of molecules is different.
\begin{figure}
\begin{center}
\centerline{\includegraphics[width=8.75cm]{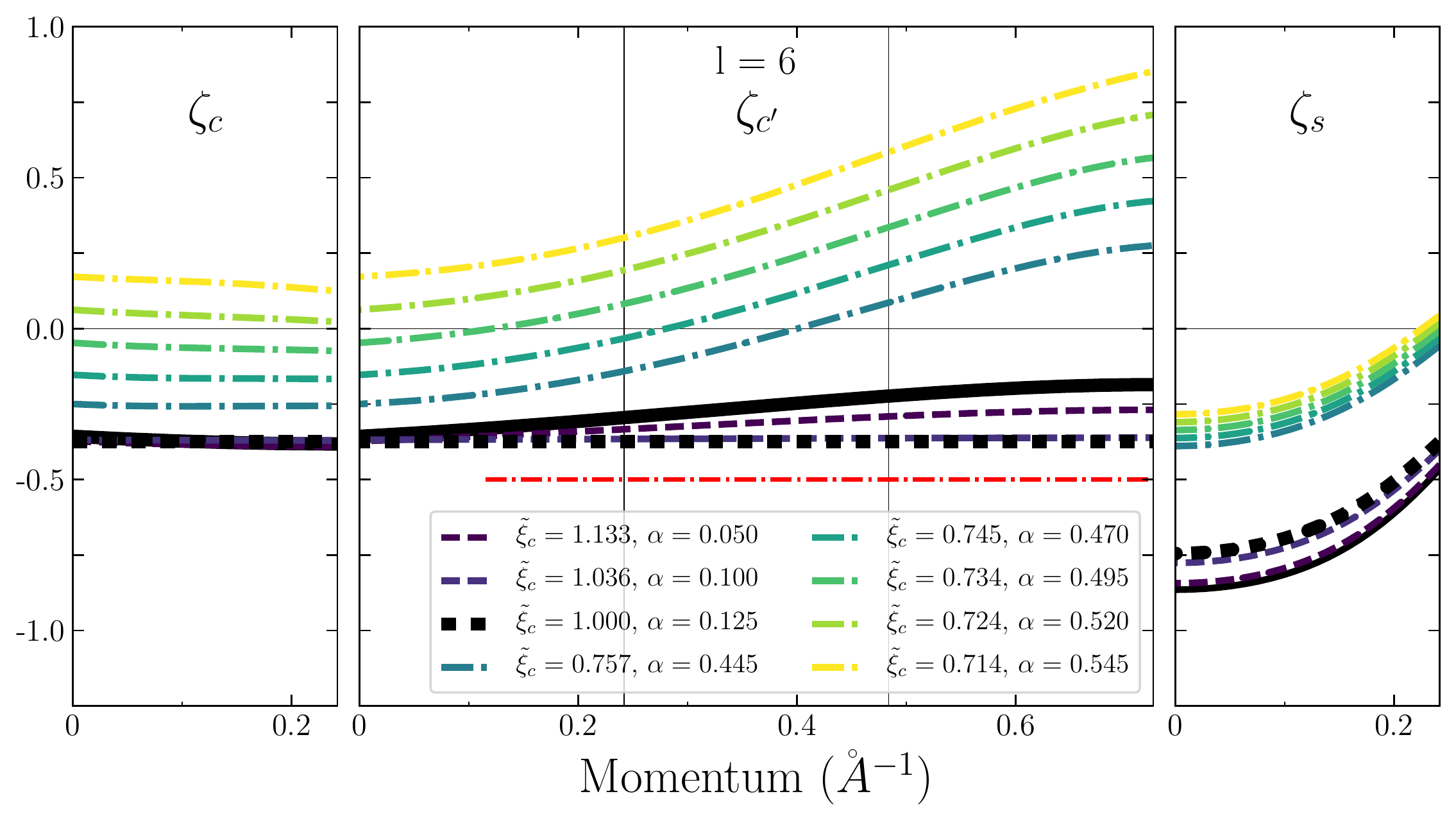}}
\caption{The exponents, Eq. (\ref{equ19}), in the
spectral function, Eq. (\ref{equ16}), that control the line shape
near the theoretical $c$, $c'$, and $s$ branch lines in Fig. \ref{figure1} 
for $n_e^Q = n_e=0.59$, $u=U/4t=1.225$, and $l=6$.
The $k$ intervals for which these exponents are negative
describe the $(k,\omega)$-plane location of the corresponding experimental 
high-energy structures in the ARPES maps of Fig. \ref{figure2} related to the 
TCNQ stacks of molecules.
This applies to the specific exponent lines corresponding to the parameter
values ${\tilde{\xi}}_c^C = 0.734$ and $\alpha_ C = 0.495$ for which
$\zeta_{c'} (k)$ crosses zero at $k = k_{cs}^Q = 0.115\,\textrm{\AA}^{-1}$
and all exponents meet the agreement criteria.
The $-1/2$ exponent dashed-dotted line refers to the $c-s$ boundary line.}
\label{figure4}
\end{center}
\end{figure}

Charge-spin separation at high-energy scales beyond the TLL is associated with the emergence
of independent $c,c',c''$ (charge) branch lines and the $s$ (spin) branch line, respectively, 
represented by solid lines in the lower panel of Fig. \ref{figure2}. Their singularities 
are controlled by momentum, interaction, and density dependent negative exponents. They are generated by creation 
of independent charge and spin fractionalized particles, respectively, that move with {\it different} group 
velocities. On the other hand, the $c-s$ boundary lines represented by 
dotted lines in that figure panel whose singularities are controlled by a classical exponent $-1/2$
are generated by creation of one charge and one spin fractionalized 
particle that propagate with the {\it same} velocity. This can thus be understood as charge-spin 
recombination. Hence in different regions of the $(k,\omega)$ plane both spectral features of different 
type emerge that can be associated with charge-spin separation and charge-spin recombination, respectively.

In this paper two Hamiltonians of the same form, Eq. (\ref{equ1}), but different parameter values
and potentials have been used to describe the related TTF and TCNQ stacks of molecules
high-energy spectral features, respectively. Indeed, it is expected that the effects of possible interaction terms coupling
both problems can be neglected in the $(k,\omega)$-plane regions where such spectral features
do not overlap. The only exception is thus the TCNQ related $c-s$ boundary line that runs in Fig. \ref{figure2}
in the interval $k \in [k_{cs}^Q,\pi/a_0]$ where $k = k_{cs}^Q = 0.115\,\textrm{\AA}^{-1}$.
It is expected that upon running in the ``TTF $(k,\omega)$-plane region'' its singularity may be
washed out or changed by such interaction terms. 

Moreover, the values of the charge parameters ${\tilde{\xi}}_c$
and $K_{\rho} = {\tilde{\xi}}_c^2/2$ and related SDS exponent $\alpha = (2 - {\tilde{\xi}}_c^2)^2/(8{\tilde{\xi}}_c^2)$ 
specific to TTF and TCNQ, respectively, were accessed via the dependence on ${\tilde{\xi}}_c$ of the exponents, Eq. (\ref{equ19}). 
Consistent with this result, the corresponding related TTF and TCNQ high-energy branch lines run in different regions of the $(k,\omega)$ plane.
However, in the low-energy TLL limit where such exponents are not valid, the corresponding branch lines 
overlap near $k\approx k_F$. An open question is the effect in that limit of the interaction terms on the 
low-energy suppression of the photoelectron intensity, $I (\omega,0)$. If such interaction terms
remain weak at low energy, then $I (\omega,0) = C_C\vert\omega\vert^{\alpha_C}+C_F\vert\omega\vert^{\alpha_F}$
where $\alpha_C \approx 0.50$ and $\alpha_F \approx 0.74$ are the ``TCNQ'' and 
``TTF'' SDS exponents, respectively, and $C_C$ and $C_F$ are constants. As noted in Sec. \ref{SECI}, in this case the 
leading contribution would be $I (\omega,0)\propto\vert\omega\vert^{\alpha}$ where $\alpha = \alpha_C 
\approx 0.50$. If otherwise, $\alpha$ will have a single value in the range $\in [0.50,0.74]$ and its expression 
will involve both $\alpha_C$ and $\alpha_F$.

Another interesting issue is the apparent less pronounced effect of the charge-spin separation in the stacks of 
TTF molecules. To address this issue, one should take into account that at density $n_e^F = 1.41>1$ the microscopic mechanisms 
that control the spectral properties are similar to those of the one-electron addition spectral function at density 
$n_e = 2 - n_e^F = 0.59<1$. For the latter quantum problem, the charge-spin separation persists within
the model, Eq. (\ref{equ1}). However, as discussed below, it acquires a different form. 
That the exponent $\zeta_{s_s} (k)$ in Eq. (\ref{equ19}) is positive and negative in Figs.  \ref{figure3} and \ref{figure4}
for the TTF and TCNQ related $s_s =s'$ and $s_s =s$ branch lines, respectively, explains why only the latter exhibits 
ARPES peak structures associated with spin degrees of freedom. 

Nonetheless, at higher energy values $\omega >0$ not shown in the sketch of Fig. 1, there is an inverted upper-band 
$s$ branch line. At its lowest energy point at $k=\pi/a_0 - k_F$, that line has for $u>0$ an energy gap relative
to the $c''$ branch line highest energy point at that $k$ value. This is an important line, since in the $u\rightarrow 0$ 
limit the whole $k>0$ one-electron spectrum stems from the $s$ branch line for $k\in [0,k_F]$ and $\omega <0$,
$c''$ branch line for $k\in [k_F,\pi/a_0 - k_F]$ and $\omega >0$, and that inverted upper-band $s$ branch line
for $k\in [\pi/a_0 - k_F,\pi/a_0]$ and $\omega >0$. Indeed, its energy gap at $k=\pi/a_0 - k_F$ vanishes 
as $u\rightarrow 0$. 

Using the relation, Eq. (\ref{equ15}), that line gives rise to a related TTF $s$ branch line whose 
exponent is negative for $k\in [\pi/a_0 - k_F + \delta p_{Fs},\pi/a_0]$. It is located at higher energies,
$\omega \in [-1.390,-1.203]$\,eV, for which there are no ARPES data in Fig. \ref{figure2}. For $k>0$ it connects the $(k,\omega)$-plane 
points $(\pi/a_0 - k_F,-2\mu)$ and $(\pi/a_0,-2\mu - W_s)$. Here $-2\mu = - 4.149\,t = - 1.203$\, eV 
and $-2\mu - W_s = - 4.793\,t = - 1.390$\, eV where $2\mu$ is twice the absolute value of the chemical potential 
(it should be distinguished from the reduced mass $\mu$ in Eq. (\ref{equA7}) of Appendix \ref{APPA})
and $W_s = 0.644\,t = 0.187$\, eV is the $s$ branch line energy bandwidth. Hence for the model, Eq. (\ref{equ1}), at electronic densities 
$n_e>1$ the charge-spin separation is associated with the singularities at the (charge) $c''$ branch line and that 
higher-energy (spin) $s$ branch line. Whether the latter line emerges at higher energy in the ARPES data
remains an open question.

Finally, an interesting issue is that the effects of the finite-range interactions have lowered the 
TCNQ transfer integral $t= 0.40$ eV within the 1D Hubbard model description \cite{Sing_03} 
to $t= 0.24$ eV. This is consistent with evidence that the observed transfer of spectral weight at $k_F$ over the entire conduction 
band width with increasing temperatures cannot be reconciled within the use of $t= 0.40$ eV
whereas it can be reconciled with a value of $t= 0.24$ eV \cite{Sing_07}.

In summary, our quantitative results confirm that the interplay of one dimensionality and finite-range interactions plays a major
role in the one-electron spectral properties of TTF-TCNQ. In particular, our results identify the specific microscopic processes
that determine the $(k,\omega)$-plane location of the corresponding ARPES structures. Specifically, we have shown that the high-energy spectral properties of TTF-TCNQ  are controlled by the scattering of charge fractionalized particles in the unitary limit of (minus) infinite scattering length. 
The representation of such processes in terms of interactions of the fractionalized 
particles has greatly simplified a complex many-electron problem.

These results  apply as well to a wide class of non-Fermi liquid and nonperturbative many-electron 1D and 
quasi-1D systems \cite{Carmelo_19} of which TTF-TCNQ is a typical example. Further,
the unitary limit of (minus) infinite scattering length also appears in neutron-neutron interactions in shells 
of neutron stars and in the scattering of ultracold atoms. That it also plays
an active role in condensed matter systems treated beyond the Fermi liquid approximation represents a 
fundamental conceptual novelty. 
Finally, our representation of the microscopic mechanisms associated with nonperturbative many-electron interactions in terms of 
fractionalized particles scattering in the unitary limit may also be useful for the further understanding of
two-dimensional strongly correlated electronic systems in which the concept of quasiparticle itself breaks down, for example
in the cases  of high-temperature superconductivity and other types of exotic superconductivity
where the microscopic mechanisms are not well understood.

%%%%%%%%%%%%%%%%%%%%%%%%%%%%%%%%%%%%%%%%%%%%%%%%%%%%%%%%%%%%%%%%%%%%%%%%%
\acknowledgements

We  thank Claus S. Jacobsen for providing the single crystals used in our ARPES studies. J.M.P.C. acknowledges 
the late Adilet Imambekov for discussions that were helpful in writing this paper. He also would like to thank Boston 
University's Condensed Matter Theory Visitors Program for support and the hospitality of MIT. 
J.M.P.C. and T.\v{C}. acknowledge the support from Funda\c{c}\~ao para a Ci\^encia e
Tecnologia (FCT) through the Grants Nos. UID/FIS/04650/2013 and PTDC/FIS-MAC/29291/2017, J.M.P.C. acknowledges
that from the FCT Grants Nos. SFRH/BSAB/142925/2018 and
POCI-01-0145-FEDER-028887, and T.\v{C}. acknowledges the
support from the National Natural Science Foundation of China Grant No. 11650110443.
%%%%%%%%%%%%%%%%%%%%%%%%%%%%%%%%%%%%%%%%%%%%%%%%%%%%%%%%%%%%%%%%%%%%%%%%%

%%%%%%%%%%%%%%%%%%%%%%%%%%%%%%%%%%%%%%%%%%%%%%%%%%%%%%%%%%%%%%%%%%%%%%%%%%
\appendix

\section{Useful MQIM-HO quantities}
\label{APPA}

The MQIM-HO renormalized phase shifts $2\pi{\tilde{\Phi}}_{c,s} (\pm 2k_F,q')$ and $2\pi{\tilde{\Phi}}_{c,c} (\pm 2k_F,q)$ 
for $q\neq \mp 2k_F$ in the exponents expressions, Eq. (\ref{equ19}), are given by, 
\begin{equation}
2\pi{\tilde{\Phi}}_{c,s} (\pm 2k_F,q')={{\tilde{\xi}}_c\over\xi_c}\,2\pi\Phi_{c,s} (\pm 2k_F,q') 
\label{equA1}
\end{equation}
and
\begin{eqnarray}
&& 2\pi{\tilde{\Phi}}_{c,c} (\pm 2k_F,q) = 2\pi{\tilde{\Phi}}_{c,c}^{{\tilde{a}}} (\pm 2k_F,q) + 2\pi{\tilde{\Phi}}_{c,c}^{R_{\rm eff}} (k_r) 
\nonumber \\
&& 2\pi{\tilde{\Phi}}_{c,c}^{{\tilde{a}}} (\pm 2k_F,q) = {\xi_c\over {\tilde{\xi}}_c}{({\tilde{\xi}}_c -1)^2\over (\xi_c -1)^2}\,2\pi\Phi_{c,c} (\pm 2k_F,q)
\nonumber \\
& & \hspace{2.6cm} = {\arctan\left({{\tilde{a}}\over L}\,2\pi\right)\over\arctan\left({a\over L}\,2\pi\right)}\,2\pi\Phi_{c,c} (\pm 2k_F,q)
\nonumber \\
& & 2\pi{\tilde{\Phi}}_{c,c}^{R_{\rm eff}} (k_r) = - \gamma \times
\nonumber \\ 
&& \arctan\left({1\over 2}R_{\rm eff}\,k_r\sin^2 \left({({\tilde{\xi}}_c -1)^2\over  {\tilde{\xi}}_c}\pi\right) + P_c (k_r)\right) 
\label{equA2}
\end{eqnarray}
respectively. Here $\gamma = -1$ and $\gamma = +1$ for one-electron removal 
and addition, respectively, which is an extension of the results of Ref. \onlinecite{Carmelo_19} to 
one-electron addition. 

The effective-range expansion obeyed by the phase shift ${\tilde{\Phi}}_c (k_r) = \gamma\,2\pi{\tilde{\Phi}}_{c,c} (\pm 2k_F,\pm 2k_F + k_r)$
where $k_r = q \mp 2k_F$ is the small relative momentum reads \cite{Carmelo_19,Bethe_49,Blatt_49,Flambaum_99,Burke_11},
\begin{equation}
\cot ({\tilde{\Phi}}_c (k_r)) = {-1\over {\tilde{a}}\,k_r} + {1\over 2}\,R_{\rm eff}\,k_r 
- P_{\rm eff}\,R_{\rm eff}^3\,k_r^3 + {\cal{O}} (k_r^5) \, .
\label{equA3}
\end{equation}
For the ${\tilde{\delta}}_c =0$ bare model it is merely given by $\cot (\Phi_c (k_r)) = {-1\over a\,k_r}$.
Here ${\tilde{a}}$ and $a$ are the renormalized and bare scattering lengths, respectively,
and $R_{\rm eff}$ is the effective range. The shape parameter $P_{\rm eff}$ and those of higher
order and the corresponding effects of $P_c (k_r)$ in Eq. (\ref{equA2})
are in the the present unitary limit negligible \cite{Carmelo_19}.

To determine the scattering lengths, one uses the phase shift $\lim_{k_r\rightarrow0}{\tilde{\Phi}}_c (k_r)$
in the expansion, Eq. (\ref{equA3}), in the thermodynamic limit. This is straightforwardly 
extended to one-electron addition $\gamma = +1$ as
\begin{eqnarray}
{\tilde{\Phi}}_c (k_r) & = & -2\pi{\tilde{\Phi}}_{c,c} (\pm 2k_F, \pm 2k_F + k_r)\vert_{k_r = \pm \gamma{2\pi\over L}}
\nonumber \\
& = & \pm \gamma {({\tilde{\xi}}_c -1)^2\pi\over {\tilde{\xi}}_c}\hspace{0.20cm}{\rm for}\hspace{0.20cm}\gamma = \pm 1 \, .
\label{equA4}
\end{eqnarray}
The use of this expression (which also applies at ${\tilde{\delta}}_c =0$ with 
${\tilde{\Phi}}_c (k_r)=-2\pi{\tilde{\Phi}}_{c,c}$
reading $\Phi_c (k_r)=-2\pi\Phi_{c,c}$) in the two above effective-range expansions 
gives the scattering lengths in the thermodynamic limit as
\begin{eqnarray}
\tilde{a} & = & - {L\over 2\pi}\tan \left({({\tilde{\xi}}_c -1)^2\pi\over {\tilde{\xi}}_c}\right)
\rightarrow - \infty 
\hspace{0.20cm}{\rm for}\hspace{0.20cm}{\tilde{\xi}}_c\neq 1\hspace{0.20cm}{\rm and}
\nonumber \\
a & = & - {L\over 2\pi}\tan \left({(\xi_c -1)^2\pi\over \xi_c}\right) \rightarrow - \infty 
\hspace{0.20cm}{\rm for}\hspace{0.20cm}\xi_c\neq 1 \, ,
\label{equA5}
\end{eqnarray}
so that the unitary limit \cite{Carmelo_19,Zwerger_12,Horikoshi_17} holds for both
one-electron removal and addition, $\gamma = \mp 1$. The ratio,
\begin{equation}
{{\tilde{a}}\over a} = {\tan (\pi({\tilde{\xi}}_c -1)^2/{\tilde{\xi}}_c)\over\tan (\pi(\xi_c -1)^2/\xi_c)} \, ,
\label{equA6}
\end{equation}
is though finite.

The potential $V_c (x)$ induced by $V_e (r)$ vanishes for large $x$ as \cite{Carmelo_19},
\begin{eqnarray}
V_c^{\rm asy} (x) & = & - {\gamma_c\over x^l} = - {C_c\over (x/2r_l)^l}\hspace{0.2cm}{\rm where}
\nonumber \\
C_c & = & {1\over (2r_l)^2\mu}\hspace{0.2cm}{\rm and}\hspace{0.2cm}\gamma_c = {(2r_l)^{l-2}\over \mu} \, .
\label{equA7}
\end{eqnarray}
Here $\mu$ is a nonuniversal reduced mass, $l>5$ is an integer determined by the large-$r$ behavior of $V_e (r)$, and $2r_l$ is 
a length scale whose $l$ dependence for the range ${\tilde{\delta}}_c > {\tilde{\delta}}_c^{(1)}$ of interest for the present problem 
is \cite{Carmelo_19},
\begin{equation}
2r_l = {3\pi a_0\over 2}\sin\left({\pi\over l-2}\right)\left({l-2\over\sqrt{2}}\right)^{2\over l-2}
{\Gamma^2 \left({2\over l-2}\right)\Gamma \left({3\over l-2}\right)\over
\Gamma \left({1\over l-2}\right)\Gamma \left({4\over l-2}\right)}  \, .
\label{equA8}
\end{equation}
Here $a_0$ is the lattice spacing and $\Gamma (z)$ is the $\Gamma$ function.

In the interval $x\in [x_0,\infty]$ where $V_c (x)<0$, the ``momentum'' $\sqrt{2\mu (-V_c (x))}$ 
obeys the sum rule \cite{Carmelo_19},
\begin{eqnarray}
\Phi & = & \int_{x_0}^{x_2}dx\sqrt{2\mu (-V_c (x))} 
\hspace{0.20cm}{\rm where}\hspace{0.20cm} x_2 = 2r_l\left({4\sqrt{2}\over\pi\theta_c}\right)^{2\over l-2} 
\nonumber \\
& & {\rm with}\hspace{0.20cm}\tan (\Phi) = - {\Delta a\over{\tilde{a}}}\cot \left({\pi\over l-2}\right) \, .
\label{equA9}
\end{eqnarray} 
Here $\theta_c =\sqrt{(\xi_c^4 - {\tilde{\xi}}_c^4/(\xi_c^4 - 1)}$ for ${\tilde{\delta}}_c\in ]0,{\tilde{\delta}}_c^{(1)}[$
and $\theta_c = 1$ for ${\tilde{\delta}}_c\in ]{\tilde{\delta}}_c^{(1)},{\tilde{\delta}}_c^{({1\over 2})}[$ and $\Delta a = a - {\tilde{a}}$.

For the interval ${\tilde{\delta}}_c \in ]{\tilde{\delta}}_c^{(1)},{\tilde{\delta}}_c^{({1\over 2})}[$ of interest for TTF-TCNQ, 
the effective range in Eq. (\ref{equA3}) is given by \cite{Carmelo_19},
\begin{equation}
R_{\rm eff} = a_0\left(1 - c_1\,\left({{\tilde{a}}\over a}\right) + c_2\,\left({{\tilde{a}}\over a}\right)^2\right) \, ,
\label{equA10}
\end{equation}
where
\begin{eqnarray}
c_1 & = & {2\over \cos\left({\pi\over l-2}\right)} {\Gamma \left({2\over l-2}\right)\Gamma \left({l-4\over l-2}\right)
\over \Gamma \left({1\over l-2}\right)\Gamma \left({l-3\over l-2}\right)}\hspace{0.20cm}{\rm and}
\nonumber \\
c_2 & = & {3\,(l+1)\over (l-1)\cos^2\left({\pi\over l-2}\right)}{\Gamma \left({3\over l-2}\right)\Gamma \left(-{l+1\over l-2}\right)
\over \Gamma \left({-1\over l-2}\right)\Gamma \left(-{l-1\over l-2}\right)} \, .
\label{equA11}
\end{eqnarray}
 
The parameter $\xi_c = \xi_c (n_e,u)$ in the $\xi_c\rightarrow {\tilde{\xi}}_c$ transformation is defined 
by the following relation and equation,
\begin{eqnarray}
\xi_c & = & \xi_c \left({\sin Q\over u}\right)
\hspace{0.20cm}{\rm where}\hspace{0.20cm}\xi_c (r)
\hspace{0.20cm}{\rm is}\hspace{0.20cm}{\rm the}\hspace{0.20cm}{\rm solution}\hspace{0.20cm}{\rm of}
\nonumber \\
&& {\rm the}\hspace{0.20cm}{\rm integral}\hspace{0.20cm}{\rm equation,}
\nonumber \\
\xi_c (r) & = & 1 + \int_{-{\sin Q\over u}}^{\sin Q\over u} d r' K (r-r')\,\xi_c (r') \, .
\label{equA12}
\end{eqnarray}
Here the kernel $K (r)$ is given by,
\begin{equation}
K (r) = {i\over 2\pi} {d\over dr}\ln {\Gamma \Bigl({1\over 2}+i{r\over 4}\Bigr)\,\Gamma
\Bigl(1-i{r\over 4}\Bigr)\over \Gamma \Bigl({1\over 2}-i{r\over 4}\Bigr)\,\Gamma
\Bigl(1+i{r\over 4}\Bigr)} \, . 
\nonumber
\end{equation}
For $n_e \in ]0,1[$ and $u\ll 1$ the limiting behavior of $\xi_c$ is,
\begin{equation}
\xi_c = \sqrt{2}\left(1 - {U\over 8t\pi\,\sin\left({\pi\over 2}n_e\right)}\right) \, .
\label{equA13}
\end{equation}

\section{Rotated-electron representation and related $c$ and $s$ representations}
\label{APPB}

Some of the results presented in this Appendix for ${\tilde{\delta}}_c =0$ have not
been presented elsewhere and are needed for the related ${\tilde{\delta}}_c >0$
results provided in Appendix \ref{APPC}.

\subsection{The general rotated-electron representation}
\label{APPB1}

The {\it rotated electron} operators,
\begin{equation}
{\tilde{c}}_{j,\sigma}^{\dag} = {\hat{U}}^{\dag}\,c_{j,\sigma}^{\dag}\,{\hat{U}}\, ,\hspace{0.20cm}
{\tilde{c}}_{j,\sigma} =
{\hat{U}}^{\dag}\,c_{j,\sigma}\,{\hat{U}} 
\, ,\hspace{0.20cm}
{\tilde{n}}_{j,\sigma} = {\tilde{c}}_{j,\sigma}^{\dag}{\tilde{c}}_{j,\sigma} \, ,
\label{equB1}
\end{equation}
and ${\tilde{n}}_{j}=\sum_{\sigma}{\tilde{n}}_{j,\sigma}$ are generated from those of the electrons by the 
unitary operator $\hat{U} = e^{\hat{S}}$. It is uniquely defined in Ref. \onlinecite{Carmelo_17}
in terms of the $4^L\times 4^L$ matrix elements between all the ${\tilde{\delta}}_c =0$ bare model energy eigenstates.
An important property is that rotated-electron single and double occupancy are good quantum numbers for 
the whole $u>0$ range. For electrons this only applies in the $u\rightarrow\infty$ limit
and thus $u^{-1}\rightarrow 0$ limit in which rotated electrons adiabatically become electrons. 

The bare model Hamiltonian ${\hat{H}}_{\cal{H}}$, Eq. (\ref{equ1}) for ${\tilde{\delta}}_c = 0$,
has in the rotated-electron operators representation an infinite number of terms
given by the Baker-Campbell-Hausdorff formula,
\begin{eqnarray}
{\hat{H}}_{\cal{H}} & = & t\,\tilde{T}_0 + {\tilde{V}}_{\cal{H}} + \delta {\hat{H}}_{\cal{H}}  
\nonumber \\
\delta {\hat{H}}_{\cal{H}} & = & t\sum_{\iota=\pm1}\tilde{T}_{\iota} + [{\hat{H}}_{\cal{H}},{\tilde{S}}\,] + {1\over2}\,[[{\hat{H}}_{\cal{H}},{\tilde{S}}\,],{\tilde{S}}\,] + ...
\nonumber \\
{\tilde{V}}_{\cal{H}} & = & U\sum_{j=1}^{L}\tilde{\rho}_{j,\uparrow}\tilde{\rho}_{j,\downarrow} \, .
\label{equB2}
\end{eqnarray}
Here $\tilde{\rho}_{j,\sigma} = \left(\tilde{n}_{j,\sigma} - {1\over 2}\right)$ and
${\tilde{S}} = {\hat{U}}^{\dag}\,\hat{S}\,{\hat{U}}= \hat{S}$. The Hamiltonian terms
${\tilde{H}}_{\cal{H}} \equiv t\,\tilde{T} + {\tilde{V}}_{\cal{H}}$ where
\begin{eqnarray}
\tilde{T} & = & \sum_{d =0,\pm 1} \tilde{T}_{d} 
\nonumber \\
\tilde{T}_{0} & = & \sum_{j=1}^L \sum_{\iota =\pm 1}(\tilde{T}_{0,j,\iota} + \tilde{T}_{0,j,\iota}^{\dag})
\nonumber \\
\tilde{T}_{+1} & = & \sum_{j=1}^L \sum_{\iota =\pm 1}\tilde{T}_{+1,j,\iota}
\hspace{0.20cm}{\rm and}\hspace{0.20cm}\tilde{T}_{-1} = \tilde{T}_{+1}^{\dag} \, ,
\label{equB3}
\end{eqnarray}
$d=0,\pm 1$ gives the change in the number of rotated-electron doubly occupied sites and
\begin{eqnarray}
\tilde{T}_{0,j,\iota} & = &  - \sum_{\sigma}\{\tilde{n}_{j,-\sigma}\,\tilde{c}_{j,\sigma}^{\dag}\,
\tilde{c}_{j+\iota,\sigma}\,\tilde{n}_{j+\iota,-\sigma} 
\nonumber \\
& + & (1-\tilde{n}_{j,-\sigma})\,\tilde{c}_{j,\sigma}^{\dag}\,\tilde{c}_{j+\iota,\sigma}\,(1-\tilde{n}_{j+\iota,-\sigma})\}
\nonumber \\
\tilde{T}_{+1,j,\iota} & = & 
- \sum_{\sigma}\{\tilde{n}_{j,-\sigma}\,\tilde{c}_{j,\sigma}^{\dag}\,\tilde{c}_{j+\iota,\sigma}\,(1-\tilde{n}_{j+\iota,-\sigma}) 
\nonumber \\
& + & \tilde{n}_{j+\iota,-\sigma}\,\tilde{c}_{j+\iota,\sigma}^{\dag}\,\tilde{c}_{j,\sigma}\,(1-\tilde{n}_{j,-\sigma})\} \, ,
\label{equB4}
\end{eqnarray}
have the same expression in terms of rotated-electron operators
as the full Hamiltoninan ${\hat{H}}_{\cal{H}}= t\,\hat{T} + {\hat{V}}_{\cal{H}}$ in terms of electron operators. 
The form of the commutators,
\begin{eqnarray}
[{\tilde{V}}_{\cal{H}},\tilde{T}_d] = d\times \tilde{T}_d
\hspace{0.20cm}{\rm where}\hspace{0.20cm} d = 0, \pm 1 \, ,
\nonumber
\end{eqnarray}
is behind all higher terms in $\delta {\hat{H}}_{\cal{H}} $, Eq. (\ref{equB2}), 
having a kinetic nature. Indeed, the expression of ${\hat{S}}={\tilde{S}}$ in the 
unitary operator $\hat{U} = e^{\hat{S}}$ only involves the three $d=0,\pm 1$ rotated kinetic operators 
$\tilde{T}_{d}$, Eqs. (\ref{equB3}) and (\ref{equB4}).

\subsection{General fractionalized particles representation from rotated-electron degrees of freedom separation}
\label{APPB2}

The rotated-electron operators, Eq. (\ref{equB1}), naturally factorize as follows
upon acting onto the full Hilbert space\cite{Carmelo_17},
\begin{eqnarray}
{\tilde{c}}_{j,\uparrow}^{\dag} & = &
f_{j,c}^{\dag}\left({1\over 2} - {\tilde{s}}^{z}_{j,s} - {\tilde{s}}^{z}_{j,\eta}\right) 
\nonumber \\
& + & (-1)^j\, f_{j,c}\left({1\over 2} +{\tilde{s}}^{z}_{j,s} + {\tilde{s}}^{z}_{j,\eta}\right) 
\nonumber \\
{\tilde{c}}_{j,\downarrow}^{\dag} & = &
(f_{j,c}^{\dag} + (-1)^j\,f_{j,c})({\tilde{s}}^{+}_{j,s} + {\tilde{s}}^{+}_{j,\eta}) \, ,
\label{equB5}
\end{eqnarray}
and ${\tilde{c}}_{j,\sigma} = ({\tilde{c}}_{j,\sigma}^{\dag})^{\dag}$ where $\sigma = \uparrow,\downarrow$.

The rotated-electron singly occupied sites separate into lattice/charge 
degrees of freedom described by the spin-less $c$ particles of creation operator $f_{j,c}^{\dag}$ and spin 
degrees of freedom associated with rotated spins $1/2$ with local operators ${\tilde{s}}^l_{j,s}$ where 
$l = z, \pm$. The rotated spins are the spins of the rotated electrons that singly occupy sites whereas their 
charges are carried by the $c$ particles.

On the other hand, the degrees of freedom of the remaining sites unoccupied and doubly occupied by
rotated electrons separate into lattice/charge degrees of freedom described by the $c$ holes associated with
the annihilation operator $f_{j,c}$ and $\eta$-spin/charge degrees of freedom 
associated with the $\eta$-spin degrees freedom of the rotated-electron unoccupied sites (up $\eta$-spin
projection) and rotated-electron doubly occupied sites (down $\eta$-spin projection). Such onsite rotated 
$\eta$-spin degrees of freedom correspond to the $l = z, \pm$ local operators ${\tilde{s}}^l_{j,\eta}$.

The related $l = z, \pm$ rotated quasi-spin operators,
\begin{eqnarray}
{\tilde{q}}^l_{j} & = & {\tilde{s}}^l_{j,s} + {\tilde{s}}^l_{j,\eta} 
\hspace{0.20cm}{\rm where}\hspace{0.20cm} l=z,\pm
\hspace{0.20cm}{\rm and}
\nonumber \\
{\tilde{q}}^-_{j} & = & ({\tilde{q}}^+_{j})^{\dag} = 
({\tilde{c}}_{j,\uparrow}^{\dag}
+ (-1)^j\,{\tilde{c}}_{j,\uparrow})\,
{\tilde{c}}_{j,\downarrow} 
\nonumber \\
{\tilde{q}}^{z}_{j} & = & ({\tilde{n}}_{j,\downarrow} - 1/2) \, ,
\label{equB6}
\end{eqnarray}
read (i) ${\tilde{q}}^l_{j} = {\tilde{s}}^l_{j,s}$ and (ii) ${\tilde{q}}^l_{j} = {\tilde{s}}^l_{j,\eta}$
for rotated-electron (i) singly and (ii) unoccupied and doubly occupied sites.

Manipulations based on Eqs. (\ref{equB5}) and (\ref{equB6}) then give
\begin{eqnarray}
f_{j,c}^{\dag} & = & (f_{j,c})^{\dag} = {\tilde{c}}_{j,\uparrow}^{\dag}\,
(1-{\tilde{n}}_{j,\downarrow}) + (-1)^j\,{\tilde{c}}_{j,\uparrow}\,{\tilde{n}}_{j,\downarrow} 
\nonumber \\
{\tilde{n}}_{j,c} & = & f_{j,c}^{\dag}\,f_{j,c}\hspace{0.20cm}{\rm for}\hspace{0.20cm} j = 1,...,L \, ,
\label{equB7}
\end{eqnarray}
where ${\tilde{n}}_{j,\sigma}$ is provided in Eq. (\ref{equB1}). Hence,
\begin{eqnarray}
{\tilde{s}}^l_{j,\eta} & = & f_{j,c}\,f_{j,c}^{\dag}\,{\tilde{q}}^l_{j}
\hspace{0.20cm}{\rm and}\hspace{0.20cm}
{\tilde{s}}^l_{j,s} = f_{j,c}^{\dag} \,f_{j,c}\,{\tilde{q}}^l_{j} \, ,
\label{equB8}
\end{eqnarray}
where $f_{j,c}\,f_{j,c}^{\dag}$ and $f_{j,c}^{\dag} \,f_{j,c}$ are suitable site projectors.

Straightforward manipulations of Eqs. (\ref{equB5})-(\ref{equB7}) give
\begin{eqnarray}
\{f^{\dag}_{j,c}\, ,f_{j',c}\} = \delta_{j,j'}\hspace{0.20cm}{\rm and}
\hspace{0.20cm}\{f_{j,c}^{\dag}\, ,f_{j',c}^{\dag}\} = \{f_{j,c}\, ,f_{j',c}\} = 0 \, .
\nonumber 
\end{eqnarray}
The operators $f^{\dag}_{j,c}$ and $f_{j,c}$ commute with ${\tilde{q}}^{l}_{j}$
for $l = z, \pm$ and ${\tilde{s}}^l_{j,s}$ and ${\tilde{s}}^l_{j,\eta}$ obey the usual 
$SU(2)$ operator algebra.

The kinetic operators $\tilde{T}_{0}$ and $\tilde{T}_{\pm 1}$, Eq. (\ref{equB3}), and 
the interaction ${\tilde{V}}_{\cal{H}}$ in Eq. (\ref{equB2}) can then be written as
\begin{eqnarray}
\tilde{T}_{0} & = & - {1\over 2}\sum_{j=1}^{L}\sum_{\iota =\pm 1}
(f_{j,c}^{\dag}f_{j+\iota ,c} + f_{j+\iota ,c}^{\dag}f_{j,c})
\nonumber \\
& \times & \left(1 + 4\,{\tilde{\vec{q}}}_{j}\cdot{\tilde{\vec{q}}}_{j+\iota }\right)
\nonumber \\
\tilde{T}_{+1} & = & (\tilde{T}_{-1})^{\dag} = - {1\over 2}\sum_{j=1}^{L}\sum_{\iota =\pm 1} (-1)^j\,f_{j,c}f_{j+\iota ,c} 
\nonumber \\
& \times & \left(1 - 4\,{\tilde{\vec{q}}}_{j}\cdot{\tilde{\vec{q}}}_{j+\iota }\right)
\nonumber \\
{\tilde{V}}_{\cal{H}} & = & {U\over 2}\sum_{j=1}^L \left({1\over 2} - f_{j,c}^{\dag}f_{j,c}\right) \, ,
\label{equB9}
\end{eqnarray}
where
\begin{eqnarray}
{\tilde{\vec{q}}}_{j}\cdot{\tilde{\vec{q}}}_{j+\iota } = {\tilde{q}}^{z}_{j}{\tilde{q}}^{z}_{j+\iota } 
+ {1\over 2}\left({\tilde{q}}^{+}_{j}{\tilde{q}}^{-}_{j+\iota } 
+ {\tilde{q}}^{-}_{j}{\tilde{q}}^{+}_{j+\iota }\right) \, .
\nonumber 
\end{eqnarray}
Hence the Hamiltonian, Eq. (\ref{equB2}), can be expressed as
\begin{eqnarray}
\hat{H}_{\cal{H}} & = &  \hat{H}_{{\cal{H}}\,c}^0 + \delta\hat{H}_{\cal{H}}^*
\hspace{0.20cm}{\rm where}
\nonumber \\
\hat{H}_{{\cal{H}}\,c}^0 & = &
- {t\over 2}\sum_{j=1}^{L}\sum_{\iota =\pm 1}
(f_{j,c}^{\dag}f_{j+\iota ,c} + f_{j+\iota ,c}^{\dag}f_{j,c}) 
\nonumber \\
& + & {U\over 2}\sum_{j=1}^L \left({1\over 2} - f_{j,c}^{\dag}f_{j,c}\right) 
\nonumber \\
\delta\hat{H}_{\cal{H}}^* & = & - 2t\sum_{j=1}^{L}\sum_{\iota =\pm 1}
(f_{j,c}^{\dag}f_{j+\iota ,c} + f_{j+\iota ,c}^{\dag}f_{j,c})\,
{\tilde{\vec{q}}}_{j}\cdot{\tilde{\vec{q}}}_{j+\iota } 
\nonumber \\
& + & \delta\hat{H}_{\cal{H}} \, .
\label{equB10}
\end{eqnarray}

\subsection{$c$ and $s$ particle representation that diagonalizes the
${\tilde{\delta}}_c =0$ model in the one-electron subspace}
\label{APPB3}

By use of $c$ particle operators labeled by momentum, which are related to the corresponding local operators by,
\begin{eqnarray}
f_{j,c}^{\dag} & = & (f_{j,c})^{\dag} = {1\over{\sqrt{L}}}\sum_{q=-\pi}^{\pi}\,e^{-iq j}\,
f_{q,c}^{\dag} \hspace{0.20cm}{\rm where}\hspace{0.20cm} j = 1,...,L \, ,
\nonumber \\
f_{q,c}^{\dag} & = & (f_{q,c})^{\dag} = {1\over{\sqrt{L}}}\sum_{j=1}^{L}\,e^{+iq j}\,f_{j,c}^{\dag} 
\nonumber \\
& & \hspace{0.20cm}{\rm for}\hspace{0.20cm} q \in [-\pi,\pi] \, ,
\label{equB11}
\end{eqnarray}
the term $\hat{H}_{{\cal{H}}\,c}^0$ in Eq. (\ref{equB10}) can be written as
\begin{eqnarray}
\hat{H}_{{\cal{H}}\,c}^0 = \sum_{q=-\pi}^{\pi}\left({U\over 4}- 2t(\cos q + u)f_{q,c}^{\dag}f_{q,c}\right) \, .
\nonumber
\end{eqnarray}
In the one-electron subspace it is ground-state normal ordered as
\begin{eqnarray}
:\hat{H}_{\cal{H}}: & = &  :\hat{H}_{{\cal{H}}\,c}^0: + :\delta\hat{H}_{\cal{H}}^*:\hspace{0.20cm}{\rm where}
\nonumber \\
:\hat{H}_{{\cal{H}}\,c}^0: & = & \sum_{q=-\pi}^{\pi}\left(-2t (\cos q-\cos 2k_F)\right):f_{q,c}^{\dag}f_{q,c}: 
\nonumber \\
:\delta\hat{H}_{\cal{H}}^*: & = & \delta\hat{H}_{\cal{H}}^* - \langle GS\vert \delta\hat{H}_{\cal{H}}^*\vert GS\rangle \, .
\label{equB12}
\end{eqnarray}
Here $\delta\hat{H}_{\cal{H}}^*$ also appears in Eq. (\ref{equB10}) and
$:f_{q,c}^{\dag}f_{q,c}: = f_{q,c}^{\dag}f_{q,c} - N_{c}^{GS} (q)$ where
in the thermodynamic limit, $N_{c}^{GS} (q) = \langle GS\vert f_{q,c}^{\dag}f_{q,c}\vert GS\rangle
= \theta (2k_F-\vert q\vert)$.

The use of the ${\tilde{\delta}}_c =0$ bare model exact Bethe-ansatz solution reveals 
that $:\delta\hat{H}_{\cal{H}}^*:$ in Eq. (\ref{equB12}) reads
\begin{eqnarray}
:\delta\hat{H}_{\cal{H}}^*: & = & \sum_{q=-\pi}^{\pi}\delta\varepsilon_c (q):f_{q,c}^{\dag}f_{q,c}:+ :\hat{H}_{{\cal{H}}\,s}:  
\hspace{0.20cm}{\rm where}
\nonumber \\
\delta\varepsilon_c (q) & = & \varepsilon_c (q) + 2t (\cos q-\cos 2k_F) \hspace{0.20cm}{\rm and}
\nonumber \\
:\hat{H}_{{\cal{H}}\,s}: & = & \sum_{q'=-k_F}^{k_F}\varepsilon_s (q'):f_{q',s}^{\dag}f_{q',s}: \, .
\label{equB13}
\end{eqnarray}
Here $\varepsilon_c (q)$ and $\varepsilon_s (q')$ are defined in Ref. \onlinecite{Carmelo_19}
and $f_{q',s}^{\dag}$ and $f_{q',s}$ are $s$ particle
creation and annihilation operators \cite{Carmelo_18,Carmelo_17}. 
For $u\rightarrow 0$ and $u\rightarrow\infty$, $\delta\varepsilon_c (q)$ reads,
\begin{eqnarray}
\delta\varepsilon_c (q) & = & - 4t\left(\sin^2 \left({q\over 2}\right) - \sin^2 k_{F}\right)
\nonumber \\
& & {\rm for}\hspace{0.20cm}q\in [-2k_{F},2k_F] 
\hspace{0.20cm}{\rm and}\hspace{0.20cm}u\rightarrow 0
\nonumber \\
& = & -2t\left(\cos (\vert q\vert - k_F) - \cos q - (\cos k_{F}-\cos 2k_{F})\right)
\nonumber \\
& & {\rm for}\hspace{0.20cm} \vert q\vert \in [2k_{F},\pi] 
\hspace{0.20cm}{\rm and}\hspace{0.20cm}u\rightarrow 0
\nonumber \\
\delta\varepsilon_c (q) & = & 0 \hspace{0.20cm}{\rm for}\hspace{0.20cm}q \in [-\pi,\pi]
\hspace{0.20cm}{\rm and}\hspace{0.20cm}u\rightarrow\infty   \, .
\nonumber
\end{eqnarray}
It vanishes for $u\rightarrow\infty$ upon rotated electrons becoming electrons.
In the one-electron subspace the diagonalized Hamiltonian $:\hat{H}_{\cal{H}}:$, Eq. (\ref{equB12}), is 
in the thermodynamic limit given by
\begin{eqnarray}
:\hat{H}_{\cal{H}}: & = & :\hat{H}_{{\cal{H}}\,c}: + :\hat{H}_{{\cal{H}}\,s}:
\nonumber \\
:\hat{H}_{{\cal{H}}\,c}: & = & \sum_{q=-\pi}^{\pi}\varepsilon_c (q) :f_{q,c}^{\dag}f_{q,c}: 
\nonumber \\
:\hat{H}_{{\cal{H}}\,s}: & = & \sum_{q'=-k_F}^{k_F}\varepsilon_s (q'):f_{q',s}^{\dag}f_{q',s}: \, .
\label{equB14}
\end{eqnarray}
The $c$ band hole and particle energy bandwidths $W_c^h = \varepsilon_c (\pm\pi)$ and $W_c^p = - \varepsilon_c (0) $, respectively,
are such that $W_c = W_c^h  + W_c^p = 4t$ for all $u>0$ values and $n_e \in ]0,1[$. 

In the low-energy TLL limit the following Hamiltonian with linearized $c$ and $s$ band
energy dispersions describes the diagonal Hamiltonian, Eq. (\ref{equB14}),
\begin{eqnarray}
:\hat{H}_{\cal{H}}: & = & :\hat{H}_{{\cal{H}}\,c}: + :\hat{H}_{{\cal{H}}\,s}:
\nonumber \\
:\hat{H}_{{\cal{H}}\,c}: & = & {\pi \,v_{Fc} \over L}\sum_{k,\iota =\pm}:\sigma_{c,\iota} (\iota k)\sigma_{c,\iota} (-\iota k): 
\nonumber \\
:\hat{H}_{{\cal{H}}\,s}: & = & {\pi \,v_{Fs} \over L}
\sum_{k,\iota =\pm}:\sigma_{s,\iota} (\iota k) \sigma_{s,\iota} (-\iota k): \, .
\label{equB15}
\end{eqnarray}
Here $v_{Fc} = v_c (2k_F)$ and $v_{Fs} = v_s (k_F)$ where 
$v_c (q) =\partial \varepsilon_c (q)\partial q$ and $v_s (q') =\partial \varepsilon_s (q')\partial q'$,
respectively, and
\begin{eqnarray}
\sigma_{c,\iota} (k) & = & \sum_p f_{\iota 2k_F + p + k,c}^{\dag} f_{\iota 2k_F + p,c}
\nonumber \\
\sigma_{s,\iota} (k) & = & \sum_{p'} f_{\iota k_F + p' + k,s}^{\dag} f_{\iota k_F + p',s} \, .
\nonumber
\end{eqnarray}
The $k$ and $p$ summations in $:\hat{H}_{{\cal{H}}\,c}:$ and
$\sigma_{c,\iota} (k)$ run in the intervals given in Table \ref{table3} and
the $k$ and $p'$ summations in $:\hat{H}_{{\cal{H}}\,s}:$ 
and $\sigma_{s,\iota} (k)$ run in the following intervals,
\begin{eqnarray}
k & \in & [-\delta p_{Fs}-p',-p']\hspace{0.20cm}{\rm for}\hspace{0.20cm}\iota = +
\nonumber \\
k & \in & [-p',\delta p_{Fc}-p']\hspace{0.20cm}{\rm for}\hspace{0.20cm}\iota = -
\nonumber \\
p' & \in & [-\delta p_{Fs},0]\hspace{0.20cm}{\rm for}\hspace{0.20cm}\iota = +
\nonumber \\
p' & \in & [0,\delta p_{Fc}]\hspace{0.20cm}{\rm for}\hspace{0.20cm}\iota = - \, .
\label{equB16}
\end{eqnarray}

\subsection{A useful $c$ and $s$ particle nondiagonal representation
in the one-electron subspace}
\label{APPB4}

There exists a uniquely defined transformation,
\begin{equation}
f_{q,c}^{\dag} \rightarrow {\bar{f}}_{q,c}^{\dag}
\hspace{0.20cm}{\rm and}\hspace{0.20cm}
f_{q,c} \rightarrow {\bar{f}}_{q,c} \, ,
\label{equB17}
\end{equation}
that leads to a representation for which the $c$ particles have interactions associated with an effective
potential $V_c^1 (x)$. It controls the one-electron matrix elements dependence
on $c$ particle/hole phase shifts. In that representation, the Hamiltonian, Eq. (\ref{equB14}), in the 
one-electron subspace is given by,
\begin{eqnarray}
:\hat{H}_{{\cal{H}}}: & = & :\hat{H}_{{\cal{H}}\,c}: + :\hat{H}_{{\cal{H}}\,s}:
\nonumber \\
:\hat{H}_{{\cal{H}}\,c}: & = & \sum_{q=-\pi}^{\pi}\varepsilon_c^1 (q):{\bar{f}}_{q,c}^{\dag}{\bar{f}}_{q,c}: 
\nonumber \\
& + & {1\over L}\sum_{k,q,q'} {\cal{V}}_c^1 (k) {\bar{f}}_{q,c}^{\dag}{\bar{f}}_{q+k,c}\,{\bar{f}}_{q',c}{\bar{f}}_{q'-k,c}^{\dag}
\nonumber \\
:\hat{H}_{{\cal{H}}\,s}: & = & \sum_{q'=-k_F}^{k_F}\varepsilon_s (q'):f_{q',s}^{\dag}f_{q',s}:  \, .
\label{equB18}
\end{eqnarray}
Here $\varepsilon_c^1 (q)$ and $v_c^1 (q) =\partial \varepsilon_c^1 (q)\partial q$ read
\begin{eqnarray}
\varepsilon_c^1 (q) & = & {\varepsilon_c (q)\over (1 + \beta_c^1)} \hspace{0.20cm}{\rm for}\hspace{0.20cm} 
q \in [-2k_F,2k_F]
\nonumber \\
& = & {W^{1 h}_c\over 8t\,\beta_c^1}\{(1+\beta_c^1)W^{1 h}_c
\nonumber \\
& - & \sqrt{\left((1+\beta_c^1)W^{1 h}_c\right)^2 - 16t\,\beta_c^1\,\varepsilon_c (q)}\}
\nonumber \\
& & \hspace{1.50cm}{\rm for}\hspace{0.20cm} \vert q\vert \in [2k_F,\pi] \, ,
\label{equB19}
\end{eqnarray}
\begin{eqnarray}
v_c^1 (q) & = & {v_c (q)\over (1 + \beta_c^1)} \hspace{0.20cm}{\rm for}\hspace{0.20cm} 
q \in [-2k_F,2k_F]
\nonumber \\
& = & {W^{1 h}_c\over\sqrt{\left((1+\beta_c^1)W^{1 h}_c\right)^2 - 
16t\,\beta_c^1\,\varepsilon_c (q)}}\,v_c (q) 
\nonumber \\
& & \hspace{2.25cm}{\rm for}\hspace{0.20cm} \vert q\vert \in [2k_F,\pi] \, ,
\label{equB20}
\end{eqnarray}
where $\beta_c^1$ is given below and
$W^{1 h}_c = \varepsilon_c^1 (\pi)$ and $W^{1 p}_c = - \varepsilon_c^1 (0)$ 
are such that $W^{1}_c = W^{1 h}_c + W^{1 p}_c = 4t$.
Inversion of Eqs. (\ref{equB19}) and (\ref{equB20}) leads to
\begin{eqnarray}
\varepsilon_c (q) & = & \left(1 + \beta_c^1\right)\varepsilon_c^1 (q) \hspace{0.20cm}{\rm for}\hspace{0.20cm} 
q \in [-2k_F,2k_F]
\nonumber \\
& = & \left(1 + \beta_c^1\left\{1 - {4t\over W^{1 h}_c}
\left({\varepsilon_c^1 (q)\over W^{1 h}_c}\right)\right\}\right)\varepsilon_c^1 (q) 
\nonumber \\
& & \hspace{2.25cm}{\rm for}\hspace{0.20cm} \vert q\vert \in [2k_F,\pi] \, ,
\label{equB21}
\end{eqnarray}
\begin{eqnarray}
v_c (q) & = & \left(1 + \beta_c^1\right)v_c^1 (q) \hspace{0.20cm}{\rm for}\hspace{0.20cm} 
q \in [-2k_F,2k_F]
\nonumber \\
& = &  \left(1 + \beta_c^1\left\{1 - {8t\over W^{1 h}_c} \left({\varepsilon_c^1 (q)\over W^{1 h}_c}\right)\right\}\right)v_c^1 (q) 
\nonumber \\
& & \hspace{2.25cm}{\rm for}\hspace{0.20cm} \vert q\vert \in [2k_F,\pi] \, .
\label{equB22}
\end{eqnarray}

In the low-energy limit, the term $:\hat{H}_{{\cal{H}}\,c}:$ of the nondiagonal Hamiltonian, Eq. (\ref{equB18}), is equivalent
to the charge TLL model,
\begin{eqnarray}
& & :\hat{H}_{{\cal{H}}\,c}:\, = {\pi \,v_{Fc}^1 \over L}\sum_{k,\iota = \pm}:\sigma_{c,\iota}^1 (\iota k)\sigma_{c,\iota}^1 (-\iota k): 
\nonumber \\
& & \hspace{1.50cm} + \,{1\over L}\sum_{k,\iota = \pm} (-{\cal{V}}_c^1 (k))\times 
\nonumber \\
& & (:\sigma_{c,\iota}^1 (\iota k)\sigma_{c,\iota}^1 (-\iota k): + 
\sigma_{c,\iota}^1 (\iota k) \sigma_{c,-\iota}^1 (-\iota k)) \, ,
\label{equB23}
\end{eqnarray}
where $v_{Fc}^1 = v_c^1 (2k_F)$,
$\sigma_{c,\iota}^1 (k) = \sum_p {\bar{f}}_{\iota 2k_F + p + k,c}^{\dag} {\bar{f}}_{\iota 2k_F + p,c}$,
and ${\cal{V}}_c^1 (k)$ is here and in Eq. (\ref{equB18}) the Fourier transform of the effective potential $V_c^1 (x)$ 
associated with the interaction of the $c$ particles/holes with $c$ mobile scattering centers.
The transformation, Eq. (\ref{equB17}), is in the low-energy TLL limit such that, 
\begin{eqnarray}
& & \sigma_{c,\iota} (k) = e^{S_c^1}\,\sigma_{c,\iota}^1 (k)\,e^{-S_c^1} 
\hspace{0.20cm}{\rm where}
\nonumber \\
& & S_c^1 = {2\pi\over L}\sum_{k>0}{\ln (1 + \alpha_c^1 (k))\over 4k} \times
\nonumber \\
& & (\sigma_{c,+}^1 (k)\sigma_{c,-}^1 (-k)
- \sigma_{c,-}^1 (k) \sigma_{c,+}^1 (-k)) \, .
\label{equB24}
\end{eqnarray}
For very small $k$, ${\cal{V}}_c^1 (k)$ in Eq. (\ref{equB23}) is given by
\begin{eqnarray}
{\cal{V}}_c^1 (k) & = & - {\pi\over 2}\,\alpha_c^1 (k)\,v_{Fc}^1 
\hspace{0.20cm}{\rm where}\hspace{0.20cm}
\alpha_c^1 (k) = \alpha_c^1 + {\cal{O}} (k^2)
\nonumber \\
{\rm and} & & \alpha_c^1 = \alpha_c^1 (0) = {4 - \xi_c^4\over\xi_c^4} \, .
\label{equB25}
\end{eqnarray}
The parameter $\alpha_c^1$ and the related parameter $\beta_c^1$ control
the ratio $v_{Fc}/v_{Fc}^1$ of the velocities
in Eqs. (\ref{equB15}) and (\ref{equB23}) as follows,
\begin{eqnarray}
\sqrt{1 + \alpha_c^1} & = & 1 + \beta_c^1 = {v_{Fc}\over v_{Fc}^1}  = {2\over\xi_c^2} 
\nonumber \\
\beta_c^1 & = & {v_{Fc} -v_{Fc}^1 \over v_{Fc}^1} 
= {2 - \xi_c^2\over\xi_c^2} \, .
\label{equB26}
\end{eqnarray}

The charge TLL described by the Hamiltonian term $:\hat{H}_{{\cal{H}}\,c}:$, Eq. (\ref{equB23}), is such that within
the notation used in the second expression given in Eq. (7) of Ref. \onlinecite{Voit_93} the equality $g_{4\rho} = g_{2\rho}$ holds 
and $g_{4\rho}$ corresponds to ${\pi\over 2}\,\alpha_c^1\,v_{Fc}^1 + {\cal{O}} (k)$ in  Eq. (\ref{equB25}). The velocities $v_{\rho}$ and
$v_F$ of that reference refer here to $v_{Fc}=v_c (2k_F)$ and $v_{Fc}^1 = v_c^1 (2k_F)$, respectively.

$v_c^1 (2k_F)$ {\it is not} the $U=0$ Fermi velocity, $v_F = 2t\sin k_F$.
Except in the $u\rightarrow 0$ limit, $v_{Fc}/v_{Fc}^1 = 2/\xi_c^2$ is different from 
$v_{Fc}/v_F= v_{Fc}/(2t\sin k_F)$. The physics of that velocity is revealed by its expression,
$v_{Fc}^1= {1\over 2}\,j_c^{\rho}$. Here, $j_c^{\rho} =\xi_c^2\,v_{Fc}$, is the elementary charge 
current that controls the metal charge stiffness $2\pi D_{\rho}^0 = 2v_{Fc}^1 = j_c^{\rho}$ in the real part of the conductivity, 
$\sigma_{\rho} (\omega) = 2\pi D_{\rho}^0 \delta (\omega) + \sigma_{\rho}^{\rm reg} (\omega)$ \cite{Carmelo_18}.
Hence ${\cal{V}}_c^1 (0) = \pi ({1\over\xi_c^4} - {1\over 4})\,2\pi D_{\rho}^0$ in Eq. (\ref{equB25}).
The related velocity $v_{Fc}^0$,
\begin{eqnarray}
v_{Fc}^0 & = & {2\over\xi_c^2}\,v_{Fc}
\hspace{0.20cm}{\rm such}\hspace{0.20cm}{\rm that}\hspace{0.20cm}
v_{Fc} = \sqrt{v_{Fc}^0\times v_{Fc}^1} 
\nonumber \\
\lim_{u\rightarrow 0}v_{Fc}^0 & = & \lim_{u\rightarrow 0}v_{Fc} = \lim_{u\rightarrow 0}v_{Fc}^1 = v_F \, ,
\label{equB27}
\end{eqnarray}
controls the compressibility, $\chi_{\rho}^0 = 2/(\pi n_e^2 v_{Fc}^0)$.
The three velocities in Eq. (\ref{equB27}) become in the $u\rightarrow 0$ limit the 
Fermi velocity, $v_F = 2t\sin k_F$.

The {\it exact} expressions of the dispersion $\varepsilon_c^1 (q)$ and potential Fourier transform 
${\cal{V}}_c^1 (k)$ in the Hamiltonian, Eq.  (\ref{equB18}), could in principle be extracted from the Bethe-ansatz 
solution. However, since for ${\tilde{\delta}}_c >0$ there is no exact solution, here we used a controlled and very good 
approximation that also applies for ${\tilde{\delta}}_c >0$ to derive the expression of $\varepsilon_c^1 (q)$ 
given in Eq. (\ref{equB19}) and relation, Eq. (\ref{equB21}).
This approximation ensures that $v_c^1 (q) = \partial\varepsilon_c^1 (q)/\partial q$ has its exact
TLL value $v_{Fc}^1 = v_c^1 (2k_F)$ uniquely defined by the relations in Eq. (\ref{equB26}).
In addition, it accounts for the $c$ band energy bandwidth $W^{1}_c=4t$
and the group velocity values $v_c^1 (0) = v_c^1 (\pm\pi) = 0$ 
remaining invariant under the transformation, Eq. (\ref{equB17}). 
That the expression of $\varepsilon_c^1 (q)$ in Eq. (\ref{equB19}) obeys all those 
exact properties is behind it being a very good approximation. 

\section{Representations for the Hamiltonian with finite-range interactions}
\label{APPC}

\subsection{Rotated-electron representation and $c$ and $s$ particle representation
that directly emerges from it}
\label{APPC1}

In the rotated-electron representation the ${\tilde{\delta}}_c >0$ Hamiltonian, Eq. (\ref{equ1}), has again 
an infinite number of terms given by the Baker-Campbell-Hausdorff formula,
\begin{eqnarray}
{\hat{H}} & = & t\,\tilde{T}_0 + {\tilde{V}} + \delta {\hat{H}}  
\nonumber \\
\delta {\hat{H}} & = & t\sum_{\iota=\pm1}\tilde{T}_{\iota} + [{\hat{H}},{\tilde{S}}\,] + {1\over2}\,[[{\hat{H}},{\tilde{S}}\,],{\tilde{S}}\,] + ... \, .
\label{equC1}
\end{eqnarray}
The $d=0,\pm 1$ kinetic operators $\tilde{T}_{d}$ and the operator ${\tilde{S}}={\hat{S}}$ have the same expressions
as for ${\tilde{\delta}}_c =0$. In contrast, the rotated-electron interaction Hamiltonian term ${\tilde{V}}$,
\begin{eqnarray}
\tilde{V} & = & U\sum_{j=1}^{L}\tilde{\rho}_{j,\uparrow}\tilde{\rho}_{j,\downarrow} 
\nonumber \\
& + & \sum_{r=1}^{L/2-1}V_e (r)
\sum_{\sigma=\uparrow,\downarrow}\sum_{\sigma'=\uparrow,\downarrow}\sum_{j=1}^{L}\tilde{n}_{j,\sigma}\tilde{n}_{j+r,\sigma'} \, ,
\label{equC2}
\end{eqnarray}
where $\tilde{\rho}_{j,\sigma} = \left(\tilde{n}_{j,\sigma} - {1\over 2}\right)$ and $\tilde{n}_{j,\sigma} = \tilde{c}_{j,\sigma}^{\dag}\, \tilde{c}_{j,\sigma}$,
has new terms associated with the finite-range interactions. Those render the commutator terms of $\delta {\hat{H}}$ 
in Eq. (\ref{equC1}) different from those of the ${\tilde{\delta}}_c =0$ bare model. In the one-electron subspace, only the 
charge term $:\hat{H}_{{\cal{H}}\,c}:$ of the normal-ordered Hamiltonian, Eq. (\ref{equB14}) of Appendix \ref{APPB}, is 
though changed by the finite-range electron interactions.

${\tilde{V}}$ in Eq. (\ref{equC2}) can be expressed {\it solely} in terms of the $c$ particle 
operators, Eq. (\ref{equB7}) of Appendix \ref{APPB}, as
\begin{eqnarray}
\tilde{V} & = & \sum_{r=1}^{L/2-1}V_e (r)\sum_{j=1}^Lf_{j,c}^{\dag}f_{j,c}\,f_{j+r,c}^{\dag}f_{j+r,c}
\nonumber \\
& + & {U\over 2}\sum_{j=1}^L \left({1\over 2} - f_{j,c}^{\dag}f_{j,c}\right) \, .
\label{equC3}
\end{eqnarray}
The Hamiltonian, Eq. (\ref{equC1}), can then be rewritten as,+
\begin{equation}
\hat{H} = \hat{H}_{c}^0 + \delta\hat{H}^* \, .
\label{equC4}
\end{equation}
The charge-only term $\hat{H}_{c}^0$ and $\delta\hat{H}^*$ are here given by
\begin{eqnarray}
\hat{H}_c^0 & = & - {t\over 2}\sum_{j=1}^{L}\sum_{\iota =\pm 1}
(f_{j,c}^{\dag}f_{j+\iota ,c} + f_{j+\iota ,c}^{\dag}f_{j,c}) 
\nonumber \\
 & + & {U\over 2}\sum_{j=1}^L \left({1\over 2} - f_{j,c}^{\dag}f_{j,c}\right)
\nonumber \\
& + & \sum_{r=1}^{L/2-1}V_e (r)\sum_{j=1}^Lf_{j,c}^{\dag}f_{j,c}\,f_{j+r,c}^{\dag}f_{j+r,c} \, ,
\nonumber \\
\delta\hat{H}^* & = & - 2t\sum_{j=1}^{L}\sum_{\iota =\pm 1}
(f_{j,c}^{\dag}f_{j+\iota ,c} + f_{j+\iota ,c}^{\dag}f_{j,c})\,
{\tilde{\vec{q}}}_{j}\cdot{\tilde{\vec{q}}}_{j+\iota } 
\nonumber \\
& + & \delta\hat{H} \, .
\label{equC5}
\end{eqnarray}
Under the operator transformation, Eq. (\ref{equB11}) of Appendix \ref{APPB},
the term $\hat{H}_c^0$ in Eq. (\ref{equC5}) reads
\begin{eqnarray}
\hat{H}_c^0 & = & \sum_{q=-\pi}^{\pi}\left({U\over 4}- 2t(\cos q + u)f_{q,c}^{\dag}f_{q,c}\right) 
\nonumber \\
& + & {1\over L}\sum_{k,q,q'} {\cal{V}}_e (k) f_{q,c}^{\dag}f_{q+k,c}\,f_{q',c}^{\dag}f_{q'-k,c} \, ,
\nonumber
\end{eqnarray}
where ${\cal{V}}_e (k)$ is the Fourier transform of $V_e (r)$.

In the one-electron subspace, the normal-ordered expression of the Hamiltonian, Eq. (\ref{equC4}), 
is given by
\begin{eqnarray}
:\hat{H}: & = & :\hat{H}_{c}^0: + :\delta\hat{H}^*:
\nonumber \\
:\hat{H}_c^0: & = & \sum_{q=-\pi}^{\pi}\left(-2t(\cos q -\cos 2k_F)\right):f_{q,c}^{\dag}f_{q,c}:
\nonumber \\
& + & {1\over L}\sum_{k,q,q'} {\cal{V}}_e (k) f_{q,c}^{\dag}f_{q+k,c}\,f_{q',c}^{\dag}f_{q'-k,c} \, .
\label{equC6}
\end{eqnarray}

In the present case of the ${\tilde{\delta}}_c > 0$ model Hamiltonian, Eq. (\ref{equ1}), the electronic potential $V_e (r)$ 
leads to some renormalization of the terms in the expansion of $\delta {\hat{H}}$ in Eq. (\ref{equC1}).
Under it $V_e (r)$ is replaced by a renormalized rotated-electron potential $V_{re} (r)$. 
All the higher kinetic operator terms in $\delta {\hat{H}}$ are renormalized by the finite-range 
potential $V_e (r)$. Hence its renormalization by such kinetic operators that leads
to $V_{re} (r)$ is a weaker higher-order effect.

For simplicity, we provide here the rotated kinetic operator terms in the expression
of the Hamiltonian term $[{\tilde{H}},{\tilde{S}}\,]$ of $\delta {\hat{H}}$ in  Eq. (\ref{equC1}).
It involves the three $d=0,\pm 1$ operators $\tilde{T}_{d}$ and 
four additional kinetic operators renormalized by the potential $V_e (r)$ 
that are given by the commutators,
\begin{equation}
\tilde{J}_0^+ = [\tilde{V},\tilde{T}_0] \, , \hspace{0.20cm}\tilde{J}_0^- = (\tilde{J}_0^+)^{\dag}
\, , \hspace{0.20cm} \tilde{J}_{\pm 1} = [\tilde{V},\tilde{T}_{\pm 1}] \, .
\label{equC7}
\end{equation}
They explicitly read
\begin{eqnarray}
\tilde{J}_0^+ & = & \sum_{r=1}^{L/2-1}V_e (r) \sum_{j=1}^L \sum_{\iota =\pm1}
(\tilde{T}_{0,j,\iota} - \tilde{T}_{0,j,\iota}^{\dag})
\nonumber \\
& \times & (\tilde{n}_{j+r} + \tilde{n}_{j-r} - \tilde{n}_{j+r+\iota} - \tilde{n}_{j-r+\iota})
\nonumber \\
\tilde{J}_{\pm 1} & = & \pm U\tilde{T}_{\pm 1} 
\pm \sum_{r=1}^{L/2-1}4V_e (r) \sum_{j=1}^L \sum_{\iota =\pm1}\tilde{T}_{\pm 1,j,\iota} 
\nonumber \\
& \times & (\tilde{n}_{j+r} + \tilde{n}_{j-r} + \tilde{n}_{j+r+\iota} + \tilde{n}_{j-r+\iota}) \, .
\label{equC8}
\end{eqnarray}

Higher kinetic operator terms in the expression of $\delta {\hat{H}}$ in Eq. (\ref{equC1})
also only involve the operators $\tilde{T}_{0,j,\iota}$ and $\tilde{T}_{\pm 1,j,\iota}$, Eq. (\ref{equB4})
of Appendix \ref{APPB}, and the operator $\tilde{n}_{j} = \sum_{\sigma}\tilde{n}_{j,\sigma} = \sum_{\sigma}\tilde{c}_{j,\sigma}^{\dag}\, \tilde{c}_{j,\sigma}$  
at sites with different relative positions. 

Accounting for the higher-order terms in the Hamiltonian term $:\delta\hat{H}^*:$
in Eqs. (\ref{equC5}) and (\ref{equC6}), one finds the following expression of the
Hamiltonian, Eq. (\ref{equ1}), in the one-electron subspace,
\begin{eqnarray}
:\hat{H}: & = & :\hat{H}_{c}: + :\hat{H}_{s}:
\nonumber \\
:\hat{H}_{c}: & = & \sum_{q=-\pi}^{\pi}\varepsilon_c (q):f_{q,c}^{\dag}f_{q,c}:
\nonumber \\
& + & {1\over L}\sum_{k,q,q'} {\cal{V}}_{re} (k) f_{q,c}^{\dag}f_{q+k,c}\,f_{q',c}^{\dag}f_{q'-k,c} 
+ \hat{H}_{re}
\nonumber \\
:\hat{H}_s: & = & \sum_{q'=-k_F}^{k_F}\varepsilon_s (q'):f_{q',s}^{\dag}f_{q',s}:  \, .
\label{equC9}
\end{eqnarray}
Here $\hat{H}_{re}$ has a complicated expression not needed for our studies and ${\cal{V}}_{re} (k)$ 
is the Fourier transform of $V_{re} (r)$. At $k=0$ it is given by Eq. (\ref{equ9}), as justified below in Sec. \ref{APPC3}.
The term $\hat{H}_{re}$ in Eq. (\ref{equC9}) prevents diagonalizing the Hamiltonian $:\hat{H}:$ 
when acting onto the one-electron subspace. This results from states generated by occupancy configurations 
of the fractionalized particles emerging from the rotated electrons not being in general energy eigenstates of the
${\tilde{\delta}}_c >0$ Hamiltonian, Eq. (\ref{equ1}).

In the singularities subspace such occupancy configurations generate states that are energy eigenstates
or have quantum overlap mainly with one energy eigenstate. The effects of $\hat{H}_{re}$ can then be neglected and 
$:\hat{H}:$ in Eq. (\ref{equC9}) is diagonalized under a uniquely-defined transformation,
\begin{equation}
f_{q,c}^{\dag} \rightarrow {\tilde{f}}_{q,c}^{\dag}
\hspace{0.20cm}{\rm and}\hspace{0.20cm}
f_{q,c} \rightarrow {\tilde{f}}_{q,c} \, .
\label{equC10}
\end{equation}
This gives the Hamiltonian expression in Eq. (\ref{equ11}). 
It is valid under the $\xi_c\rightarrow {\tilde{\xi}}_c$ transformation for 
$\xi_c \in [\xi_c^0,\sqrt{2}[$ and $\xi_c^0 = \sqrt{1 + W_c^p/4t}$. This
refers to $u$ and $n_e$ values within the unitary-limit MQIM-HO 
regime for which the expression provided in Eq. (\ref{equ12}) for ${\tilde{\varepsilon}}_c (q)$ is a very good approximation.
The corresponding expression of ${\tilde{v}}_c (q) = \partial {\tilde{\varepsilon}}_c (q)/\partial q$ is
\begin{eqnarray}
{\tilde{v}}_c (q) & = & \left(1 + \beta_c\right)v_c (q) \hspace{0.20cm}{\rm for}\hspace{0.20cm} 
q \in [-2k_F,2k_F]
\nonumber \\
& = &  \left(1 + \beta_c \left\{1 - {8t\over W_c^h} \left({\varepsilon_c (q)\over W_c^h}\right)\right\}\right)v_c (q) 
\nonumber \\
& & \hspace{2.25cm}{\rm for}\hspace{0.20cm} \vert q\vert \in [2k_F,\pi] \, .
\label{equC11}
\end{eqnarray}
Here $v_c (q) = \partial \varepsilon_{c} (q)/\partial q$ and $\beta_c$ in Eqs. (\ref{equ12}) and (\ref{equC11}) 
is given in Eq. (\ref{equ13}), as justified below in Sec. \ref{APPC3}.
The energy bandwidth ${\tilde{W}}_c = {\tilde{\varepsilon}}_c (\pm\pi) - {\tilde{\varepsilon}}_c (0)=4t$
remains invariant under the $\xi_c\rightarrow{\tilde{\xi}}_c$ transformation whereas
${\tilde{W}}_c^h = {\tilde{\varepsilon}}_c (\pm\pi)$ and 
${\tilde{W}}_c^p = - {\tilde{\varepsilon}}_c (0)$ slightly decrease and increase, respectively, relatively to their 
${\tilde{\delta}}_c = 0$ bare value.

The controlled approximation used in Appendix \ref{APPB} to derive the energy dispersion 
$\varepsilon_c^1 (q)$, Eq. (\ref{equB19}) of that Appendix, is used to obtain the 
expressions for the $c$ band energy dispersion given in Eq. (\ref{equ12}) and below for other
related $c$ particle representations. It ensures that the corresponding group velocity has the correct 
TLL value at the $c$ band Fermi points, $q=\pm 2k_F$, accounts for the $c$ band energy bandwidth 
reading $4t$ for all such representations, and the $c$ band velocity vanishing both at 
$q=0$ and $q=\pm\pi$ for $u>0$. 

That in the case of the expressions, Eqs. (\ref{equ12}) and (\ref{equC11}),
the corresponding Fermi velocity at $q=2k_F$ increases relative to $v_{Fc}$
under the transformation is behind they being a good approximation provided that $\xi_c \in [\xi_c^0,\sqrt{2}[$ in the 
$\xi_c\rightarrow {\tilde{\xi}}_c$ transformation. In all remaining cases considered in the
following, the Fermi velocity at $q=2k_F$ decreases relative to $v_{Fc}$. The corresponding $c$ dispersion 
and velocity expressions are then a very good approximation for all $\xi_c$ and ${\tilde{\xi}}_c$ intervals.

\subsection{$c$ and $s$ representations for the Fourier transforms of three related potentials}
\label{APPC2}

The three $c$ particle representations associated with the potentials
$V_{re} (r)$, ${\tilde{V}}_c^1 (x)$, $V_c (x)$ considered in Secs. \ref{SECIVB} and \ref{SECIVC}
and their Fourier transforms ${\cal{V}}_{re} (k)$, ${\tilde{\cal{V}}}_c^1 (k)$, ${\cal{V}}_c (k)$, respectively, 
are such that,
\begin{eqnarray}
\lim_{{\tilde{\delta}}_c\rightarrow 0}{\tilde{V}}_c^1 (x) & = & V_c^1 (x)
\nonumber \\
\lim_{{\tilde{\delta}}_c\rightarrow 0}V_c (x) & = & \lim_{{\tilde{\delta}}_c\rightarrow 0} V_{re} (r) = 0
\hspace{0.20cm}{\rm and}\hspace{0.20cm}{\rm thus}  
\nonumber \\
\lim_{{\tilde{\delta}}_c\rightarrow 0}{\tilde{\cal{V}}}_c^1 (k) & = & {\cal{V}}_c^1 (k)
\nonumber \\
\lim_{{\tilde{\delta}}_c\rightarrow 0}{\cal{V}}_c (k) & = & \lim_{{\tilde{\delta}}_c\rightarrow 0} {\cal{V}}_{re} (k) = 0 \, .
\label{equC12}
\end{eqnarray}
Here ${\cal{V}}_c^1 (k)$ is the Fourier transform, Eq. (\ref{equB25}) of Appendix \ref{APPB}, 
of the ${\tilde{\delta}}_c =0$ bare potential $V_c^1 (x)$. Such Fourier transforms can be 
expanded in powers of $k$ as
\begin{eqnarray}
{\check{\cal{V}}} (k) & = & \int_{-\infty}^{\infty} dz\,e^{-i k z}\,V (\vert z\vert) 
= 2\int_{0}^{\infty} dy\,e^{-i k y}\,V (y) 
\nonumber \\
& = & \sum_{l=0}^{\infty} k^{2l} \,{\check{\cal{C}}}_l \hspace{0.20cm}{\rm where}
\nonumber \\
{\check{\cal{C}}}_l & = & {2 (-1)^l\over (2l)!}\int_{0}^{\infty} dy\,y^{2l}\,V (y) \, .
\label{equC13}
\end{eqnarray}
For small $k$ they are thus given by
\begin{equation}
{\check{\cal{V}}} (k) = {\check{\cal{V}}} (0) + {\cal{O}} (k^2) \, .
\label{equC14}
\end{equation}
The general notation used in Eqs. (\ref{equC13}) and (\ref{equC14}) for 
the three $c$ particle representations is such that
\begin{eqnarray}
V (y) & = & V_{re} (r)\vert_{r=y} \, ,\hspace{0.10cm}{\check{\cal{C}}} = {\cal{C}}_{re,l}
\hspace{0.10cm}{\rm for}\hspace{0.10cm}{\check{\cal{V}}} (k)={\cal{V}}_{re} (k) 
\nonumber \\
V (y) & = & {\tilde{V}}_c^1 (x)\vert_{x=y} \, ,\hspace{0.10cm}{\check{\cal{C}}}_l = {\tilde{\cal{C}}}_{c,l}^1
\hspace{0.10cm}{\rm for}\hspace{0.10cm}{\check{\cal{V}}} (k)={\tilde{\cal{V}}}_c^1 (k) \, ,
\nonumber \\
V (y) & = & V_c (x)\vert_{x=y} \, ,\hspace{0.10cm}{\check{\cal{C}}} = {\cal{C}}_{c,l}
\hspace{0.10cm}{\rm for}\hspace{0.10cm}{\check{\cal{V}}} (k)={\cal{V}}_c (k) \, ,
\label{equC15}
\end{eqnarray}
respectively. The Hamiltonian in the one-electron subspace, Eq. (\ref{equC9}), has expressions in all three $c$ particle 
representations of general form,  
\begin{eqnarray}
:\hat{H}: & = & :\hat{H}_{c}: + :\hat{H}_{s}:
\nonumber \\
:\hat{H}_{c}: & = & \sum_{q=-\pi}^{\pi}{\check{\varepsilon}}_c (q):{\check{f}}_{q,c}^{\dag}{\check{f}}_{q,c}: 
\nonumber \\
& + & {1\over L}\sum_{k,q,q'} {\check{\cal{V}}} (k)
{\check{f}}_{q,c}^{\dag}{\check{f}}_{q+k,c}\,{\check{f}}_{q',c}{\check{f}}_{q'-k,c}^{\dag}
+ \check{H}_{re}
\nonumber \\
:\hat{H}_{s}: & = & \sum_{q'=-k_F}^{k_F}\varepsilon_s (q'):f_{q',s}^{\dag}f_{q',s}: \, .
\label{equC16}
\end{eqnarray}
Here,
\begin{eqnarray}
{\check{f}}_{q,c}^{\dag} & = & f_{q,c}^{\dag} \, ; \hspace{0.10cm}
{\check{\varepsilon}}_c (q) = \varepsilon_c (q)\hspace{0.10cm}{\rm for}\hspace{0.10cm}{\check{\cal{V}}} (k)={\cal{V}}_{re} (k) 
\nonumber \\
{\check{f}}_{q,c}^{\dag} & = & {\bar{f}}_{q,c}^{\dag} \, ; \hspace{0.10cm}
{\check{\varepsilon}}_c (q) = {\tilde{\varepsilon}}_c^1 (q)\hspace{0.10cm}{\rm for}\hspace{0.10cm}{\check{\cal{V}}} (k)={\tilde{\cal{V}}}_c^1 (k) 
\nonumber \\
{\check{f}}_{q,c}^{\dag} & = & {\breve{f}}_{q,c}^{\dag} \, ; \hspace{0.10cm}
{\check{\varepsilon}}_c (q) = {\breve{\varepsilon}}_c (q)\hspace{0.10cm}{\rm for}\hspace{0.10cm}{\check{\cal{V}}} (k)={\cal{V}}_c (k) \, ,
\label{equC17}
\end{eqnarray}
and $\check{H}_{re}$ refers to $\hat{H}_{re}$, $\bar{H}_{re}^1$, and $\breve{H}_{re}$, respectively.

In all three $c$ particle representations, the Hamiltonian expressions, Eq. (\ref{equC16}), 
can be diagonalized in the singularities subspace under uniquely defined transformations,
\begin{equation}
{\check{f}}_{q,c}^{\dag} \rightarrow {\tilde{f}}_{q,c}^{\dag}
\hspace{0.20cm}{\rm and}\hspace{0.20cm}
{\check{f}}_{q,c} \rightarrow {\tilde{f}}_{q,c} \, ,
\label{equC18}
\end{equation}
that lead to the Hamiltonian in Eq. (\ref{equ11}). 

The $c$ particle energy dispersions ${\check{\varepsilon}}_c (q)$ in Eq. (\ref{equC16}) and corresponding group velocities ${\check{v}}_c (q)$
can be expressed in terms of the corresponding bare ${\tilde{\delta}}_c =0$ quantities as
\begin{eqnarray}
{\check{\varepsilon}}_c (q) & = & {(1 + \beta_c)\over (1 + {\check{\beta}}_c)}\,\varepsilon_c (q) \hspace{0.20cm}{\rm for}\hspace{0.20cm} 
q \in [-2k_F,2k_F]
\nonumber \\
& = & {{\check{W}}_c^h\over 8t}{(1 + \beta_c)\over({\check{\beta}}_c -  \beta_c)}\times \{{(1+{\check{\beta}}_c)\over(1 + \beta_c)}\, {\check{W}}_c^h
\nonumber \\
& - & \sqrt{\left({(1+{\check{\beta}}_c)\over(1 + \beta_c)}\,{\check{W}}_c^h\right)^2 - 16t\,{({\check{\beta}}_c -  \beta_c)\over(1 + \beta_c)}\,\varepsilon_c (q)}\}
\nonumber \\
& & \hspace{1.50cm}{\rm for}\hspace{0.20cm} \vert q\vert \in [2k_F,\pi] \, .
\label{equC19}
\end{eqnarray}
\begin{eqnarray}
{\check{v}}_c (q) & = & {(1 + \beta_c)\over (1 + {\check{\beta}}_c)}\,v_c (q) \hspace{0.20cm}{\rm for}\hspace{0.20cm} 
q \in [-2k_F,2k_F]
\nonumber \\
& = &  {{\check{W}}_c^h\over \sqrt{\left({(1+{\check{\beta}}_c)\over(1 + \beta_c)}\,{\check{W}}_c^h\right)^2 
- 16t\,{({\check{\beta}}_c -  \beta_c)\over(1 + \beta_c)}\,\varepsilon_c (q)}}\,v_c (q) 
\nonumber \\
& & \hspace{2.25cm}{\rm for}\hspace{0.20cm} \vert q\vert \in [2k_F,\pi] \, .
\label{equC20}
\end{eqnarray}
Here ${\check{W}}_c^h = {\check{\varepsilon}}_c (\pi)$, ${\tilde{\beta}}_c^1$ and ${\breve{\beta}}_c$ are obtained in the following, 
and $\beta_c$ is given by Eq. (\ref{equ13}), as confirmed below in Sec. \ref{APPC3} .
That ${\check{\beta}}_c = \beta_c$ and ${\check{W}}_c^h = W_c^h$ for ${\check{\cal{V}}} (k)={\cal{V}}_{re} (k)$
ensures that ${\check{\varepsilon}}_c (q)$ and ${\check{v}}_c (q)$ in Eqs. (\ref{equC19}) and
(\ref{equC20}) are the ${\tilde{\delta}}_c =0$ bare quantities
$\varepsilon_c (q)$ and $v_c (q)$, respectively.  
 
Three TLL Hamiltonians with the following general form describe in the low-energy limit 
$:\hat{H}_{c}:$ in Eq. (\ref{equC16}),
\begin{eqnarray}
&& :\hat{H}_{c}: = {\pi \,{\check{v}}_{Fc} \over L}\sum_{k,\iota = \pm}:{\check{\sigma}}_{c,\iota} (\iota k){\check{\sigma}}_{c,\iota} (-\iota k): 
\nonumber \\
& &  \hspace{0.85cm}+\,{1\over L}\sum_{k,\iota = \pm} (-{\check{\cal{V}}} (k))\times
\nonumber \\
& & (:{\check{\sigma}}_{c,\iota} (\iota k) {\check{\sigma}}_{c,\iota} (-\iota k):
+ {\check{\sigma}}_{c,\iota} (\iota k){\check{\sigma}}_{c,-\iota} (-\iota k)) 
\hspace{0.20cm}{\rm where}
\nonumber \\
& & {\check{\sigma}}_{c,\iota} (k) = \sum_p {\check{f}}_{\iota 2k_F + p + k,c}^{\dag} {\check{f}}_{\iota 2k_F + p,c} \, .
\label{equC21}
\end{eqnarray}
Here $k$ and $p$ run in the intervals given in Table \ref{table3},
the $c$ particle operators are provided in Eq. (\ref{equC17}), and
${\check{v}}_{Fc} = {\check{v}}_{c} (2k_F)$ refers for the three representations 
to $v_{Fc} = v_c (2k_F)$, ${\tilde{v}}_{Fc}^1 = {\bar{v}}_c^1 (2k_F)$, and ${\breve{v}}_{Fc} = {\breve{v}}_c (2k_F)$.

In the low-energy limit, the transformations, Eq. (\ref{equC18}), are such that
\begin{eqnarray}
& & {\tilde{\sigma}}_{c,\iota} (k) = e^{{\check{S}}}\,{\check{\sigma}}_{c,\iota} (k)\,e^{-{\check{S}}} 
\hspace{0.20cm}{\rm where}
\nonumber \\
& & {\check{S}} = {2\pi\over L}\sum_{k>0}{\ln (1 + {\check{\alpha}} (k))\over 4k} \times 
\nonumber \\
& & ({\check{\sigma}}_{c,+} (k){\check{\sigma}}_{c,-} (-k)
- {\check{\sigma}}_{c,-} (k) {\check{\sigma}}_{c,+} (-k)) \, ,
\label{equC22}
\end{eqnarray}
and ${\check{S}}$ stands for $S_{re}$, ${\tilde{S}}_c^1$, and ${\breve{S}}_c$, respectively.
Known TLL procedures then lead for very small $k$ to
\begin{eqnarray}
{\check{\cal{V}}} (k) & = & {\check{\iota}}\,{\pi\over 2}\,{\check{\alpha}} (k)\,{\check{v}}_{Fc} 
\hspace{0.20cm}{\rm where}
\nonumber \\
{\check{\alpha}} (k) & = & {\check{\alpha}} + {\cal{O}} (k^2)
\hspace{0.20cm}{\rm and}\hspace{0.20cm}{\check{\alpha}} = {\check{\alpha}} (0) \, ,
\label{equC23}
\end{eqnarray}
and
\begin{equation}
\sqrt{1 + {\check{\alpha}}} = 1 + {\check{\beta}} = {{\tilde{v}}_{Fc}\over {\check{v}}_{Fc}}
\hspace{0.20cm}{\rm and}\hspace{0.20cm}
{\check{\beta}} = {{\tilde{v}}_{Fc} - {\check{v}}_{Fc}\over {\check{v}}_{Fc}} \, ,
\label{equC24}
\end{equation}
for ${\tilde{\delta}}_c \in [0,{\tilde{\delta}}_c^{(1)}[\,;]{\tilde{\delta}}_c^{(1)},{\tilde{\delta}}_c^{({1\over 2})}[$.
Here ${\check{\alpha}} (k)$ stands for $\alpha_{re} (k)$, ${\tilde{\alpha}}_c^1 (k)$, and ${\breve{\alpha}}_c (k)$
and ${\check{\iota}}$ reads ${\check{\iota}}=-1$ for ${\tilde{\alpha}}_c^1$ and ${\breve{\alpha}}_c$ and ${\check{\iota}}=+1$
for $\alpha_c$. One then finds for ${\tilde{\cal{V}}}_c^1 (k)$ and ${\cal{V}}_c (k)$,
\begin{eqnarray}
{\tilde{\alpha}}_c^1 & = & {4- {\tilde{\xi}}_c^4\over {\tilde{\xi}}_c^4} \, ; \hspace{0.20cm}
{\tilde{\beta}}_c^1 = {{\tilde{v}}_{Fc} - {\tilde{v}}_{Fc}^1\over {\tilde{v}}_{Fc}^1} = {2 - {\tilde{\xi}}_c^2\over {\tilde{\xi}}_c^2}
\nonumber \\
{\breve{\alpha}}_c & = & {\xi_c^4 - {\tilde{\xi}}_c^4\over {\tilde{\xi}}_c^4} \, ; \hspace{0.20cm}
{\breve{\beta}}_c = {{\tilde{v}}_{Fc} - {\breve{v}}_{Fc}\over {\breve{v}}_{Fc}} = {\xi_c^2 - {\tilde{\xi}}_c^2\over {\tilde{\xi}}_c^2}
\nonumber \\
& & {\rm for}\hspace{0.20cm}{\tilde{\delta}}_c \in [0,{\tilde{\delta}}_c^{(1)}[\,;]{\tilde{\delta}}_c^{(1)},{\tilde{\delta}}_c^{({1\over 2})}[ \, .
\label{equC25}
\end{eqnarray}

Combining the above results gives
\begin{equation}
{{\tilde{v}}_{Fc}\over {\tilde{v}}_{Fc}^0} = {{\tilde{\xi}}_c^2\over 2} \, ;
\hspace{0.20cm}
{{\tilde{v}}_{Fc}\over {\tilde{v}}_{Fc}^1} = {2\over{\tilde{\xi}}_c^2} \, ;
\hspace{0.20cm}
{{\breve{v}}_{Fc}\over {\tilde{v}}_{Fc}^1} = {2\over\xi_c^2} \, ;
\hspace{0.20cm}
{{\tilde{v}}_{Fc}\over {\breve{v}}_{Fc}} = \left({\xi_c\over {\tilde{\xi}}_c}\right)^2 \, ,
\label{equC26}
\end{equation}
where ${\tilde{v}}_{Fc}^0$ such that 
${\tilde{v}}_{Fc} = \sqrt{{\tilde{v}}_{Fc}^0\times {\tilde{v}}_{Fc}^1}$
controls the renormalization of the compressibility
in Eq. (\ref{equ6}). Hence,
\begin{eqnarray}
{{\tilde{v}}_{Fc}\over {\tilde{v}}_{Fc}^1} = 
{{\breve{v}}_{Fc}\over {\tilde{v}}_{Fc}^1} & \times & {{\tilde{v}}_{Fc}\over {\breve{v}}_{Fc}} 
\hspace{0.20cm}{\rm and}\hspace{0.20cm}
{2\over {\tilde{\xi}}_c^2} = {2\over\xi_c^2}\times \left({\xi_c\over {\tilde{\xi}}_c}\right)^2 
\nonumber \\
{\rm for} & & {\tilde{\delta}}_c \in [0,{\tilde{\delta}}_c^{(1)}[\,;]{\tilde{\delta}}_c^{(1)},{\tilde{\delta}}_c^{({1\over 2})}[ 
\hspace{0.20cm}{\rm and}
\nonumber \\
{{\breve{v}}_{Fc}\over {\tilde{v}}_{Fc}^1}\vert_{{\tilde{\delta}}_c = {\tilde{\delta}}_c ^{\,\,\breve{}}} & = & 
{{\tilde{v}}_{Fc}\over {\breve{v}}_{Fc}}\vert_{{\tilde{\delta}}_c = {\tilde{\delta}}_c ^{\,\,\breve{}}}  = {2\over\xi_c^2} \, ,
\label{equC27}
\end{eqnarray}
where ${{\tilde{v}}_{Fc}\over {\breve{v}}_{Fc}} < {{\breve{v}}_{Fc}\over {\tilde{v}}_{Fc}^1}$
for ${\tilde{\delta}}_c < {\tilde{\delta}}_c ^{\,\,\breve{}}$,
${{\tilde{v}}_{Fc}\over {\breve{v}}_{Fc}} > {{\breve{v}}_{Fc}\over {\tilde{v}}_{Fc}^1}$
for ${\tilde{\delta}}_c > {\tilde{\delta}}_c ^{\,\,\breve{}}$,
and ${\tilde{\delta}}_c^{\,\,\breve{}}$ and ${\tilde{\xi}}_c^{\,\,\breve{}}$ are defined in Eq. (\ref{equ10}).

In the singularities subspace the Hamiltonians read
\begin{eqnarray}
:\hat{H}: & = & :\hat{H}_{c}: + :\hat{H}_{s}:
\nonumber \\
:\hat{H}_{c}: & = & \sum_{q=-\pi}^{\pi}{\check{\varepsilon}}_c (q):{\check{f}}_{q,c}^{\dag}{\check{f}}_{q,c}: + \hat{V}_{c,\gamma}
\nonumber \\
\hat{V}_{c,-1} & = & {1\over L}\sum_{\iota =\pm1}\sum_{k,p,q} {\check{\cal{V}}} (k)
\nonumber \\
& \times & {\check{f}}_{\iota 2k_F + p,c}^{\dag}{\check{f}}_{\iota 2k_F + p+k,c}\,{\check{f}}_{q,c}{\bar{f}}_{q-k,c}^{\dag}
\nonumber \\
\hat{V}_{c,+1} & = & {1\over L}\sum_{\iota =\pm1}\sum_{k,p,q} {\check{\cal{V}}} (k)
\nonumber \\
& \times & {\check{f}}_{\iota 2k_F + p+k,c}{\check{f}}_{\iota 2k_F + p,c}^{\dag}\,{\bar{f}}_{q-k,c}^{\dag}{\check{f}}_{q,c}
\nonumber \\
:\hat{H}_{s}: & = & \sum_{q'=-k_F}^{k_F}\varepsilon_s (q'):f_{q',s}^{\dag}f_{q',s}: \, .
\label{equC28}
\end{eqnarray}
Here the $k$, $p$, $q$ summations in $\hat{V}_{c,\gamma}$ run in 
intervals provided in Table \ref{table3} and ${\check{\cal{V}}} (k)$ is
given in Eq. (\ref{equC23}).

\subsection{Enhancement parameters associated with the potential $V_{re} (r)$ that controls
the renormalization of the one-electron spectrum}
\label{APPC3}

Diagonalizing the low-energy TLL Hamiltonians, Eq. (\ref{equC21}), under the 
transformations, Eq. (\ref{equC22}), gives
\begin{eqnarray}
:\hat{H}_{c}: & = & {\pi \,{\tilde{v}}_{Fc}\over L}\sum_{k,\iota}:{\tilde{\sigma}}_{c,\iota} (\iota k){\tilde{\sigma}}_{c,\iota} (-\iota k): 
\hspace{0.20cm}{\rm where}
\nonumber \\
& & {\tilde{\sigma}}_{c,\iota} (k) = \sum_p {\tilde{f}}_{\iota 2k_F + p + k,c}^{\dag} {\tilde{f}}_{\iota 2k_F + p,c} \, .
\label{equC29}
\end{eqnarray}
This is equivalent in the low-energy limit to the $c$ term of the Hamiltonian in the singularities subspace, Eq. (\ref{equ11}). 

At ${\tilde{\delta}}_c = 0$ the following boundary conditions hold,
\begin{eqnarray}
{\tilde{v}}_{Fc} & = & {\breve{v}}_{Fc} = v_{Fc} \hspace{0.20cm}{\rm and}\hspace{0.20cm}
{\tilde{v}}_{Fc}^0 = v_{Fc}^0 = {2\over\xi_c^2}\,v_{Fc}
\nonumber \\
{\tilde{v}}_{Fc}^1 & = & v_{Fc}^1 = {\xi_c^2\over 2}\,v_{Fc}
\hspace{0.20cm}{\rm at}\hspace{0.20cm}{\tilde{\delta}}_c = 0 \, .
\label{equC30}
\end{eqnarray}

Upon increasing ${\tilde{\delta}}_c$ within the interval ${\tilde{\delta}}_c \in [0,{\tilde{\delta}}_c^{\,\,\breve{}}-\delta]$
where ${\tilde{\delta}}_c^{\,\,\breve{}}$ is defined in Eq. (\ref{equ10}) and $\delta \ll {\tilde{\delta}}_c^{\,\,\breve{}}$, 
the velocity ${\tilde{v}}_{Fc}$ increases whereas the velocities ${\breve{v}}_{Fc}$ and ${\tilde{v}}_{Fc}^1$ remain 
unchanged. The exact relations, Eq. (\ref{equC26}), then imply that the velocity ${\tilde{v}}_{Fc}^0$ increases
and the following expressions in terms of the ${\tilde{\delta}}_c$-independent velocity $v_{Fc}$ apply, 
\begin{eqnarray}
{\tilde{v}}_{Fc} & = & \left({\xi_c\over {\tilde{\xi}}_c}\right)^2 v_{Fc} 
\hspace{0.20cm}{\rm and}\hspace{0.20cm}{\breve{v}}_{Fc} = v_{Fc} 
\nonumber \\
{\tilde{v}}_{Fc}^0 & = & {2\over {\tilde{\xi}}_c^2}\,\left({\xi_c\over {\tilde{\xi}}_c}\right)^2 v_{Fc} 
\nonumber \\
{\tilde{v}}_{Fc}^1 & = & v_{Fc}^1 = {\xi_c^2\over 2}\,v_{Fc}
\hspace{0.20cm}{\rm for}\hspace{0.20cm}{\tilde{\delta}}_c < {\tilde{\delta}}_c^{\,\,\breve{}} - \delta \, .
\label{equC31}
\end{eqnarray}

That the $c$ band energy bandwidth, $W_c = {\tilde{W}}_c = 4t$, remains invariant under the $\xi_c\rightarrow{\tilde{\xi}}_c$ 
transformation imposes constraints to the deformation caused to the $c$ band energy
dispersion by the finite-range interaction effects. The compressibility in Eq. (\ref{equ6}) is largest in the ${\tilde{\delta}}_c=0$ limit and 
tends to be suppressed by the finite-range interactions. The degree of the above deformation is limited by the
largest compressibility reached at ${\tilde{\delta}}_c = 0$, $\chi_{\rho}^0 = 2/(\pi n_e^2 v_{Fc}^0)$, through the 
inequality ${\tilde{v}}_{Fc} \leq v_{Fc}^0$. Here $v_{Fc}^0$ is the velocity defined in Eq. (\ref{equB27}) of Appendix \ref{APPB}.
That inequality limits the corresponding degree of enhancement of several renormalized quantities as follows,
\begin{eqnarray}
{\tilde{v}}_{Fc} & \leq & v_{Fc}^0\hspace{0.20cm}{\rm and}\hspace{0.20cm}{\rm thus}\hspace{0.20cm}
{{\tilde{v}}_{Fc}\over v_{Fc}} \leq \left({\xi_c\over {\tilde{\xi}}_c^{\,\,\breve{}}}\right)^2
= {2\over \xi_c^2} 
\nonumber \\
\alpha_c & \leq & {\xi_c^4 - ({\tilde{\xi}}_c^{\,\,\breve{}})^4\over ({\tilde{\xi}}_c^{\,\,\breve{}})^4} = {4-\xi_c^4\over\xi_c^4} 
\nonumber \\
\beta_c & \leq & {\xi_c^2 - ({\tilde{\xi}}_c^{\,\,\breve{}})^2\over ({\tilde{\xi}}_c^{\,\,\breve{}})^2} = {2-\xi_c^2\over\xi_c^2} \, ,
\label{equC32}
\end{eqnarray}
where ${\tilde{\xi}}_c^{\,\,\breve{}}$ is defined in Eq. (\ref{equ10}).

From the use of the $\xi_c$ behavior in Eq. (\ref{equA13}) of Appendix \ref{APPA} for $u\ll 1$,
one finds that in that limit the inequalities, Eq. (\ref{equC32}), imply the following inequality for 
$\int_0^{\infty}dr V_{re} (r)$ and ${\cal{V}}_{re} (k)$ at $k=0$, Eq. (\ref{equ9}), 
that {\it explicitly involves} the $c$ band energy bandwidth, $W_c = {\tilde{W}}_c = 4t$, 
\begin{equation}
\int_0^{\infty}dr V_{re} (r) = {1\over 2}{\cal{V}}_{re} (0) \leq {U\over W_c}\,t = {U\over 4} \hspace{0.20cm}{\rm for}\hspace{0.20cm}u\ll 1 \, .
\label{equC33}
\end{equation}
The physics behind this inequality is associated with properties of the potentials $V_{e} (r)$ 
and $V_{re} (r)$, such that $V_e (r)\propto U$ and $V_{re} (r)\propto U$. 
That inequality is directly controlled by the $u\rightarrow 0$ $c$ particle density of 
states at the Fermi level, ${\cal{D}}_c (\epsilon_F) = L/[2\pi\,t\,\sin \left({\pi n_e\over 2}\right)]$.
Indeed, in the $u\ll 1$ limit the $\alpha_c$ and $\beta_c$ inequalities in Eq. (\ref{equC32}) read
\begin{eqnarray}
\alpha_c & \leq & {U\over L}\,{\cal{D}}_c (\epsilon_F) = {U\over 2t\,\pi\sin \left({\pi n_e\over 2}\right)}
\hspace{0.20cm}{\rm and}
\nonumber \\
\beta_c & = & {{\tilde{v}}_{Fc} - v_{Fc}\over v_{Fc}} \leq 
{U\over 2L}\,{\cal{D}}_c (\epsilon_F) =
{U\over 4t\,\pi\sin \left({\pi n_e\over 2}\right)}
\nonumber \\
& & {\rm for}\hspace{0.20cm}u\ll 1 \, .
\label{equC34}
\end{eqnarray}

For the whole $u>0$ range, upon increasing ${\tilde{\delta}}_c$ within the interval 
${\tilde{\delta}}_c \in [0,{\tilde{\delta}}_c^{\,\,\breve{}}+\delta]$, 
where $\delta\ll {\tilde{\delta}}_c^{\,\,\breve{}}$,
the renormalized $c$ band Fermi velocity ${\tilde{v}}_{Fc}$ is enhanced from ${\tilde{v}}_{Fc} = v_{Fc}$ at
${\tilde{\delta}}_c =0$ to its maximum value, ${\tilde{v}}_{Fc} = (\xi_c/{\tilde{\xi}}_c^{\,\,\breve{}})^2 \,v_{Fc} = v_{Fc}^0$, 
at ${\tilde{\delta}}_c ={\tilde{\delta}}_c^{\,\,\breve{}}+\delta$. Upon further increasing ${\tilde{\delta}}_c$ 
above ${\tilde{\delta}}_c^{\,\,\breve{}}+\delta$, the velocity ${\tilde{v}}_{Fc}$ remains having
its maximum value. 

The exact relations provided in Eq. (\ref{equC26}) then impose that upon further increasing ${\tilde{\delta}}_c$ 
above ${\tilde{\delta}}_c^{\,\,\breve{}}+\delta$, the velocity ${\tilde{v}}_{Fc}^0$ further increases and
both the velocities ${\tilde{v}}_{Fc}^1$ and ${\breve{v}}_{Fc}$ decrease, the corresponding velocities expressions in terms 
of the ${\tilde{\delta}}_c$-independent velocity $v_{Fc}$ reading,
\begin{eqnarray}
{\tilde{v}}_{Fc}^0 & = & \left({2\over\xi_c\,{\tilde{\xi}}_c}\right)^2 v_{Fc} 
\nonumber \\
{\tilde{v}}_{Fc} & = & {2\over\xi_c^2}\,v_{Fc} \hspace{0.20cm} {\rm and}
\hspace{0.20cm} {\breve{v}}_{Fc} = {2\over\xi_c^2}\left({{\tilde{\xi}}_c\over\xi_c}\right)^2 v_{Fc}
\nonumber \\
{\tilde{v}}_{Fc}^1 & = & \left({{\tilde{\xi}}_c\over\xi_c}\right)^2 v_{Fc}  
\hspace{0.20cm}{\rm for}\hspace{0.20cm}{\tilde{\delta}}_c >{\tilde{\delta}}_c^{\,\,\breve{}} + \delta \, .  
\label{equC35}
\end{eqnarray}

One finds two regimes, determined by the initial bare charge parameter $\xi_c$ value in the $\xi_c\rightarrow{\tilde{\xi}}_c$
transformation,
\begin{eqnarray}
{\rm Regime\hspace{0.10cm}1} \rightarrow \xi_c \in [2^{1\over 4},\sqrt{2}[
\hspace{0.15cm}{\rm and}\hspace{0.15cm}
{\rm Regime\hspace{0.10cm}2} \rightarrow \xi_c \in ]1,2^{1\over 4}] \, .
\nonumber
\end{eqnarray}

The velocity relations given in Eqs. (\ref{equC30}), (\ref{equC31}), and  (\ref{equC35}) then uniquely determine that
\begin{eqnarray}
&& \alpha_c = \lim_{k\rightarrow 0}\alpha_c (k) = {\xi_c^4- {\tilde{\xi}}_c^4\over {\tilde{\xi}}_c^4}
\nonumber \\
&& \sqrt{1 + \alpha_c} = (1 + \beta_c) = {{\tilde{v}}_{Fc}\over v_{Fc}} = \left({\xi_c\over{\tilde{\xi}}_c}\right)^2
\nonumber \\
&& \beta_c = {{\tilde{v}}_{Fc} - v_{Fc}\over v_{Fc}} = {\xi_c^2 - {\tilde{\xi}}_c^2\over {\tilde{\xi}}_c^2} 
\nonumber \\
{\rm for} & & {\tilde{\delta}}_c \in [0,{\tilde{\delta}}_c^{\,\,\breve{}}-\delta]
\hspace{0.20cm}{\rm in}\hspace{0.20cm}{\rm regime\hspace{0.10cm}1}
\nonumber \\
{\rm for} & & {\tilde{\delta}}_c \in [0,{\tilde{\delta}}_c^{(1)}[\,;]{\tilde{\delta}}_c^{(1)},{\tilde{\delta}}_c^{\,\,\breve{}}-\delta]
\hspace{0.20cm}{\rm in}\hspace{0.20cm}{\rm regime\hspace{0.10cm}2} \, ,
\label{equC36}
\end{eqnarray}
\begin{eqnarray}
{\rm and} && \alpha_c = \lim_{k\rightarrow 0}\alpha_c (k) = {\xi_c^4 - ({\tilde{\xi}}_c^{\,\,\breve{}})^4\over ({\tilde{\xi}}_c^{\,\,\breve{}})^4} =
{4 - \xi_c^4\over \xi_c^4} 
\nonumber \\
&& \sqrt{1 + \alpha_c} = (1 + \beta_c) = {{\tilde{v}}_{Fc}\over v_{Fc}} = \left({\xi_c\over {\tilde{\xi}}_c^{\,\,\breve{}}}\right)^2 = {2\over \xi_c^2}
\nonumber \\
&& \beta_c = {{\tilde{v}}_{Fc} - v_{Fc}\over v_{Fc}} = {\xi_c^2 - ({\tilde{\xi}}_c^{\,\,\breve{}})^2\over ({\tilde{\xi}}_c^{\,\,\breve{}})^2} =
{2 - \xi_c^2\over \xi_c^2} 
\nonumber \\
{\rm for} & & {\tilde{\delta}}_c \in [{\tilde{\delta}}_c^{\,\,\breve{}}+\delta,{\tilde{\delta}}_c^{(1)}[\,;]{\tilde{\delta}}_c^{(1)},{\tilde{\delta}}_c^{({1\over 2})}[
\hspace{0.20cm}{\rm in}\hspace{0.20cm}{\rm regime\hspace{0.10cm}1}
\nonumber \\
{\rm for} & & {\tilde{\delta}}_c \in [{\tilde{\delta}}_c^{\,\,\breve{}}+\delta,{\tilde{\delta}}_c^{({1\over 2})}[
\hspace{0.20cm}{\rm in}\hspace{0.20cm}{\rm regime\hspace{0.10cm}2} \, .
\label{equC37}
\end{eqnarray}
The only property needed for our studies of the ${\tilde{\xi}}_c$ dependent quantities in the small interval 
${\tilde{\delta}}_c \in [{\tilde{\delta}}_c^{\,\,\breve{}} - \delta,{\tilde{\delta}}_c^{\,\,\breve{}} + \delta]$
is that their derivative with respect to ${\tilde{\xi}}_c$ has no discontinuity in it. 

Finally, we discuss the relation in the $u\ll 1$ limit of the parameter $\alpha_c$ in
Eqs. (\ref{equ9}), (\ref{equC36}), and (\ref{equC37}) to the {\it Coulomb enhancement parameter} 
considered in the studies of Ref. \onlinecite{Takahashi_84}. For $u\ll 1$, one has that $\delta\approx 0$ 
and $\alpha_c$ reads
\begin{eqnarray}
& & \alpha_c = 
{4\left(1 - {U\over L}\,{\cal{D}}_c (\epsilon_F)\right) - {\tilde{\xi}}_c^4\over {\tilde{\xi}}_c^4}
\hspace{0.20cm} {\rm for} \hspace{0.20cm} {\tilde{\xi}}_c \in [0,{\tilde{\delta}}_c^{\,\,\breve{}}]
\nonumber \\
& & = {U\over L}\,{\cal{D}}_c (\epsilon_F) \hspace{0.20cm} {\rm for} \hspace{0.20cm} 
{\tilde{\delta}}_c \in [{\tilde{\delta}}_c^{\,\,\breve{}},{\tilde{\delta}}_c^{(1)}[\,;
]{\tilde{\delta}}_c^{(1)},{\tilde{\delta}}_c^{({1\over 2})}[ \, .
\label{equC38}
\end{eqnarray}
Here ${\tilde{\delta}}_c^{\,\,\breve{}} = U/[4\sqrt{2}\,\pi\,t\sin\left({\pi n_e\over 2}\right)]$, 
${\tilde{\delta}}_c^{(1)} = \sqrt{2} - 1- {\tilde{\delta}}_c^{\,\,\breve{}}$, 
${\tilde{\delta}}_c^{({1\over 2})} = \sqrt{2} - {1/2} - {\tilde{\delta}}_c^{\,\,\breve{}}$,
and ${\cal{D}}_c (\epsilon_F) = L/[2\pi\,t\,\sin \left({\pi n_e\over 2}\right)]$.

The interval, ${\tilde{\xi}}_c \in [0,{\tilde{\delta}}_c^{\,\,\breve{}}]$,  
is extremely small, its width vanishing in the $u\rightarrow 0$ limit. 
In the complementary interval, ${\tilde{\delta}}_c \in [{\tilde{\delta}}_c^{\,\,\breve{}},{\tilde{\delta}}_c^{(1)}[\,;
]{\tilde{\delta}}_c^{(1)},{\tilde{\delta}}_c^{({1\over 2})}[$, of physical interest for TTF-TCNQ, the
parameter $\alpha_c$ that here controls changes in charge quantities 
reads $\alpha_c =  U/[2\pi\,t\,\sin \left({\pi n_e\over 2}\right)]$. 
Since for $u\rightarrow 0$ the charge and spin degrees of freedom recombine, in the $u\ll 1$ limit $\alpha_c$ also 
controls changes in spin quantities. Provided that $L$ is replaced by the Avogadro 
number $N_0$, it equals the Coulomb enhancement parameter 
$\alpha = U\,D (\epsilon_F)/N_0=U/[2\pi\,t\,\sin \left({\pi n_e\over 2}\right)]$ used to
study other properties of TTF-TCNQ in Ref. \onlinecite{Takahashi_84}.
[It should be distinguished from the SDS exponent $\alpha$ in Eq. (\ref{equ6}).]

The charge and spin quantities deviation effects result in the $u\ll 1$ limit from a small perturbation
to the ${\tilde{\delta}}_c =0$ bare model caused by a charge and spin probe, respectively.
Here it refers to the small $r>0$ potential $V_e (r) \propto U$,
the parameter $\alpha_c\ll 1$, Eq. (\ref{equC38}), controlling the resulting 
enhancement in the $u\ll 1$ limit of the $k=0$ Fourier transform 
${\cal{V}}_{re} (k) = {\pi\over 2}\,\alpha_c (k)\,v_{Fc}$, Eqs. (\ref{equ9}) and (\ref{equC33}).

On the other hand, the Coulomb enhancement parameter $\alpha = U/[2\pi\,t\,\sin \left({\pi n_e\over 2}\right)]\ll 1$ 
given in the nonnumbered equation appearing after Eq. (11) of Ref. \onlinecite{Takahashi_84} controls in the $u\ll 1$ limit 
the $k=0$ value $\kappa (0) = 1/\sqrt{1 + \alpha}$ of the factor $\kappa (k)$ in the Korringa relation.
That parameter is behind a slightly larger value of the $u\rightarrow 0$ Fermi velocity $v_F$ relative to the spin diffusion velocity 
$v_F^{SD}$ through a $u\ll 1$ relation, $v_F = \sqrt{1 + \alpha}\,v_F^{SD}$. 
Such a velocity deviation results from a small spin perturbation to the ${\tilde{\delta}}_c =0$ bare model 
that gives rise in the $u\ll 1$ limit to the Knight shift. 

For $u\ll 1$, such spin quantities can be expressed in terms of the parameter $\alpha_c$,
Eq. (\ref{equC38}), as $\kappa (0) = 1/\sqrt{1 + \alpha_c}$ and
$v_F = \sqrt{1 + \alpha_c}\,v_F^{SD}$, similarity to the present problem relation
${\tilde{v}}_{Fc} = \sqrt{1 + \alpha_c}\,v_{Fc}$.
The equality of the two parameters only holds though for $u\ll 1$.

%%%%%%%%%%%%%%%%%%%%%%%%%%%%%%%%%%%%%%%%%%%%%%%%%%%%%%%%%%%%%%%%%%%%%%%%%%

\end{document}